%% file: main.tex
\documentclass[11pt]{article}
\pdfoutput=1
\usepackage{caption}
\usepackage[titles]{tocloft}
\usepackage{jheppub}
\usepackage{makecell}
\usepackage[dvipsnames]{xcolor}
\usepackage{amsmath}
\usepackage{mathrsfs}

\usepackage{tikz}
\usetikzlibrary{decorations.pathmorphing,decorations.pathreplacing,calligraphy,calc,cd,external,arrows.meta,patterns}

\usepackage[export]{adjustbox}
\usepackage{hyperref}
\usepackage{bbm}
\usepackage{mathtools}
\usepackage{empheq}
\usepackage{nccmath}
\usepackage{blkarray}
\usepackage{esint}
\usepackage{microtype}
\usepackage{slashed}
\usepackage{physics}
\usepackage{amsfonts}
\usepackage{mathdots}
\usepackage{parskip}

\let\oldFootnote\footnote
\newcommand\nextToken\relax

\renewcommand\footnote[1]{%
    \oldFootnote{#1}\futurelet\nextToken\isFootnote}

\newcommand\isFootnote{%
    \ifx\footnote\nextToken\textsuperscript{,}\fi}
    
\hypersetup{
    pdfencoding=unicode,
	colorlinks=true,
	urlcolor=Maroon,
	linkcolor=RoyalBlue,
	citecolor=Maroon,
	pdftitle={Gravitational waveforms from restriction theory and rapid-decay homology},
	pdfauthor={Giacomo Brunello, Vsevolod Chestnov, Giulio Crisanti, Mathieu Giroux, Sid Smith},
	pdfdisplaydoctitle=true,
	pdfstartview=FitH,
	linktocpage=true
}

\definecolor{charcoal}{HTML}{343837}
\definecolor{maroon}{RGB}{128,0,0}

\def\D{\mathrm{D}}
\def\rd{\mathrm{d}}

\def\leq{\leqslant}

\def\Re{\mathrm{Re}\,}

\def\e{\mathrm{e}}
\def\d{\mathrm{d}}

\DeclareMathOperator*{\sumint}{%
\mathchoice%
  {\ooalign{$\displaystyle\sum$\cr\hidewidth$\displaystyle\int$\hidewidth\cr}}
  {\ooalign{\raisebox{.14\height}{\scalebox{.7}{$\textstyle\sum$}}\cr\hidewidth$\textstyle\int$\hidewidth\cr}}
  {\ooalign{\raisebox{.2\height}{\scalebox{.6}{$\scriptstyle\sum$}}\cr$\scriptstyle\int$\cr}}
  {\ooalign{\raisebox{.2\height}{\scalebox{.6}{$\scriptstyle\sum$}}\cr$\scriptstyle\int$\cr}}
}

\definecolor{darkgreen}{rgb}{0.0, 0.4, 0.0}

\usepackage[titles]{tocloft}
\usepackage{dirtytalk}
\usepackage{changepage}
\usepackage{subcaption}
\usepackage{float}
\usepackage{enumitem}  
\usepackage{stackengine}
\usepackage{bm}

 \DeclarePairedDelimiterXPP\EV[1]{E}(){}{
 
 #1}
 \DeclarePairedDelimiterXPP\Var[1]{V}(){}{
 
 #1}

\newcommand{\ampl}{\mathcal{M}}

\newcommand{\cW}{\mathcal{W}}
\def\unitn{\hat{\textbf{n}}}

\usepackage{cleveref}
\usepackage{nicematrix}
\newcommand{\namedref}[2]{\hyperref[#2]{#1~\ref*{#2}}}
\newcommand{\secref}[1]{\namedref{Section}{#1}}

\newcommand{\appref}[1]{\namedref{Appendix}{#1}}

\newcommand{\figref}[1]{\namedref{Figure}{#1}}

\definecolor{green1}{HTML}{3D792A}
\definecolor{cyan1}{HTML}{37cdaa}
\definecolor{blue1}{HTML}{5d7ac4}
\definecolor{red1}{HTML}{d0482a}
\definecolor{purple1}{HTML}{845ea8}
\definecolor{orange1}{HTML}{e07229}
\definecolor{yellow1}{HTML}{edcb52}
\definecolor{red}{HTML}{921818}
\definecolor{purple}{HTML}{53047A}
\definecolor{yellow}{HTML}{f4e097}
\definecolor{gr}{gray}{0.7}
\newcommand{\gr}[1]{{\color{gr}#1}}
\newcommand{\red}[1]{{\color{red}#1}}

\newcommand{\eps}{\varepsilon}
\newcommand{\mzero}{\gr{\cdot}}

\newcommand{\mZero}{\gr{0}}
\newcommand{\mId}{\mathbb{1}}

\newcommand{\vecJ}{\mathbf{J}}
\newcommand{\vecF}{\mathbf{F}}
\newcommand{\ord}{\mathrm{ord}}

\makeatletter
    \newcommand*\bigcdot{\mathpalette\bigcdot@{1}}
    \newcommand*\smtimes{\mathpalette\smtimes@{.7}}
    \newcommand*\bigcdot@[2]{\mathbin{\vcenter{\hbox{\scalebox{#2}{$\m@th#1\bullet$}}}}}
    \newcommand*\smtimes@[2]{\mathbin{\vcenter{\hbox{\scalebox{#2}{$\m@th#1\times$}}}}}
\makeatother

\usepackage[bb=boondox]{mathalfa}
\usepackage{amssymb}

\newcommand{\brk}[1]{(#1)}

\newcommand{\bigbrk}[1]{\bigl(#1\bigr)}
\newcommand{\Bigbrk}[1]{\Bigl(#1\Bigr)}

\newcommand{\sbrk}[1]{[#1]}

\newcommand{\brc}[1]{\{#1\}}

\newcommand{\supbrk}[1]{^{\brk{#1}}}

\usetikzlibrary{fadings}
\tikzfading[name=fade right,
right color=transparent!0,
left color=transparent!100]

\colorlet{choral}{RoyalBlue}
\colorlet{darkred}{Maroon}
\definecolor{neworange}{HTML}{E99F0C}
\colorlet{Orange}{neworange}
\colorlet{orange}{neworange}
\colorlet{darkorange}{neworange}
\tikzset{
    partial ellipse/.style args={#1:#2:#3}{
        insert path={+ (#1:#3) arc (#1:#2:#3)}
    }
}
\tikzset{ shorten <>/.style={ shorten >=#1, shorten <=#1 } }

\tikzset{gradRtoB/.style={
    postaction={
        decorate,
        decoration={
            markings,
            mark=at position \pgfdecoratedpathlength-0.5pt with {\arrow[blue,line width=#1] {}; },
            mark=between positions 0 and \pgfdecoratedpathlength-0pt step 0.5pt with {
                \pgfmathsetmacro\myval{multiply(divide(
                    \pgfkeysvalueof{/pgf/decoration/mark info/distance from start}, \pgfdecoratedpathlength),100)};
                \pgfsetfillcolor{RoyalBlue!\myval!Maroon!};
                \pgfpathcircle{\pgfpointorigin}{#1};
                \pgfusepath{fill};}
}}}}

\tikzset{
    vector/.style={
        decoration={snake, aspect=0.75, mirror, segment length=2mm},
        decorate
    },
	photon/.style={decorate, decoration={snake, amplitude=1pt, segment length=6pt}
	}
}
\title{Gravitational waveforms from restriction theory and rapid-decay homology}
\usepackage{orcidlink}
\definecolor{orcidlogocol}{named}{Maroon}
\author[\!a,b,c,d,\orcidlink{0009-0004-4788-738X}]{Giacomo Brunello,}\emailAdd{giacomo.brunello@sns.it}
\author[\!e,i,\orcidlink{0000-0001-7067-0315}]{Vsevolod Chestnov,}\emailAdd{vsevolod.chestnov@maths.ox.ac.uk}
\author[\!f,\orcidlink{0009-0009-3053-2394}]{Giulio Crisanti,}\emailAdd{g.crisanti@ed.ac.uk}
\author[\!g,h,\orcidlink{0000-0002-2672-634X}]{Mathieu Giroux,}\emailAdd{giroux@ias.edu}
\author[\,a,b,f,\orcidlink{0009-0007-7799-0136}]{\\ Sid Smith}\emailAdd{sid.smith@ed.ac.uk}
\affiliation{$^a$Dipartimento di Fisica e Astronomia, Universita di Padova, Via Marzolo 8, 35131 Padova, Italy}
\affiliation{$^b$INFN, Sezione di Padova,
Via Marzolo 8, I-35131 Padova, Italy.}
\affiliation{$^c$Institut de Physique Théorique, CEA, CNRS, Université Paris-Saclay, F-91191 Gif-sur-Yvette cedex, France}
\affiliation{$^d$Scuola Normale Superiore, Piazza dei Cavalieri 7, 56126, Pisa, Italy and INFN Sezione di Pisa, Largo
Pontecorvo 3, 56127 Pisa, Italy}
\affiliation{$^e$ Dipartimento di Fisica e Astronomia, Universit\`a di Bologna
    e INFN, Sezione di Bologna,
    via Irnerio 46, I-40126 Bologna, Italy.}
\affiliation{$^f$Higgs Centre for Theoretical Physics, University of Edinburgh, James Clerk Maxwell Building,Peter Guthrie Tait Road, Edinburgh, EH9 3FD, United Kingdom}
\affiliation{$^g$Department of Physics, Columbia University, 538 West 120th Street, New York, NY 10027, U.S.}
\affiliation{$^h$Institute for Advanced Study, Einstein Drive, Princeton, NJ 08540, USA}
\affiliation{$^i$Mathematical Institute, University of Oxford, OX2 6GG, UK}
\abstract{
We present a systematic framework for computing frequency-domain gravitational waveforms from relativistic binary scattering in different asymptotic regimes.
The method yields a controlled series expansion that can in principle be extended to arbitrary order in the relevant kinematic parameter.
By combining differential-equation techniques with restriction theory and algebraic-geometry methods for impact-parameter-space Fourier integrals, we derive recursion relations that generate the leading-order (tree-level) waveform in both the soft-emission and post-Newtonian regimes,
establishing a proof of principle for extending the approach to higher-loop computations.
Finally, following constraints from rapid-decay homology, we show that the Fourier integrals underlying the waveform satisfy $\varepsilon$-form differential equations mixing Bessel- and exponential-type kernels, marking a first step toward uncovering the analytic structure of the exact solution.
}

\begin{document} 

\addtocontents{toc}{\protect\thispagestyle{empty}}

\maketitle

\thispagestyle{empty}

\setcounter{page}{3}
\allowdisplaybreaks
\input{Sections/Introduction}

\input{Sections/KMOC}
\input{Sections/waveform_tree}
\input{Sections/Method}
\input{Sections/Soft_expansion}

\input{Sections/PN_expansion}
\input{Sections/Canonical_form}
\input{Sections/Conclusions}

\acknowledgments
We would like to thank Donato Bini, Stefano De Angelis, Claude Duhr, David A. Kosower, Pierpaolo Mastrolia, Saiei-Jaeyeong Matsubara-Heo, Jessen Munch, Tiziano Peraro, Andrzej Pokraka, Lorenzo Tancredi, and Raj Patil for comments on the draft and/or for valuable discussions at various stages of this work.
G.B. research is supported by the Italian MIUR under contract 20223ANFHR (PRIN2022), by the European Research Council, under grant
ERC-AdG-885414 Ampl2Einstein, by the Università Italo-Francese, under grant Vinci, and by the Amplitudes INFN scientific initiative.
V.C. research is supported by the European Research Council (ERC) under the European Union's Horizon Europe research and innovation program grant agreement 101040760 (ERC Starting Grant \emph{FFHiggsTop}), and by the ERC Synergy Grant MaScAmp 101167287. 
G.C. research is supported by the United Kingdom Research and Innovation grant UKRI FLF MR/Y003829/1.
M.G. is supported by the U.S. Department of Energy (DOE) grant No. DE-SC0011941.
S.S. research is partially supported by the Amplitudes INFN scientific initiative.
Views and opinions expressed are those of the authors only and do not necessarily reflect those of the European Union or the European Research Council. Neither the European Union nor the granting authority can be held responsible for them.

\appendix

\input{Sections/appA_BoundarySoft}
\input{Sections/appC_method}

\bibliographystyle{JHEP}
\bibliography{refs}

\end{document}

%% file: Sections/Introduction.tex
{\section{Introduction}\label{sec:introduction}}
Gravitational-wave physics is entering an era in which precision measurements are becoming possible.
In the ongoing fourth LIGO-Virgo-KAGRA observing run, roughly two hundred candidate gravitational-wave events have been reported cumulatively~\cite{LIGOScientific:2025hdt}.

Accurate waveform models (used to predict the amplitude and phase evolution of the radiation emitted by coalescing compact binaries such as black holes or neutron stars) play an increasingly central role in both the detection and interpretation of these signals. As detector sensitivities continue to improve, and especially with the advent of future observatories such as LISA~\cite{Caprini:2019egz}, the Einstein Telescope~\cite{Punturo:2010zz}, and Cosmic Explorer~\cite{Reitze:2019iox}, one naturally anticipates that gravitational-wave astronomy will progressively rely on waveform models of correspondingly higher accuracy.

Framing the relativistic massive binary (two-body) problem within quantum field theory has led to significant advances in recent years.
In this framework, the compact objects responsible for the observed signals are modeled as heavy, point-like particles (with or without spin) interacting through graviton exchange; an approximation that can be justified by a certain hierarchy between the mass and the characteristic size of each body as well as their orbital separation or impact parameter.

Different setups based on effective field theories (EFTs), scattering amplitudes, and worldline approaches have enabled progress both in the weak-field, slow-motion post-Newtonian (PN) regime and in the weak-field, relativistic post-Minkowskian (PM) regime, corresponding respectively to expansions in $v/c$ and in Newton's constant $G$.

In the PN regime, the synergy between EFT frameworks~\cite{Goldberger:2004jt,Goldberger:2009qd,%
Foffa:2013qca,Porto:2016pyg} and multi-loop Feynman calculus~\cite{Foffa:2016rgu}
enabled the computation of conservative Hamiltonians at 
$4$PN~\cite{Foffa:2019yfl,Foffa:2019rdf,Blumlein:2020pog} 
and $5$PN~\cite{Foffa:2019hrb,Blumlein:2021txe,Porto:2024cwd} orders. Similarly, in the PM regime, scattering-amplitude and 
worldline methods~\cite{Kalin:2020mvi,Mogull:2020sak} have led to the computation of 
relativistic dynamics at 
$2$PM~\cite{Cheung:2018wkq,Bjerrum-Bohr:2018xdl},
$3$PM~\cite{Bern:2019nnu,Bern:2019crd,
Parra-Martinez:2020dzs,Cheung:2020gyp,DiVecchia:2021bdo,Brandhuber:2021eyq,Kalin:2020fhe}, 
$4$PM~\cite{Dlapa:2021npj,Bern:2021dqo,Dlapa:2022lmu,%
Jakobsen:2023ndj,Jakobsen:2023hig,Damgaard:2023ttc}, 
and, more recently, $5$PM at first self-force order~\cite{
Bern:2023ccb,Driesse:2024xad,Bern:2024adl,%
Driesse:2024feo,Dlapa:2025biy}.

Merely all these developments are underpinned by major conceptual and technical progress in the computation of multi-loop Feynman integrals, notably through integration-by-parts (IBP) identities~\cite{Tkachov:1981wb,Chetyrkin:1981qh,%
Laporta:2000dsw,Smirnov:2008iw,Lee:2012cn,Maierhofer:2017gsa,Peraro:2019svx,Wu:2023upw,Smith:2025xes} as well as intersection theory~\cite{Mastrolia:2018uzb,Frellesvig:2019kgj,Frellesvig:2019uqt,Frellesvig:2020qot,Caron-Huot:2021xqj,Caron-Huot:2021iev,Chestnov:2022xsy,Fontana:2023amt,Brunello:2023rpq,Brunello:2024tqf} for integral reduction and the differential-equation (DE) method~\cite{Kotikov:1990kg,KOTIKOV1991123,Bern:1993kr,Gehrmann:1999as,Argeri:2007up} for their evaluation.
Among these, the reformulation of DEs into an $\varepsilon$-factorized form~\cite{Henn:2013pwa} stands out as a way to expose the analytic structure of the integral family by organizing the solutions as uniform-weight iterated integrals~\cite{Chen:1977oja,Goncharov:1998kja}, and provides in an increasing number of cases a systematic path to analytic or high-precision numerical evaluation (see, e.g.,~\cite{Chicherin:2020oor,Henn:2025xrc} for recent applications to five- and six-gluon two-loop amplitudes).

These impressive developments have extended the reach of perturbative methods well beyond particle physics, notably into gravitational-wave physics, where direct comparisons with numerical relativity now show agreement even in regions approaching the merger~\cite{Rettegno:2023ghr}.

While these comparisons validate perturbation theory deeper and deeper into the strong-field regime, a complementary line of research has pursued a first-principles derivation of gravitational waveforms from scattering-amplitude building blocks. In this formulation, the classical waveform is extracted not from the late-time dynamics of the binary source (as would be the case in an \emph{exclusive} or ``in-out'' scattering process) but from its \emph{inclusive} (``in-in'') on-shell scattering data.

In this way, the leading-order (tree-level) time-domain waveform, originally derived by Kovacs and Thorne~\cite{Kovacs:1978eu}, has since been reproduced within the observable-based Kosower--Maybee--O'Connell (KMOC) formalism~\cite{Kosower:2018adc,Cristofoli:2021vyo} and through analytic continuations of the five-point S-matrix~\cite{Caron-Huot:2023ikn}. Related derivations using worldline methods can be found in~\cite{Jakobsen:2021smu,Mougiakakos:2021ckm,Jakobsen:2021lvp,Riva:2021vnj}.

The evaluation of the time- and frequency-domain gravitational waveforms has since been extended to next-to-leading order (one-loop) in momentum 
space~\cite{Brandhuber:2023hhy,Herderschee:2023fxh,Georgoudis:2023lgf,Caron-Huot:2023vxl,Bohnenblust:2023qmy}, 
including both linear~\cite{Bohnenblust:2023qmy} and quadratic spin effects~\cite{Bohnenblust:2025gir}. 
These results have recently been cross-checked against the multipolar post-Minkowskian 
(PM) formalism~\cite{Blanchet:1985sp,Blanchet:1989ki,Blanchet:2013haa} after a suitable reorganization
of the amplitude-based expressions~\cite{Bini:2023fiz,Bini:2024rsy}.

To streamline the extraction of classical physics that we are interested in, it is common to Fourier-transform with respect to the momentum transfer $q$, trading $q$ for the impact parameter $b$; the closest relative transverse separation of the incoming bodies. This yields a mixed $(\omega,b)$ representation, which we will hereafter refer to as the ``frequency-domain waveform in impact-parameter space'' or simply the \emph{frequency-domain waveform}.

Carrying out this Fourier transform is, however, highly non-trivial (even at the lowest order in perturbation theory) and motivates the analytic and asymptotic methods developed in this work to compute the of the frequency-domain waveform. 

Previous amplitude-based computations~\cite{Jakobsen:2021smu,Mougiakakos:2021ckm,Jakobsen:2021lvp, Brandhuber:2023hhy,Herderschee:2023fxh,Georgoudis:2023lgf,Caron-Huot:2023vxl,Bohnenblust:2023qmy}
have typically treated the Fourier transform as a step separate from the reduction of the amplitude
to momentum-space master integrals. This separation leads to a proliferation of Fourier integrals
with non-trivial special-function kernels, obscuring the underlying structure and making the
integration challenging even numerically.

A different strategy was proposed in~\cite{Brunello:2024ibk,Brunello:2025todo}, building on the
algebraic properties of the Fourier transform~\cite{Brunello:2023fef}. In this approach, the
frequency-domain waveform is interpreted as a twisted period integral, and IBP
identities are derived directly for the Fourier integrals. In practice, a minimal joint basis of
such integrals is obtained via Fourier integrals IBPs and then evaluated using a hybrid method: the
loop-momentum (i.e., phase independent) sector is solved through differential equations, while the residual Fourier integrals
are computed ``brute force'' by contour deformation and direct integration.

While this approach provides analytic next-to-leading-order results~\cite{Brunello:2025todo}, its systematic extension to higher loops is expected to be considerably more challenging. The Fourier kernels become increasingly intricate, and no general-purpose evaluation algorithm is currently available. Moreover, extracting high-order asymptotics in specific kinematic limits (e.g., soft or PN) from fully relativistic expressions requires delicate region analyses and nested series expansions under the integral, whose complexity grows rapidly with loop order. Nonetheless, results from the soft and PN regimes indicate a remarkably simple underlying function space at next-to-next-to-leading order~\cite{Bini:2024ijq,Georgoudis:2025vkk}.

Motivated by these considerations, we develop a systematic method for computing the integrals
that enter the frequency-domain waveform.
The method combines analytic studies of the differential equations satisfied by the master integrals with the powerful framework of restriction theory~\cite{Chestnov:2023kww}, enabling asymptotic expansions (e.g., soft and post-Newtonian) to be extracted directly at the level of the differential system. 
A differential system in a given kinematic limit $z_1\to0$ can be studied following three steps:
\begin{enumerate}
    \item 
    \textit{Splitting into regions}: The system's \textit{residue matrix} encodes the behavior at $z_1\to 0$. \textit{Eigenvalues} denote different \textit{kinematic regions} of the solution. For each region, the dimension of each eigenspace denotes the number of master integrals (\textit{eigenvectors}) that contribute to the leading behavior in that region. 
    \item 
    \textit{Leading order solution}: The solution in the limit is obtained by solving simpler DE systems, the \textit{restrictions} of the original one to the eigenspaces of the
    various eigenvalue.
    \item 
    \textit{Higher order recursion}: Once the leading behavior is known, the full solution can be obtained recursively by matrix multiplication through algebraic relations arising from the $z_1$ differential system.
\end{enumerate}
This approach nicely minimizes the proliferation of complicated (and unnecessary!) Fourier kernels and is expected to generalize naturally to higher loops. At tree-level, we apply the method to obtain the PN expansion up to $\mathcal{O}(v^{30})$, confirming agreement with lower-order results in the literature~\cite{Bini:2024rsy}. A similar result is obtained in the soft graviton regime, where agreement with the literature has also been verified~\cite{Georgoudis:2023eke,Bini:2024rsy}.

To achieve these results, we additionally rely on several technical tools from algebraic geometry. 
First, we use a companion-matrix reformulation of the differential equations \cite{Brunello:2024tqf}, 
which enables fully algebraic recursive solutions. 
We further simplify square-root structures using Gröbner bases, 
as implemented in the package \texttt{SPQR}~\cite{Chestnov:2025todo}, 
and make extensive use of finite-field reconstruction 
techniques~\cite{vonManteuffel:2014ixa,Peraro:2016wsq,Peraro:2019svx}.

Complementarily, we derive the $\varepsilon$-form satisfied by the Fourier master
integrals. Guided by constraints from rapid-decay homology~\cite{Pham1985LaDD,MatsubaraHeo2017OnTR},
we identify Bessel functions and exponentials (along with integrals of thereof) in the
$\varepsilon$-form integration kernels. This suggests that the tree-level impact-parameter waveform can be expressed as an iterated integral over this class of functions, whose mathematical structure remains to be fully understood. To the best of our knowledge, this is the first appearance of such functions in the context of an $\varepsilon$-factorized differential equation, setting them apart from their polylogarithmic, elliptic, and Calabi--Yau counterparts observed in Feynman integrals.

\paragraph{Organization of the paper.}
We begin in \secref{sec:KMOC} with a review of the KMOC framework for computing
gravitational waveforms from scattering amplitudes.
In \secref{sec:waveform_tree}, we construct the leading-order frequency-domain waveform in terms of
Fourier master integrals and derive the corresponding system of differential equations.
The restriction method introduced in \secref{sec:method} is then used to solve this system in specific
kinematic limits.
Its application to the soft limit is presented in \secref{sec:soft_expansion}, where we find agreement
with existing results, and to the post-Newtonian regime in \secref{sec:PN_expansion}, where the expansion is pushed to orders far beyond the state-of-the art.
In \secref{sec:canonical} investigates the existence of an $\varepsilon$-form for the differential
system and identifies Bessel functions and exponentials in its solution's integration kernels. Finally, we collect complementary technical details in the various Appendices. The results obtained in this paper are also provided in a computer-readable format on the GitHub repository~\cite{repo}.

\paragraph{Conventions.} 
We use the mostly-minus metric $(+,-,-,-)$. The dimension of spacetime is denoted
by $\D=4-2\varepsilon$, with $\varepsilon$ being the usual dimensional regulator.
Finally, we use ``hat'' notation as follows to collect powers of $2\pi$ in the integration measure and delta-functions as:
\begin{equation}
    \hat{\dd}^\D q \equiv \frac{\dd^\D q}{(2\pi)^\D}\,, \quad
    \hat{\delta}^{n}(q) \equiv (2\pi)^{n} \delta^{n}(q)\,.
\end{equation}
We also distinguish between the spacetime dimension $\D = 4 - 2\varepsilon$ of the integration (loop) variables 
and the spin (or polarization) dimension $\D_s$. 
Using the dimensional-reduction scheme~\cite{Siegel:1979wq}, we keep all external kinematics strictly four-dimensional 
while contracting spin indices in a $\D_s$-dimensional metric; throughout, we take $\D_s = 4$ 
(see~\cite{Gnendiger:2017pys} for a review).

%% file: Sections/KMOC.tex
{\section{Gravitational waveforms from scattering amplitudes}\label{sec:KMOC}}
In this section we review the necessary background on the gravitation waveform and some of its relations to scattering amplitudes. We also spell out our kinematic conventions.
\subsection{Background review}
The gravitational waveform emitted by a binary system (e.g., of scattering black holes) can be computed from scattering amplitude techniques using the Kosower--Maybee--O’Connell (KMOC) formalism~\cite{Kosower:2018adc,Cristofoli:2021vyo}.  In this framework, physical observables can be defined as the difference of the expectation values of a quantum operator $\hat{\mathcal{O}}$ between the asymptotic final state $\vert \psi \rangle_{\rm out}$ and initial state $\vert \psi \rangle_{\rm in}$ of the process:
\begin{equation}\label{eq:difference}
     \langle \Delta\hat{\mathcal{O}} \rangle  \equiv \ _{\rm out}\langle \psi \vert \hat{\mathcal{O}}\vert \psi \rangle _{\rm out }- \ _{\rm in}\langle \psi \vert \hat{\mathcal{O}}\vert \psi \rangle _{\rm in } \, . 
\end{equation}
The initial state
\begin{equation}\label{eq:stateDef}
|\psi \rangle_{\rm in} \equiv
\int\prod_{i=1}^2 \dd\Phi(p_i) \phi_i(p_i)
\e^{i b_i \cdot p_i} | p_1, p_2 \rangle_{\rm in}\,,
\end{equation}
is defined as the usual superposition of two-particle momentum eigenstates
$|p_1, p_2\rangle_{\mathrm{in}}\equiv a^\dagger(p_2)a^\dagger(p_1)\ket{0}$,
weighted by the single-particle Lorentz-invariant on-shell phase-space measures
\begin{equation}\label{eq:LIPSdef}
\dd \Phi (p_i) \equiv
\theta(p_i^0)\,\hat{\delta}(p_i^2 - m_i^2)\,\hat{\d}^\D p_i
\qquad (i=1,2)\,.
\end{equation}
The wavepackets $\phi_i(p_i)$ are chosen so that the state $|\psi \rangle_{\rm in}$ is normalizable and represents two well-separated particles at past infinity.
In particular, $\phi_i(p_i)$ is peaked around some classical momentum $p_i\to P_i$ (i.e., have support only close to the physical mass shell $P_i^2 = m_i^2$).\footnote{For the formal/axiomatic details on the construction of such multi-particle Fock space, see \cite{Haag:1958vt,ruelle1961}.}
The vectors $b_i^\mu$ specify the spatial separation of the individual wavepackets, and their difference 
\begin{equation}\label{eq:impactPdef}
    b^\mu \equiv b_1^\mu - b_2^\mu\,,
\end{equation}
parametrizes the (spacelike) impact parameter of the collision. To get the final state, we then let the system evolve via the (unitary) $S$-matrix operator
\begin{equation}
   \vert  \psi \rangle_{\rm out}  \equiv S \, \vert  \psi \rangle_{\rm in} \, .
\end{equation}
The difference \eqref{eq:difference} in the operator expectation value can then be expressed as:
\begin{equation}
     \langle \Delta\hat{\mathcal{O}} \rangle  =  \ _{\rm in}\langle \psi \vert S^{\dagger}[\hat{\mathcal{O}},S]\vert \psi \rangle _{\rm in } \, \qquad (\text{since $SS^{\dagger}=\mathbb{1}$})\,. 
     \label{eq:expectation_value}
\end{equation}
Inserting the explicit expression for the initial states we finally get:
\begin{equation}
	\label{eq:quantum_KMOC_exp_value}
	\langle{\Delta \hat{\mathcal{O}}}\rangle = \!\int\! \prod_{i=1}^{2}\biggl[\dd \Phi(p_i)\dd \Phi(p_i') \ \phi_i(p_i) \phi_i^*(p_i^\prime) \
	\e^{i b_i \cdot\left(p_i-p_i^\prime\right)}\biggr] \
	\langle p_1^\prime, p_2^\prime | S^\dagger [\hat{\mathcal{O}}, S] | p_1, p_2 \rangle \ .
\end{equation}
\paragraph{Hierarchy in length scales and the classical limit of \eqref{eq:quantum_KMOC_exp_value}.}
Our main assumption is that the observer’s distance $r$ from the binary system is the largest length scale in the problem. 
Accordingly, we assume the hierarchy
\begin{equation}\label{eq:hierarchy}
\frac{1}{m_i} \ll G m_i \ll |b| \ll r\,,
\end{equation}
which organizes the characteristic length scales that control the scattering. The first inequality, $1/m_i \ll G m_i$, ensures that each body is much heavier than the Planck mass,
\begin{equation}
m_i \gg M_{\rm Pl} \equiv \frac{1}{\sqrt{G}} = \ell_{\rm Pl}^{-1}\,,
\end{equation}
so that its Compton wavelength is negligible compared with the Planck length, $1/m_i \ll \ell_{\rm Pl}$.\footnote{For contrast, there are two opposite regimes. 
If $m \ll M_{\rm Pl}$ (as for an electron), the Compton wavelength greatly exceeds the Schwarzschild radius $r_s \sim G m$: the region where the particle’s gravity would matter is far smaller than its intrinsic quantum uncertainty, so gravity is negligible. 
Conversely, if $m \sim M_{\rm Pl}$, then $1/m \sim r_s \sim \ell_{\rm Pl}$ and quantum fluctuations of both the source and spacetime become comparable, signaling the onset of the quantum-gravity regime.}
The inequality $1/m_i \ll G m_i$ therefore selects macroscopic bodies,  such as neutron stars or astrophysical black holes,  that can be consistently described as classical sources of gravity (even though they may exhibit semiclassical effects like Hawking radiation \cite{Hawking:1975vcx,Aoude:2024sve}).

The second inequality, $r_{s,i}\sim G m_i \ll |b|$, places the interaction firmly in the large-impact-parameter regime, ensuring that no common strong-field region or apparent horizon forms. Indeed, in this regime, the metric can be expressed as a small perturbation $h_{\mu\nu}$ around flat spacetime,
\begin{equation}\label{eq:metricPert}
    g_{\mu\nu} = \eta_{\mu\nu} + 
    \kappa
    \, h_{\mu\nu}\,, \qquad (\kappa=\sqrt{32\pi G})
\end{equation}
and even near the point of closest approach it remains small: 
$h_{\mu\nu} = \mathcal{O}\!\big(G (m_1 + m_2)/|b|\big)$.
The associated curvature-scale consistently behaves as
\begin{equation}
\mathcal{R} \sim \partial^2 h 
= \mathcal{O}\!\Bigg(\frac{G (m_1 + m_2)}{|b|^3}\Bigg)
\quad\Rightarrow\quad
\mathcal{R}\,\ell_{\rm Pl}^2 
\sim 
\frac{G(m_1 + m_2)}{|b|}
\left(\frac{\ell_{\rm Pl}}{|b|}\right)^{\!2} 
\ll 1\,.
\end{equation} 

As a consequence of \eqref{eq:hierarchy}, the source wavepackets $\phi_i(p_i)$ in \eqref{eq:quantum_KMOC_exp_value} thus behave as extended, coherent lumps whose internal phases remain aligned, each moving as a single classical body of spatial width $\Delta x_i \gg 1/m_i$ (as will be discussed below), without spreading due to quantum uncertainty.
In momentum space, each lump is sharply peaked around its classical momentum $\simeq m_i u_i$, where $u_i$ is the corresponding four-velocity.

To make this picture more quantitative, first note that a stationary-phase analysis of \eqref{eq:quantum_KMOC_exp_value} shows that the integral is dominated by momentum differences of order
\begin{equation}\label{eq:stationary}
q_i \equiv p_i - p_i' \sim \frac{1}{|b|}\,.
\end{equation}
The stationary-phase condition fixes the region of $q_i$ that contributes dominantly through the oscillatory factor $\e^{i b\cdot q_i}$, but it does not constrain how rapidly the prefactors $\phi_i(p_i)\phi_i^*(p_i')$ vary across that region. Whether one can neglect the $q_i$-dependence of the wavepacket therefore depends on how ``smooth'' $\phi_i$ is on the scale $|q_i|\!\sim\!1/|b|$, or equivalently on the comparison between this stationary-phase scale and the intrinsic momentum width $\sigma_i$ of the $i^{\text{th}}$ wavepacket, which controls the rate of variation of $\phi_i(p_i)$ in momentum space. 
To ensure that the wavepacket is well localized in both position and momentum space relative to the stationary-phase scale \eqref{eq:stationary}, we choose $\sigma_i$ to lie within the parametric window
\begin{equation}
\frac{1}{|b|} \ll \sigma_i \ll m_i\,.
\label{eq:packet-window}
\end{equation}
The upper bound in \eqref{eq:packet-window} ensures that the energy spread of $\phi_i(p_i)$ remains small compared to its rest mass, so that its spatial width $\Delta x_i \sim 1/\sigma_i$ is much larger than its Compton wavelength $1/m_i$, and the packet behaves as a classical lump of size $\Delta x_i$ throughout the scattering process.

The lower bound in \eqref{eq:packet-window} ensures that the wavepacket remains nearly the same over the stationary-phase region 
$|q_i|\sim 1/|b|$ dominating the integral \eqref{eq:quantum_KMOC_exp_value}.  In other words, its relative variation there is small, and heuristically scales as
$\Delta\phi_i/\phi_i \sim |q_i|/\sigma_i \sim 1/(|b|\sigma_i)\ll 1$. By contrast, $\phi_i(p_i)$ changes appreciably only when $p_i$ is displaced from its central value 
$P_i$ by an amount of order $\sigma_i$.  
In particular, over the characteristic interval $\Delta p_i^\mu \sim \sigma_i$, 
the wavepacket changes by an amount comparable to its value, $\Delta\phi_i \sim \phi_i$.\footnote{A prototypical example of this is  the one-dimensional Gau\ss ian wavepacket centered at $P$ with width $\sigma$: $\phi(p) = \exp\!\left[-i\frac{(p - P)^2}{2\sigma^2}\right]$. Its derivative is $\partial_p \phi(p) = -i\,\frac{p - P}{\sigma^2}\,\phi(p)$ and so a small momentum displacement $\Delta p$ induces a change $ \Delta \phi \;\approx\; \partial_p \phi\, \Delta p\;=\; -i\,\frac{(p - P)\,\Delta p}{\sigma^2}\,\phi(p)$. If the displacement is of the characteristic size $|\Delta p|\sim \sigma$ and we evaluate near the peak where $|p - P|\sim \sigma$, we find $|\Delta \phi| \;\sim\; \phi(p)$, that is, the wavepacket changes by an amount comparable to its value.} 
This implies that each Lorentz component of the momentum derivatives scale parametrically as
\begin{equation}
    \partial_{p_i^\mu}\phi_i \sim \frac{\phi_i}{\sigma_i}\,,
    \qquad
    \partial_{p_i^\mu}\partial_{p_i^\nu}\phi_i\sim \frac{\phi_i}{\sigma_i^2}\,, 
\end{equation}
and so on and up to $\mathcal{O}(1)$ factors.
Expanding $\phi_i(p_i')$ in powers of $q_i$ thus gives
\begin{equation}
\phi_i(p_i')=\phi_i(p_i - q_i)
= \phi_i(p_i) - q_i^\mu \partial_\mu \phi_i(p_i)
+ \tfrac{1}{2} q_i^\mu q_i^\nu \partial_\mu\partial_\nu \phi_i(p_i) + \ldots\,,
\end{equation}
so that the relative correction between $\phi_i(p_i')$ and $\phi_i(p_i)$ is parametrically small:
\begin{equation}
\frac{\phi_i(p_i')-\phi_i(p_i)}{\phi_i(p_i)} 
=\mathcal{O}\!\left(\frac{|q_i|}{\sigma_{i}}\right)
+\mathcal{O}\!\left(\frac{|q_i|^2}{\sigma_{i}^2}\right)+\ldots
= \mathcal{O}\!\left(\frac{1}{|b|\,\sigma_{i}}\right)\,,
\end{equation}
given that \eqref{eq:stationary} and \eqref{eq:packet-window} imply $|q_i|/\sigma_{i} \sim (|b|\,\sigma_{i})^{-1} \ll 1$. 
Therefore, it is consistent, at leading order in $(|b|\,\sigma_{i})^{-1}$, to neglect the $q_i$-dependence of the wavepacket in \eqref{eq:quantum_KMOC_exp_value} and set $\phi_i(p_i') \simeq \phi_i(p_i)$.

The expectation value in the \emph{classical} limit of~\eqref{eq:quantum_KMOC_exp_value} can thus be written as:
\begin{equation}
	\label{eq:classical_KMOC_exp_value}
	\langle \Delta \hat{\mathcal{O}}\rangle_{\text{cl.}}\equiv \!\int\! \prod_{i=1}^{2}\biggl[\hat{\dd}^\D\! q_i\, \theta(p_i^0-q_i^0)\,\hat{\delta}(2 p_i \cdot q_i - q_i^2)\, \e^{i b_i \cdot q_i} \biggr]_{p_i} \ \langle p_1{-}q_1,p_2{-}q_2 |S^\dagger [\hat{\mathcal{O}}, S]| p_1,p_2 \rangle \,,
\end{equation}
where we used the shorthand $[X]_{p_i}\equiv [\dd \Phi(p_i) |\phi(p_i)|^2\, X]$. A priori, \eqref{eq:classical_KMOC_exp_value} is a valid formula for any well-defined operator in quantum field theory. In what follows, we focus on the particular case where it represents the gravitational radiation measured, out of the classical scattering of a massive binary system, at null infinity.
\paragraph{Gravitational waveform.}
The gravitational waveform is obtained from \eqref{eq:classical_KMOC_exp_value} by considering the operator
\begin{align}
\label{eq:waveform_operator}
    \hat{\mathcal{O}}_{\mathcal{W}} & \equiv \lim_{x\to \mathscr{I}^+}\, \kappa \,\varepsilon_{h}^{\mu\nu} h_{\mu\nu}\,,
\end{align}
where $h_{\mu\nu}$ is the linearized metric field from \eqref{eq:metricPert}, and $\varepsilon_{h}^{\mu\nu}$ is the polarization tensor of helicity $h$ associated to emitted graviton. Since it is ultimately measured at future null infinity $\mathscr{I}^+$, this object is best viewed as a function of the retarded time $u=t-r$ as well as the angles on the celestial sphere. The contraction with $\varepsilon_{h}^{\mu\nu}$ isolates the physical, radiative polarization components of the gravitational field and defines a manifestly gauge-invariant observable (at least under small gauge transformations that vanish at infinity \cite{Cristofoli:2022phh,Elkhidir:2024izo}).

In a quantum field theory, observables can always be expressed in terms of creation and annihilation operators, which act as the building blocks of the Fock-space operator algebra. Assuming the polarization to be normalised as $\varepsilon_h(-k)=\varepsilon_{-h}(k)=\varepsilon_h^*(k)$ and 
$\varepsilon_h(k)\!\cdot\! \varepsilon_{h'}^*(k)=-\delta_{h h'}$,
the waveform operator of \eqref{eq:waveform_operator} can be thus written as a linear combination of a single creation and annihilation operator at future null infinity ($\hat{b}\equiv \hat{a}_{\text{out}}$):
\begin{equation}
    \label{eq:waveform_operator_linear}
	\hat{\mathcal{O}}_{\mathcal{W}} = \!\int\! \dd \Phi(k) \left[ \e^{- i k \cdot x} \hat{b}_{-h} (k) + \e^{+ i k \cdot x} \hat{b}_{h}^\dagger (k)\right] \,,
\end{equation}
with $\dd \Phi (k)$ denoting the graviton massless version of the LIPS measure in \eqref{eq:LIPSdef}. 
At null infinity, where $x^0 = t \to \infty$ and $|\textbf{x}| \equiv r \to \infty$ 
while the retarded time $u$ is kept fixed, the Fourier transform from $k^\mu = \omega (1,\unitn')$ to $x^\mu = (t, r\unitn)$ over this Lorentz-invariant measure simplifies considerably:
\begin{align}
\int \frac{\hat{\dd}^{\D-1}\textbf{k}}{2|\textbf{k}|}\, 
    \e^{-i k \cdot x} f(\textbf{k})
&= \int_0^{\infty} \hat{\dd}\omega\, |\omega|^{\D-3}\, 
    \e^{-i \omega u}
    \int
    \frac{\hat{\dd}^{\D-2}\unitn'}{2}\,
    \e^{-i \omega r (1 - \unitn \cdot \unitn')} 
    f(\omega \unitn') \notag
    \\
&\simeq \frac{-i}{2(2\pi)^{\frac{\D-2}{2}}\, r^{\frac{\D-2}{2}}}
   \int_0^{\infty} \hat{\dd}\omega\,
   (-i\omega)^{\frac{\D-4}{2}}\,
   \e^{-i \omega u}\, f(\omega \unitn)
   \qquad (\omega r \gg 1)\,.
\label{eq:LipsMassless}
\end{align}
Here, the second line follows from the fact that, due to rapid oscillations, 
the Fourier transform is dominated by the stationary phase where the graviton three momenta points in the radial spatial direction of $x^\mu$, i.e., $\unitn'=\unitn$. 
The overall factors arise from the Gau\ss ian integral 
$\int \dd^{\D-2} \, \unitn\,\e^{-i \omega r \unitn^2 / 2}=\e^{-i\pi(\D-2)/4}\left(\frac{2\pi}{\omega r}\right)^{\frac{\D-2}{2}}$. 

Keeping the spacetime dimension $\D$ generic in \eqref{eq:LipsMassless} serves as a way to regularize potential infrared divergences in the graviton frequency while consistently retaining any $\varepsilon/\varepsilon$ effects, which should include those generating the BMS transformations described in~\cite{Bini:2024rsy}.

Using ~\eqref{eq:waveform_operator}, ~\eqref{eq:waveform_operator_linear} as well as ~\eqref{eq:LipsMassless},
the expansion of \eqref{eq:classical_KMOC_exp_value} near null infinity reduces to
\begin{equation}
\begin{split}
	\langle \Delta \hat{\mathcal{O}}_{\mathcal{W}}\rangle_{\text{cl.}}&\simeq
    \frac{-i}{2(2\pi)^{\frac{\D-2}{2}}\, r^{\frac{\D-2}{2}}} \!\int\limits_0^\infty\! \hat{\dd} \omega (-i\omega)^{\frac{\D-4}{2}} \, \!\int\! \prod_{i=1}^{2}\biggl[\hat{\dd}^\D\! q_i\, \theta(p_i^0{-}q_i^0)\,\hat{\delta}(2 p_i {\cdot} q_i {-} q_i^2)\, \e^{i b_i \cdot q_i} \biggr]_{p_i}\\
    &\ 
    \times \left( \e^{- i \omega u} \langle p_1^\prime, p_2^\prime |S^{\dagger}[\hat{b}_{-h}(k), S]| p_1, p_2\rangle
	-  \e^{+i \omega u} \langle p_1^\prime, p_2^\prime |S^{\dagger}[\hat{b}_{+h}^\dagger(k), S]| p_1, p_2\rangle\right) \, . 
\end{split}
\label{eq:waveform_time_0}
\end{equation}
Accordingly, for a \emph{fixed} kinematic configuration of the initial states, the quantity of interest is $\widetilde{\mathcal{W}}(u,\unitn)$, given by~\eqref{eq:waveform_time_0} with the $p_i$ subscript in the square brackets omitted. 

$\widetilde{\mathcal{W}}(u,\unitn)$ can be related to scattering amplitudes by rewriting the $S$-matrix operator in terms of the matrix elements of transfer matrix $T$:
\begin{equation}
    S = \mathbb{1} + i\,  T \, , \qquad  \langle f \vert T \vert i \rangle \, = \hat{\delta}^{\D}(p_f -p_i)\, \mathcal{M}(i\to f)+\ldots\,,
\end{equation}
where ``$+\ldots$'' accounts for (possibly) disconnected/spectator terms (see \cite[(3.9)]{Caron-Huot:2023vxl} for example) and the delta-function accounts for overall momentum conservation.

Using this relation, the time-domain waveform can be rewritten as:
\begin{equation}
    \label{eq:timedomain}
    \widetilde{\mathcal{W}}_h (u,\unitn) = \frac{1}{2(2\pi)^{\frac{\D-2}{2}}\, r^{\frac{\D-2}{2}}} 
    \!\int_0^\infty\! \hat{\dd} \omega (-i\omega)^{\frac{\D-4}{2}} \left[ \e^{- i \omega u} 
    \,\mathcal{W}_{h} (\omega,\unitn) + \mathrm{c.c.}\right]\,,
\end{equation}
where the waveform in the frequency domain (what we compute below) reads as:
\begin{equation}
    \label{eq:frequencydomain}
        \cW_{h} (\omega,\unitn)  =
         \kappa 
         \!\int\! \dd \mu
        \,\Big[
        \ampl(p_1 p_2\to p_1^\prime p_2^\prime k_{-h}) 
         +
         \sumint_X \ampl\big(p_1 p_2 \rightarrow X k_{-h}\big) \ampl^*\big(p_1^{\prime} p_2^{\prime} \rightarrow X\big) \Big]\,.
\end{equation}
For space consideration, we introduced in 
~\eqref{eq:frequencydomain} the overall 
on-shell measure $\dd \mu$:
\begin{equation}\label{eq:dmu}
    \dd\mu = \prod_{i=1}^{2}\biggl[\hat{\dd}^\D\! q_i \, \hat{\delta}(2 p_i \cdot q_i - q_i^2)\, \e^{i b_i \cdot q_i} \biggr] \hat{\delta}^{(\D)}(q_1+q_2-k)\,,
\end{equation}
as well as the sum over all possible intermediate on-shell states $X$
\begin{equation}
    \sumint_X \equiv \sum_{X} \prod_{i \in X} \int \hat{\rd}^{\D}\ell_i\, \theta(\ell_i^0) \hat{\delta}(\ell_i^2+M_i^2)\,.
    \label{eq:inclsum}
\end{equation}
Both terms in the integrand of \eqref{eq:frequencydomain} can be graphically represented as:
\begin{equation}
    \adjustbox{valign=c}{\begin{tikzpicture}[line width=1]
\begin{scope}[xshift=0
]
    \node[] at (0,0) {$\langle p_1^\prime, p_2^\prime |S^{\dagger}[\hat{b}_{-h}(k), S]| p_1, p_2\rangle=$};
\end{scope}
\begin{scope}[xshift=60]
\draw[line cap=round] (2,0.3) -- (0.8,0.3) node[above] {\small$1'$};
\draw[->] (2,0.3) -- (1,0.3);
\draw[line cap=round] (2,-0.3) -- (0.8,-0.3) node[below] {\small$2'$};
\draw[->] (2,-0.3) -- (1,-0.3);
\draw[line cap=round] (2,0.3) -- (3.2,0.3) node[above] {\small$1$};
\draw[-<] (2,0.3) -- (3,0.3);
\draw[line cap=round] (2,-0.3) -- (3.2,-0.3) node[below] {\small$2$};
\draw[-<] (2,-0.3) -- (3,-0.3);
\draw[line cap=round, photon] (2,0) -- (1,1) node[left] {\small$k$};
\draw (2,0.3) -- (0.95,0.3);
\draw (2,-0.3) -- (0.95,-0.3);
\filldraw[fill=gray!5, line width=1.2, line width=1.3pt](2,0) circle (0.6) node {$i \mathcal{M}$};
\draw[<-,color=gray,line cap=round] (-0.5+2,-0.8) -- (0.5+2,-0.8) node[below,midway] {\small$\text{time}$};
\end{scope}
\begin{scope}[xshift=206-40]
    \node[] at (0,0) {$+$};
\end{scope}
\begin{scope}[xshift=255-40]
\draw[line cap=round] (0,0.3) -- (-1.2,0.3) node[above] {\small$1'$};
\draw[->] (0,0.3) -- (-1,0.3);
\draw[line cap=round] (0,-0.3) -- (-1.2,-0.3) node[below] {\small$2'$};
\draw[->] (0,-0.3) -- (-1,-0.3);
\draw[line cap=round] (2,0.3) -- (3.2,0.3) node[above] {\small$1$};
\draw[-<] (2,0.3) -- (3,0.3);
\draw[line cap=round] (2,-0.3) -- (3.2,-0.3) node[below] {\small$2$};
\draw[-<] (2,-0.3) -- (3,-0.3);
\draw[line cap=round,photon] (2,0) -- (1,1) node[left] {\small$k$};
\filldraw[fill=gray!30](0,-0.3) rectangle (2,0.3);
\draw (2,0.3) -- (0.95,0.3);
\draw (2,-0.3) -- (0.95,-0.3);
\draw[] (1,0) node {$X$};
\filldraw[fill=gray!5, line width=1.2, line width=1.3pt](0,0) circle (0.6) node {$- i \mathcal{M}^*$};
\draw[->,color=gray,line cap=round] (-0.5,-0.8) -- (0.5,-0.8) node[below,midway] {\small$\text{time}$};
\filldraw[fill=gray!5, line width=1.2, line width=1.3pt](2,0) circle (0.6) node {$ i \mathcal{M}$};
\draw[<-,color=gray,line cap=round] (-0.5+2,-0.8) -- (0.5+2,-0.8) node[below,midway] {\small$\text{time}$};
\draw[dashed,orange,line cap=round] (1,1.2) -- (1,-0.8);
\end{scope}
\end{tikzpicture}}\,.
    \label{eq:5ptdecomp}
\end{equation}
The first term on the right-hand side of \eqref{eq:5ptdecomp} is a standard (i.e., time-ordered or ``in-out'')
five-point scattering amplitude with an on-shel graviton emitted in the final state. 
The second term, sometimes referred to as the ``\emph{cut term},''  is of a different nature as it is \emph{quadratic} in the amplitude (i.e., out-of-time-ordered or ``in-in'').

Mathematically, the role of the ``\emph{cut term}'' is to restore the proper $i\varepsilon$ prescription by converting the relevant Feynman propagators in the amplitude into their retarded counterparts~\cite{Caron-Huot:2023vxl}. Physically, it thus enforces the correct notion of causality in the observable, removes the spurious classically singular, or ``superclassical,'' contributions that would otherwise appear in the classical limit~\eqref{eq:hierarchy}, and it ensures the infrared divergences in the waveform exponentiate as required~\cite{Caron-Huot:2023vxl,Correia:2024yfx,Caron-Huot:2025ens}.

Now that the observable we wish to compute has been identified, let us explicitly specify the kinematic variables on which it depends.
\subsection{Kinematic invariants}
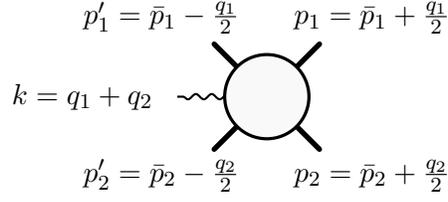
\begin{figure}
\centering
    \begin{tikzpicture}[scale=0.7,thick,
baseline={([yshift=-0.4ex]current bounding box.center)}]
    \draw[line width=2, line cap=round] (-1,1) -- (0,0) -- (1,1)  (-1,-1) -- (0,0) -- (1,-1);
    \draw[photon] (-0.5,0) node[]{} -- +(-1.2,0);
    \filldraw[fill=gray!5, line width=1.2](0,0) circle (0.8);
    \node[] at (2,1.5) {$p_1=\bar{p}_1 + \frac{q_1}{2}$};
    \node[] at (-2,1.5) {$p_1'=\bar{p}_1 - \frac{q_1}{2}$};
    \node[] at (2,-1.5) {$p_2=\bar{p}_2 + \frac{q_2}{2}$};
    \node[] at (-2,-1.5) {$p_2'=\bar{p}_2 - \frac{q_2}{2}$};
    \node[] at (-3.5,0) {$k=q_1+q_2$};
\end{tikzpicture}
\caption{Kinematics of the five-point process.
}
\label{fig:waveform_kinematics}
\end{figure}
The five-point process in \figref{fig:waveform_kinematics} depends on four momenta: those of two incoming massive particles $p_1$ and $p_2$, and the two mismatch momenta $q_1$ and $q_2$. The latter are related to the momentum of the emitted graviton $k$ and the integration momentum $q$ as:
\begin{equation}
    q = q_1 \, \qquad k= q_1+q_2 \, . 
\end{equation}
The massive momenta satisfy the on-shell conditions $p_i^2=m_i^2$ and are parametrised as
\begin{equation}
    p_i = m_i u_i \, , \qquad u_i^2 = 1 \, , 
\end{equation}
with $u_i$ the corresponding four-velocities.
In the classical limit described in \eqref{eq:hierarchy}, it is useful to introduce the shifted momenta
\begin{equation}
     \bar{p}_i = p_i + \frac{q_i}{2}
     = m_i \bar{u}_i = m_i u_i + \frac{q_i}{2} \, ,
\end{equation}
where the momentum transfer $q_i$ is of order $1/|b|$.
Since 
\begin{equation}
    \bar{u}_i - u_i = \frac{q_i}{2m_i} \sim \frac{1}{m_i |b|} \ll 1 \, ,
\end{equation}
for $m_i |b|\gg 1$, their difference is subleading in the classical expansion, and we will drop the bar notation henceforth. 
In momentum space, the independent kinematic invariants of the process are thus defined as:
\begin{equation}
\begin{aligned}
    u_1 {\cdot} u_2 &= \gamma \,, &  u_i {\cdot} q_i & = 0 \,,&  u_1 {\cdot} q_2 & = w_1 \,, &  u_2 {\cdot} q_1 & = w_2 \,,
     &u_i^2 &= 1 \,, & q_1{\cdot} q_2 &=\frac{-(q_1^2{+}q_2^2)}{2} \,.
\end{aligned}
\end{equation}
The physical region for the scattering process in momentum space is specified by
\begin{equation}
		\gamma > 1\,,\qquad w_i > 0\,,\qquad -q_i^2 > 0\,.
  \label{eq:kinematics1}
\end{equation}
After performing the Fourier transform to impact-parameter space, the dependence on the momentum transfer $q$ is replaced by the impact parameter $b$, and the independent kinematic invariants of the process are then:
\begin{equation}
\begin{aligned}
    u_1 \cdot u_2 &= \gamma \,, &  u_i \cdot k & = w_i \,, 
     &u_i^2 &= 1 \,, & u_i \cdot b &= 0 \,, & k^2 &= 0 \,.
       \label{eq:kinematics2}
\end{aligned}
\end{equation}
The physical region for the scattering process in the frequency domain is specified by
\begin{equation}
		\gamma > 1\,,\qquad w_i > 0\,,\qquad -b^2 > 0\,,\qquad b \cdot k\in \mathbb{R}\, .
  \label{eq:kinematics3}
\end{equation}

%% file: Sections/waveform_tree.tex
{\section{Tree-level frequency-domain gravitational waveform}\label{sec:waveform_tree}}
As a proof-of-principle example for the computational methods developed in the next sections, we will now focus on the tree-level (leading-order) gravitational waveform.
In this case, the cut term (the rightmost term in \eqref{eq:5ptdecomp}) vanishes, and ~\eqref{eq:frequencydomain} becomes the Fourier transform of the classical tree-level five-point amplitude
$\ampl^{(0)}_5\!\bigl(\bar{p}_1+\tfrac{q_1}{2},\, \bar{p}_2+\tfrac{q_2}{2}\to \bar{p}_1-\tfrac{q_1}{2},\, \bar{p}_2-\tfrac{q_2}{2};\, k\bigr)$. After integrating the overall momentum-conserving delta-function in \eqref{eq:dmu}, we find, using \eqref{eq:impactPdef},\footnote{The additional numerical factor $1/4$ comes from the on-shell measure \eqref{eq:dmu}.}
\begin{equation}
    \label{eq:frequencydomain_tree}
        \cW_h^{(0)}(\omega,\unitn) = \frac{\e^{i 
        \omega (1,\unitn)
        \cdot b_2}}{4}
     \int \hat{\rd}^{\D}q\,\hat{\delta}(u_1\cdot q_1)
     \hat{\delta}(u_2\cdot (k-q_1)) \,\e^{i q_1\cdot b}\Big[
        \ampl^{(0)}_5(\bar{p}_1,\bar{p}_2; k)+0\Big] \,.
\end{equation}
To lighten the above notation, we used the shorthand $\mathcal{M}^{(L)}_{2n+m}(\bar{p}_1,\ldots,\bar{p}_n; k_1,\ldots, k_m)$ to denote the classical amplitude with massive legs $\{\bar{p}_1,\ldots,\bar{p}_n\}$ and external gravitons $\{k_1,\ldots,k_m\}$.

Following the method developed in~\cite{Brunello:2024ibk,Brunello:2025todo}, 
the tree-level integrand is constructed using generalized unitarity~\cite{Bern:1994zx,Kosower:2011ty,Mastrolia:2012an,Ita:2015tya}. 
Considering generalized cuts corresponding to a single on-shell graviton exchange between the two massive legs,
with poles at $q_{i=1,2}^2 = 0$, 
the amplitude can be written as
\begin{equation}\label{eq:cuts}
\begin{split}
    \hspace{-0.3cm}
  \ampl^{(0)}_5(\bar{p}_1,\bar{p}_2; k)
  &= \sumint_{q_2,h}\ampl_{4\,h}^{(0)}(\bar{p}_1;-q_2,k)
    \ampl_{3,-h}^{(0)}(\bar{p}_2;q_2)
    \cup
    \sumint_{q_1,h}\ampl_{3\,h}^{(0)}(\bar{p}_1;q_1)
    \ampl_{4,-h}^{(0)}(\bar{p}_2;-q_1,k)
    \\&=\adjustbox{valign=c}{
   \tikzset{every picture/.style={line width=0.75pt}} 
\begin{tikzpicture}[x=0.75pt,y=0.75pt,yscale=1.2,xscale=-1.2]
\draw[line cap = round] (273,150.86) .. controls (273.93,148.69) and (275.48,148.08) .. (277.65,149.01) .. controls (279.81,149.94) and (281.36,149.33) .. (282.29,147.17) .. controls (283.22,145) and (284.77,144.39) .. (286.94,145.32) .. controls (289.1,146.25) and (290.65,145.64) .. (291.59,143.48) .. controls (292.52,141.31) and (294.07,140.7) .. (296.24,141.63) .. controls (298.4,142.56) and (299.95,141.95) .. (300.88,139.79) .. controls (301.81,137.62) and (303.36,137.01) .. (305.53,137.94) .. controls (307.69,138.87) and (309.24,138.26) .. (310.18,136.1) .. controls (311.12,133.94) and (312.67,133.33) .. (314.83,134.26) -- (319.43,132.43)  -- (319.43,132.43) node[left]{$k$};
\draw[line cap = round]   (130,79.86) .. controls (132.17,78.94) and (133.72,79.56) .. (134.64,81.73) .. controls (135.56,83.9) and (137.1,84.52) .. (139.27,83.6) .. controls (141.44,82.68) and (142.99,83.31) .. (143.91,85.48) .. controls (144.83,87.65) and (146.37,88.27) .. (148.54,87.35) .. controls (150.71,86.43) and (152.25,87.05) .. (153.18,89.22) .. controls (154.09,91.39) and (155.64,92.02) .. (157.81,91.1) .. controls (159.98,90.18) and (161.52,90.8) .. (162.45,92.97) .. controls (163.38,95.14) and (164.92,95.76) .. (167.09,94.84) .. controls (169.26,93.92) and (170.81,94.55) .. (171.72,96.72) .. controls (172.65,98.89) and (174.19,99.51) .. (176.36,98.59) -- (178.43,99.43) -- (178.43,99.43) node[left]{$k$};
\draw[line cap = round]    (80,80) -- (180,80) ;
\draw[line cap = round]    (80,151) -- (180,151) ;
\draw    (130,79.86) .. controls (131.67,81.53) and (131.67,83.19) .. (130,84.86) .. controls (128.33,86.53) and (128.33,88.19) .. (130,89.86) .. controls (131.67,91.53) and (131.67,93.19) .. (130,94.86) .. controls (128.33,96.53) and (128.33,98.19) .. (130,99.86) .. controls (131.67,101.53) and (131.67,103.19) .. (130,104.86) .. controls (128.33,106.53) and (128.33,108.19) .. (130,109.86) .. controls (131.67,111.53) and (131.67,113.19) .. (130,114.86) .. controls (128.33,116.53) and (128.33,118.19) .. (130,119.86) .. controls (131.67,121.53) and (131.67,123.19) .. (130,124.86) .. controls (128.33,126.53) and (128.33,128.19) .. (130,129.86) .. controls (131.67,131.53) and (131.67,133.19) .. (130,134.86) .. controls (128.33,136.53) and (128.33,138.19) .. (130,139.86) .. controls (131.67,141.53) and (131.67,143.19) .. (130,144.86) .. controls (128.33,146.53) and (128.33,148.19) .. (130,149.86) -- (130,150.86) -- (130,150.86) ;
\draw  [line cap = round,dash pattern={on 4.5pt off 4.5pt}, color=orange]  (80.54,115.43) -- (179.46,115.29) ;
\draw  [fill=gray!5  ,fill opacity=1 ][line width=1.5]  (106.79,79.86) .. controls (106.79,67.04) and (117.18,56.64) .. (130,56.64) .. controls (142.82,56.64) and (153.21,67.04) .. (153.21,79.86) .. controls (153.21,92.68) and (142.82,103.07) .. (130,103.07) .. controls (117.18,103.07) and (106.79,92.68) .. (106.79,79.86) -- cycle;
\node at (130,79.86) {$\ampl_{4\,h}^{(0)}$};
\draw  [fill=gray!5  ,fill opacity=1 ][line width=1.5]  (106.79,150.86) .. controls (106.79,138.04) and (117.18,127.64) .. (130,127.64) .. controls (142.82,127.64) and (153.21,138.04) .. (153.21,150.86) .. controls (153.21,163.68) and (142.82,174.07) .. (130,174.07) .. controls (117.18,174.07) and (106.79,163.68) .. (106.79,150.86) -- cycle ;
\node at (130,150.86) {$\ampl_{3\,-h}^{(0)}$};
\draw[line cap = round]    (223,80) -- (323,80) ;
\draw[line cap = round]    (223,151) -- (323,151) ;
\draw    (273,79.86) .. controls (274.67,81.53) and (274.67,83.19) .. (273,84.86) .. controls (271.33,86.53) and (271.33,88.19) .. (273,89.86) .. controls (274.67,91.53) and (274.67,93.19) .. (273,94.86) .. controls (271.33,96.53) and (271.33,98.19) .. (273,99.86) .. controls (274.67,101.53) and (274.67,103.19) .. (273,104.86) .. controls (271.33,106.53) and (271.33,108.19) .. (273,109.86) .. controls (274.67,111.53) and (274.67,113.19) .. (273,114.86) .. controls (271.33,116.53) and (271.33,118.19) .. (273,119.86) .. controls (274.67,121.53) and (274.67,123.19) .. (273,124.86) .. controls (271.33,126.53) and (271.33,128.19) .. (273,129.86) .. controls (274.67,131.53) and (274.67,133.19) .. (273,134.86) .. controls (271.33,136.53) and (271.33,138.19) .. (273,139.86) .. controls (274.67,141.53) and (274.67,143.19) .. (273,144.86) .. controls (271.33,146.53) and (271.33,148.19) .. (273,149.86) -- (273,150.86) -- (273,150.86) ;
\draw[line cap = round, dash pattern={on 4.5pt off 4.5pt}, color=orange]  (223.54,115.43) -- (322.46,115.29) ;
\draw  [fill=gray!5  ,fill opacity=1 ][line width=1.5]   (249.79,79.86) .. controls (249.79,67.04) and (260.18,56.64) .. (273,56.64) .. controls (285.82,56.64) and (296.21,67.04) .. (296.21,79.86) .. controls (296.21,92.68) and (285.82,103.07) .. (273,103.07) .. controls (260.18,103.07) and (249.79,92.68) .. (249.79,79.86) -- cycle ;
\node[yshift=-1pt] at (273,79.86) {$\ampl_{3\,-h}^{(0)}$};
\draw  [fill=gray!5  ,fill opacity=1 ][line width=1.5] (249.79,150.86) .. controls (249.79,138.04) and (260.18,127.64) .. (273,127.64) .. controls (285.82,127.64) and (296.21,138.04) .. (296.21,150.86) .. controls (296.21,163.68) and (285.82,174.07) .. (273,174.07) .. controls (260.18,174.07) and (249.79,163.68) .. (249.79,150.86) -- cycle ;
\node at (273,150.86) {$\ampl_{4\,h}^{(0)}$};
\draw[->]    (80,80) -- (96.43,80) node[below,midway]{$p_2$};
\draw[->]    (80,151) -- (96.43,151) node[above,midway]{$p_1$};
\draw[->]    (223,151) -- (239.43,151) node[above,midway]{$p_1$};
\draw[->]    (223,80) -- (239.43,80) node[below,midway]{$p_2$};
\draw[>-]    (163.57,80) -- (180,80) node[below,midway]{$p_2'$};
\draw[>-]    (163.57,151) -- (180,151) node[above,midway]{$p_1'$};
\draw[>-]    (306.57,151) -- (323,151)  node[above,midway]{$p_1'$};
\draw[>-]    (306.57,80) -- (323,80) node[below,midway]{$p_2'$};
\draw[<-,color=gray]    (136.43,107.43) -- (136.43,123.43) node[left]{$q_1$};
\draw[->,color=gray]    (279.43,107.43) -- (279.43,123.43) node[left]{$q_2$};
\draw (207,120) node [anchor=north west][inner sep=0.75pt]    {$\cup $};
\end{tikzpicture}
    }
\end{split}
\end{equation}
where 
$\sumint_{q_i,h}\equiv\sum_h\int\hat{\rd}^{\D}q_i\,\hat{\delta}(q_i^2)$ 
denotes the helicity sum-integral over each internal cut graviton. 
The operation ``$A\cup B$'' merges the two cuts $A$ and $B$
by adding them while subtracting the double counting of the common poles at $q_{i=1,2}^2=0$ (see~\cite{Bern:2011qt} for a detailed treatment).

In this case, the merging procedure described above is straightforward 
once the amplitude is expressed in terms of the following tensor structures:
\begin{equation}\label{eq:Tdef}
    \mathbf{T}
    = \bigl(
        (u_{1}\!\cdot\!F_k\!\cdot\!u_{2})^{2},\,
        (u_{1}\!\cdot\!F_k\!\cdot\!u_{2})(b\!\cdot\!F_k\!\cdot\!u_{1}),\,
        (b\!\cdot\!F_k\!\cdot\!u_{1})^{2}
      \bigr)\,.
\end{equation}
where $F_k^{\mu\nu}=k^{[\mu}\varepsilon_k^{\nu]}$ is the linearized graviton field-strength tensor.  

The tree-level building blocks entering the cuts in \eqref{eq:cuts} can be systematically expanded in the classical limit using the heavy-mass expansion \cite{Brandhuber:2021eyq}, before summing over the graviton polarizations.  
Explicitly, they are given by the three-point amplitude,
\begin{equation}
\label{eq:tree_classical_3pts}
    \ampl_3^{(0)}(\bar{p}_i; k)
    = \kappa\, m_i^2\, (\varepsilon_k \cdot u_i)^2\,,
\end{equation}
and the gravitational Compton amplitude,
\begin{equation}
\label{eq:tree_classical_compton}
    \ampl_4^{(0)}(\bar{p}_i; k_1, k_2)
    = \kappa^2 \left[
        i \pi\, m_i^3\, \delta(u_i \cdot k_1)\, 
        (\varepsilon_{k_1} \cdot u_i)^2 (\varepsilon_{k_2} \cdot u_i)^2
        - m_i^2\, 
        \frac{(u_i \cdot F_1 \cdot F_2 \cdot u_i)^2}{(k_1 \cdot u_i)^2 q^2}
    \right]\,.
\end{equation}
We notice that this contribution has two terms with different mass scalings. 
In this case, we select only the non-factorizable term proportional to $m_i^2$ when summing over graviton helicities.  
For completeness, note that the factorizable (in \eqref{eq:tree_classical_3pts}) piece is proportional to $m_i^3\delta(u_i\!\cdot\!k_1)$. 
Since $u_i$ is timelike (see \eqref{eq:kinematics2}), the simultaneous constraints $k_1^2=0$ and $u_i\!\cdot\!k_1=0$ admit only the trivial solution $k_1^\mu=0$ 
(e.g., in the rest frame $u_i=(1,\mathbf{0})$ one has $u_i\!\cdot\!k_1=k_1^0=|\mathbf{k}_1|=0$). 
Physically, this term thus corresponds to a configuration with vanishing radiated momentum and does not contribute to the emission of finite-energy gravitons observed at future null infinity.

Combining these ingredients, the tree-level five-point amplitude appearing in \eqref{eq:frequencydomain_tree} takes the following form in the classical limit \cite{Brunello:2025todo}
\begin{equation}
\begin{aligned}
\label{eq:tree_integrand_complete}
    \ampl^{(0)}_5(\bar{p}_1,\bar{p}_2; k)
    = \frac{m_1^2 m_2^2 \kappa^3}{q_1^2 q_2^2 w_1^2}
    \biggl[ &
        (q_1 \cdot F_k \cdot u_1)^2
        \left( y^2 - \frac{1}{\D_s - 2} \right)
        \\
        & + \frac{(q_1 \cdot F_k \cdot u_1)(u_1 \cdot F_k \cdot u_2)}{w_2}
        \left( -y (q_2^2 + 2 w_1 w_2) + \frac{q_2^2}{\D_s - 2} \right)
        \\
        & + \frac{(u_1 \cdot F_k \cdot u_2)^2}{4 w_2^2}
        \left( (q_2^2 y + 2 w_1 w_2)^2 - \frac{q_2^4}{\D_s - 2} \right)
    \biggr]\,,
\end{aligned}
\end{equation}
where the factors of $\tfrac{1}{\D_s-2}$ arise from the trace subtraction in the graviton
polarization sum $\sumint_{q_i,h}$ that builds the propagator numerator $\Pi_{\mu\nu,\rho\sigma}$; for instance, in de~Donder gauge, $\Pi_{\mu\nu,\rho\sigma} \propto \eta_{\mu\rho}\eta_{\nu\sigma}+\eta_{\mu\sigma}\eta_{\nu\rho}
-\tfrac{2}{\D_s-2}\eta_{\mu\nu}\eta_{\rho\sigma}$. (This tensor structure is completely fixed, up to an overall factor and terms $\propto q_i^\mu$, by Lorentz symmetry \cite{Balasubramanian:2021act}, as the physical helicities transform in the symmetric-traceless rank-2 representation of the (massless) little group $\mathrm{SO}(\D_s-2)$, the subgroup of $\mathrm{SO}(\D_s-1,1)$ that leaves a null momentum $q_i^\mu$ invariant.

\subsection{Master integral decomposition}
Starting from the amplitude in \eqref{eq:tree_integrand_complete}, 
we perform a tensor-to-scalar decomposition~\cite{Anastasiou:2023koq} 
in four-dimensional external kinematics~\cite{Peraro:2019cjj,Peraro:2020sfm}, 
expressing the waveform in \eqref{sec:waveform_tree} as a linear combination of tensor structures~$\mathbf{T}$  multiplied by scalar form factors~$\mathbf{W}^{(0)}$:
\begin{equation}
    \cW_h^{(0)}(\omega,\unitn)
    = \mathbf{T}\!\cdot\!\mathbf{W}^{(0)}\,,
\end{equation}
with $\mathbf{T}$ as defined in \eqref{eq:Tdef}.
The scalar integrals appearing in the form factors $\mathbf{W}^{(0)}$ belong to the following Fourier integral family:
\begin{equation}
    \mathcal{I}(n_{1},n_{2},n_{3}) = \int \hat{\rd}^{\D}q\,\frac{(iq\cdot b)^{-n_{3}}}{(q^{2})^{n_{1}}((q-k)^{2})^{n_{2}}}\hat{\delta}(u_{1}\cdot q)\hat{\delta}(u_{2}\cdot (q-k))\ \e^{iq\cdot b}\,.
    \label{eq:integral_family}
\end{equation}
We first make the integral dimensionless by rescaling all Lorentz vectors 
with suitable powers of $\sqrt{-b^{2}}$. 
Next, we decompose each vector into components parallel and orthogonal 
to the plane defined by the incoming four-velocities $\{u_{1},u_{2}\}$:
\begin{equation}\label{eq:dimLessVar}
    q^{\mu} = \frac{q_{\parallel}^{\mu} + q_{\perp}^{\mu}}{\sqrt{-b^{2}}}\,, 
    \qquad 
    k^{\mu} = \frac{k_{\parallel}^{\mu} + k_{\perp}^{\mu}}{\sqrt{-b^{2}}}\,, 
    \qquad 
    b^{\mu} = \sqrt{-b^{2}}\, \hat{b}^{\mu}\,.
\end{equation}
By definition, the impact parameter $b^{\mu}$ (and therefore $\hat{b}^{\mu}$) 
is orthogonal to both particle velocities (see \eqref{eq:kinematics2}). 
Hence it lies entirely within the two-dimensional subspace transverse 
to the plane spanned by $\{u_{1},u_{2}\}$. 
The momentum $q^{\mu}$ can thus be decomposed into a longitudinal component 
along this plane and a transverse component orthogonal to it. 
It is convenient to parametrize the longitudinal part as
\begin{equation}
    q_{\parallel}^{\mu} = x_{1} u_{1}^{\mu} + x_{2} u_{2}^{\mu}\,.
\end{equation}

The delta-functions in~\eqref{eq:integral_family} fix the longitudinal variables $x_{i}$, 
allowing us to express the integral family in the compact form
\begin{subequations}
    \begin{align}
        \mathcal{I}(n_{1},n_{2},n_{3}) 
        &= \frac{(-b^{2})^{n_{1}+n_{2}+1-\frac{\D}{2}}}{\sqrt{\gamma^{2}-1}}
           \, I(n_{1},n_{2},n_{3})\,, 
           \\
           \label{eq:newFam}
        I(n_{1},n_{2},n_{3}) 
        &= \int
           \frac{(iq_{\perp}\!\cdot\!\hat{b})^{-n_{3}}
           \e^{iq_{\perp}\cdot \hat{b}_{\perp}}\, 
           \hat{\rd}^{\D-2}q_{\perp}}
           {(q_{\perp}^{2}-\hat{w}_{2}^{2})^{n_1}
            ((q_{\perp}-k_{\perp})^{2}-\hat{w}_{1}^{2})^{n_2}}
           \;\ni\;
           \text{FT}\!\Bigg[
           \adjustbox{valign=c}{
           \tikzset{every picture/.style={line width=0.75pt}}
           \begin{tikzpicture}[x=0.75pt,y=0.75pt,yscale=-0.9,xscale=0.9]
           \draw   (180,116) .. controls (180,104.95) and (195.67,96) .. (215,96) .. controls (234.33,96) and (250,104.95) .. (250,116) .. node[xshift=-19,yshift=-10]{$m_2=\hat w_2$} node[xshift=-19,yshift=30]{$m_1=\hat w_1$} controls (250,127.05) and (234.33,136) .. (215,136) .. controls (195.67,136) and (180,127.05) .. (180,116) -- cycle ;
           \draw    (180,116) -- (156,116) ;
           \draw[<-]    (170,116) -- (156,116) node[above,midway]{$k_\perp$};
           \draw    (274,116) -- (250,116) ;
           \draw  [draw opacity=0][fill={rgb, 255:red, 0; green, 0; blue, 0 },fill opacity=1 ] (177.5,116) .. controls (177.5,114.62) and (178.62,113.5) .. (180,113.5) .. controls (181.38,113.5) and (182.5,114.62) .. (182.5,116) .. controls (182.5,117.38) and (181.38,118.5) .. (180,118.5) .. controls (178.62,118.5) and (177.5,117.38) .. (177.5,116) -- cycle ;
           \draw  [draw opacity=0][fill={rgb, 255:red, 0; green, 0; blue, 0 },fill opacity=1 ] (247.5,116) .. controls (247.5,114.62) and (248.62,113.5) .. (250,113.5) .. controls (251.38,113.5) and (252.5,114.62) .. (252.5,116) .. controls (252.5,117.38) and (251.38,118.5) .. (250,118.5) .. controls (248.62,118.5) and (247.5,117.38) .. (247.5,116) -- cycle ;
           \end{tikzpicture}}
           \Bigg]\,.
    \end{align}
\end{subequations}
The relevant kinematic variables are
\begin{equation}\label{eq:newkin}
    \hat{w}_{i} = \frac{w_{i}}{\sqrt{\gamma^{2}-1}}, 
    \qquad 
    k_{\perp}^{2} = \hat{w}_{1}^{2}+\hat{w}_{2}^{2}-2\gamma\hat{w}_{1}\hat{w}_{2},
    \qquad 
    \hat{b}^{2}=-1\,.
\end{equation}
This representation makes it manifest that the new integral family \eqref{eq:newFam} has the form 
of a scalar one-loop two-point Feynman integral with unequal propagator ``masses'' 
$\{\hat{w}_{1},\hat{w}_{2}\}$ and external momentum $k_{\perp}^{\mu}$,
multiplied by a Fourier exponential in the integration measure.
Let us notice that the transverse momentum $q_\perp^\mu$ is spacelike, so we can effectively write:
\begin{equation}\label{eq:BOLDQ}
    q_{\perp}^2= -\mathbf{q}_{\perp}^2\, .
\end{equation}

The integral family in \eqref{eq:newFam} can be analyzed as a twisted period integral~\cite{Brunello:2023fef}. 
In particular, one can derive a set of IBP identities for it~\cite{Brunello:2024ibk,Brunello:2025todo}, which, in dimensional regularization, are generated by the familiar expression
\begin{equation}
    0
    = \int
    \hat{\rd}^{\D-2} q_{\perp}\;
      \pdv{}{q_{\perp}^{\mu}}
      \!\left[
        \frac{v^{\mu}(iq_{\perp}\!\cdot\!\hat{b})^{-n_{3}}}
        {(q_{\perp}^{2}-\hat{w}_{2}^{2})^{n_{1}}
         ((q_{\perp}-k_{\perp})^{2}-\hat{w}_{1}^{2})^{n_{2}}}
        \e^{iq_{\perp}\cdot\hat{b}}
      \right]\!,
\end{equation}
where $v^{\mu}\!\in\!\{\hat{b}^{\mu},k_{\perp}^{\mu}\}$. 
For brevity, when no ambiguity arises, we omit the ``$\perp$'' subscripts in what follows.

Varying the seed indices $(n_{1},n_{2},n_{3})$ generates a linear system 
relating the integrals $I(n_{1},n_{2},n_{3})$. 
Restricting to $n_{3}\leq0$, the system closes on a basis of six 
independent (master) integrals, which we choose as
\begin{equation}\label{eq:startingBasis}
    \mathbf{J}
    = \bigl(
        I(1,0,0),\,
        I(1,0,-1),\,
        I(0,1,0),\,
        I(0,1,-1),\,
        I(1,1,0),\,
        I(1,1,-1)
      \bigr)^{\!\top}\!.
\end{equation}

The leading-order waveform can then be written compactly as
\begin{equation}
    \cW_{h}^{(0)}(\omega,\unitn)
    = \mathbf{T}\cdot\mathcal{C}\cdot\mathbf{J}\,,
\end{equation}
where $\mathcal{C}$ is a $3\times6$ matrix of IBP coefficients. 
The entries of $\mathcal{C}$ are rational functions of the kinematic variables 
and are provided in the ancillary file  \texttt{IBP\_coefficients.m} in the repository~\cite{repo}.

\subsection{Differential equations for the master Fourier integrals\label{subsec:family}}
We now evaluate the master integrals defined in~\eqref{eq:startingBasis} 
using the method of differential equations~\cite{Kotikov:1990kg,KOTIKOV1991123,Bern:1993kr,Gehrmann:1999as,Argeri:2007up}
(see also~\cite{Brunello:2023fef} for an introduction in the context of Fourier integrals). 
To this end, we differentiate all master integrals with respect to the external invariants 
$\{\hat{w}_{1},\hat{w}_{2},\gamma,\hat{b}\!\cdot\!k\}$ 
and express the resulting derivatives as linear combinations of the same basis 
using the IBP relations mentioned above. 
This procedure yields a coupled system of differential equations for the master integrals,
\begin{equation}\label{eq:DE}
    \partial_{a}\mathbf{J} = \Omega_{a}\cdot\mathbf{J}\,,
    \qquad
    a \in \{\hat{w}_{1},\hat{w}_{2},\gamma,\hat{b}\!\cdot\!k\}\,,
    \qquad
    \Omega_{a} = \Omega_{a,0} + \varepsilon\,\Omega_{a,1}\,.
\end{equation}

The matrices $\Omega_{a}$ exhibit a triangular structure with $2\times2$ blocks along the diagonal, 
as illustrated schematically below, and depend linearly on $\varepsilon$:
\begin{equation}\label{eq:DEstructure}
    \Omega_a =
    \begin{bNiceArray}[margin = 5pt]{cccccc}
        \CodeBefore
            \rectanglecolor{red!20}{1-1}{2-2}
            \rectanglecolor{red!20}{3-3}{4-4}
            \rectanglecolor{red!20}{5-5}{6-6}
        \Body
            \red{\bigcdot} & \red{\bigcdot} & \mzero & \mzero & \mzero & \mzero \\
            \red{\bigcdot} & \red{\bigcdot} & \mzero & \mzero & \mzero & \mzero \\
            \mzero & \mzero & \red{\bigcdot} & \red{\bigcdot} & \mzero & \mzero \\
            \mzero & \mzero & \red{\bigcdot} & \red{\bigcdot} & \mzero & \mzero \\
            \red{\bigcdot} & \red{\bigcdot} & \red{\bigcdot} & \red{\bigcdot} & \red{\bigcdot} & \red{\bigcdot} \\
            \red{\bigcdot} & \red{\bigcdot} & \red{\bigcdot} & \red{\bigcdot} & \red{\bigcdot} & \red{\bigcdot}
    \end{bNiceArray}\,.
\end{equation}
The explicit form of the matrices $\Omega_{a}$ is provided in the ancillary file \texttt{DE\_waveform\_tree.m} in the repository~\cite{repo}.

Without further manipulation (see~\secref{sec:canonical}), 
the differential system in~\eqref{eq:DE} cannot, in general, be solved throughout the entire kinematic space.
Fortunately, for most phenomenological applications, 
a complete analytic solution is unnecessary. 
Instead, it is often sufficient to analyze the behavior of the system 
in specific physically relevant regimes. 
In the following, we develop a systematic framework for studying 
these differential equations in distinct kinematic limits 
and apply it to the soft and post-Newtonian expansions of the gravitational waveform. 

\paragraph{Basis properties.} For later convenience, it is useful to record here some of the relations between the basis elements $\mathbf{J}$ in \eqref{eq:startingBasis}. In fact, we can find that there are relations between the first two blocks in  \eqref{eq:DEstructure} ($J_{1,2}$ and $J_{3,4}$)
\begin{equation}
    \begin{aligned}
    J_{3} =\e^{i\hat{b} \cdot k_\perp} 
    J_{1}\big|_{\hat{w}_{2}\rightarrow \hat{w}_{1}}\,,\qquad
    J_{4} 
    = \hat{b}^{\mu} \pdv{}{\hat{b}^{\mu}} 
      J_3 = \left[i\hat{b} \cdot k_\perp \, 
      \e^{i\hat{b} \cdot k_\perp} 
       J_{1}
       +\e^{i\hat{b} \cdot k_\perp} 
       J_{2}\right]\Big|_{\hat{w}_{2}\rightarrow \hat{w}_{1}} \,,
       \end{aligned}
\end{equation}
such that
\begin{equation}\label{eq:relationB1B2}
    \begin{bmatrix}
        J_{3} \\ 
        J_{4}
    \end{bmatrix}
    =\e^{i\hat{b} \cdot k_\perp}
    \begin{bmatrix}
        1 & 0 \\ 
        i\hat{b} \cdot k_\perp & 1
    \end{bmatrix}\cdot
    \begin{bmatrix}
        J_{1} \\ 
        J_{2}
    \end{bmatrix}
    \bigg|_{\hat{w}_{2}\rightarrow \hat{w}_{1}} \, .
\end{equation}

Next, we discuss how the solutions of~\eqref{eq:DEstructure} 
can be obtained in specific kinematic limits, 
making use of restriction theory and companion-matrix techniques.

%% file: Sections/Method.tex
{\section{Kinematic limits from differential equations}\label{sec:method}}

In this section we present a systematic procedure to construct series
expansions of gravitational waveforms in the kinematic limits of interest
(e.g. soft and post-Newtonian). The method relies on restriction theory
for Pfaffian systems of (\emph{not} necessarily $\varepsilon$-factorized) differential equations~\cite{Chestnov:2023kww, haraoka2020linear}. In a given limit, the solution of
a differential equation splits into different regions, which appear as eigenvalues of the residue matrix.
The full system can be restricted to lower-dimensional subsystems corresponding to each region.
The leading behavior can be obtained by solving these subsystems.
Higher-order terms follow from a linear recursion implemented by simple matrix multiplications.
We review only the ingredients needed for this work; for a comprehensive
treatment see~\cite{Chestnov:2023kww}.\footnote{
    See~\cite{Dulat:2014mda, Mastrolia:2017pfy, Chestnov:2022alh} for examples from Feynman integral calculus that motivated the development of restriction theory. 
}

For the reader’s convenience, we illustrate the procedure in detail on a simple univariate example at the end of this section.

\subsection{Restriction theory for differential equations\label{subsec:restriction}}
Let us consider a system of differential equations~\labelcref{eq:DE}, 
referred to as a \textit{Pfaffian} (or, in a loose way, a Gau\ss --Manin) system, 
for a set of integrals $J_{1},\dots,J_{\nu}$ 
in the variables $z_{1},\dots, z_{n}$,
\begin{equation}
    \partial_{a}\vecJ = \Omega_{a}\cdot\vecJ\,.
\end{equation}
We are interested in finding the solution $\vecJ$ as a series expansion 
in the limit $z_{1}\rightarrow0$.
A Pfaffian system is said to be in \textit{normal form}, if the matrices expand as
\begin{subequations}
    \begin{align}
    \label{eq:normal_singular}
    \Omega_{1} &= \sum_{m=-1}^{\infty}\Omega_{1}\supbrk{m}z_{1}^{m}
    \,,
    \\
    \label{eq:normal_rest}
    \Omega_{i} &= \sum_{m=0}^{\infty}\Omega_{i}\supbrk{m}z_{1}^{m}
    \,,
    \quad i=2,\ldots,n\,.
\end{align}
\end{subequations}
That is, in the limit $z_1 \rightarrow 0$, the differential equation matrix
$\Omega_{1}$ has only a simple pole singularity, and all the other matrices are finite.
For the purposes of this work, we will assume that our Pfaffian system is
already in normal form, or can be put into normal form by a suitable
gauge transformation $\vecJ=G\cdot\vecJ'$.
In this case, the solution vector $\vecJ$ can be written as a generic asymptotic expansion of the form:
\begin{align}
    \vecJ
    =
    \sum_{\lambda \in S} z_1^\lambda \>
    \sum_{l = 0}^{r_\lambda - 1} \log^l\brk{z_1} \>
    \vecJ^{\brk{\lambda, l}}\brk{z_1, \ldots, z_n}
    \>,
    \label{eq:asymptotic_expansion_generic}
\end{align}
where $S$ is a finite list of $\eps$-dependent polynomials, and the
coefficients $\vecJ^{\brk{\lambda, l}}$ are holomorphic functions of the
expansion variable $z_1$. The different polynomial powers $\lambda$ correspond to
different \textit{regions} in the asymptotic expansion, while the logarithmic
powers $l$ arise when the residue matrix $\Omega_{1}\supbrk{-1}$ has Jordan blocks
of size $r_\lambda$ larger than $1 \times 1$.
For the Pfaffian systems arising in this work, 
the series expansions of the solution vector $\vecJ$ 
will be free of the logarithmic terms \brk{meaning $r_\lambda = 1$ 
for every $\lambda \in S$}, 
so to simplify notation we drop the corresponding index and write
\begin{align}\label{eq:sol_eig}
    \vecJ = \sum_{\lambda \in S} z_{1}^{\lambda} \>
    \vecJ\supbrk{\lambda}\brk{z_{1},\dots,z_{n}}
    =
    \sum_{\lambda \in S}
    \sum_{k=0}^\infty
        z_{1}^{\lambda + k}
        \>
        \vecJ_{k}\supbrk{\lambda}\brk{z_{2},\dots,z_{n}}
        \>.
\end{align}
Next, our goal is to understand how to determine the $\vecJ_k^{\brk{\lambda}}$ 
coefficients order-by-order in the $z_1$-expansion.

\subsubsection{
Splitting into regions
\label{subsubsec:LOtheory}}
The absence of logarithmic factors in the expansion~\labelcref{eq:asymptotic_expansion_generic} 
is equivalent to the property of the residue matrix~$\Omega_{1}\supbrk{-1}$ being diagonalizable.
This means that the Jordan decomposition of the matrix contains only blocks of $1 \times 1$ size:\footnote{
    In other words, the geometric and algebraic multiplicity of each eigenvalue $\lambda$ coincide. 
}
\begin{align}
    \mathrm{JordanDecomposition}\bigbrk{\Omega_{1}\supbrk{-1}}
    &=
    \mathrm{diag}\brk{\,
        \underbracket[0.4pt]{\lambda_{1},\ldots,\lambda_{1}}_{N_{\lambda_1}},
        \underbracket[0.4pt]{\lambda_{2},\ldots,\lambda_{2}}_{N_{\lambda_2}},
        \ldots
    }\,.
\end{align}
The set of unique eigenvalues\footnote{
    The eigenvalues of $\Omega\supbrk{-1}_1$ are the roots of its characteristic polynomial, so their explicit determination may be impractical in general. However, the normal form of the Pfaffian system can be constructed without evaluating the full Jordan decomposition~\cite[Appendix B.3]{Chestnov:2023kww}. In our case, the eigenvalues turn out to be simple linear functions of the dimensional regulator~$\eps$.
} corresponds to different regions in the
asymptotic expansion of the solution vector $\vecJ$
\begin{align}
    S = \mathrm{Eigenvalues}\bigbrk{\Omega_{1}\supbrk{-1}} &= \brc{
        \lambda_{1}, \lambda_{2}, \ldots, \lambda_{\abs{S}}
    }\,,
\end{align}
and thus completely characterizes the structure of the solution 
vector~\labelcref{eq:sol_eig} at the leading order in $z_1$-expansion.

The multiplicity $N_{\lambda}$ of each eigenvalue $\lambda$ gives the number of
independent solutions associated with that region. 

For each eigenvalue $\lambda \in S$, the multiplicity $N_{\lambda}$ denotes 
the dimension of the corresponding eigenspace, and we can restrict the full 
Pfaffian system to a smaller subsystem of size $N_{\lambda} \times N_{\lambda}$ that describes 
only the corresponding region of the solution.
\subsubsection{Leading order solution \label{subsubsec:LOsol}}
To showcase this mechanism, let us consider the leading-order terms in the
$z_{1} \rightarrow 0$ limit of the differential equation~\labelcref{eq:normal_singular}:
\begin{align}
    \sum_{\lambda\in S}z_{1}^{\lambda} \Bigbrk{
        \frac{\lambda}{z_1} \vecJ_{0}\supbrk{\lambda}
        -
        \frac{1}{z_1}\Omega_{1}\supbrk{-1} \cdot \vecJ_{0}\supbrk{\lambda}+\mathcal{O}\brk{z_{1}^{0}}
    } = 0\,.
\end{align}
This leads to linear constraint on the solution vector $\vecJ_{0}\supbrk{\lambda}$ for each eigenvalue $\lambda \in S$:
\begin{align}
    \Bigbrk{\lambda\mId_{\nu}-\Omega_{1}\supbrk{-1}}
    \cdot
    \vecJ_{0}\supbrk{\lambda} &= 0 
    \,,
    \label{eq:shifted_residue_matrix}
\end{align}
Hence, for each eigenvalue $\lambda\in S$ we need to solve
the differential equation system:
\begin{align}
    \partial_{i}\vecJ_{0}\supbrk{\lambda}
    &=
    \Omega_{i}\supbrk{0}\cdot\vecJ_{0}\supbrk{\lambda}
    \,,
    \quad i=2,\ldots,n
    \,,
    \label{eq:restricted_rest}
\end{align}
restricted to the nullspace of the $\bigbrk{\lambda \mId_{\nu}-\Omega_{1}\supbrk{-1}}$ 
matrix.
Such linear constraint can be encoded in terms of a $(\nu-N_{\lambda})\times\nu$
matrix $R\supbrk{\lambda}$, whose rows form a basis of the
aforementioned nullspace:
\begin{align}
    \label{eq:symm}
    R\supbrk{\lambda}\cdot\vecJ_{0}\supbrk{\lambda} = 0\,.
\end{align}
In practice $R\supbrk{\lambda}$ can be constructed by extracting
independent rows of $\bigbrk{\lambda\mId_{\nu}-\Omega_{1}\supbrk{-1}}$.
This restriction effectively reduces the size of the system from $\nu \times \nu$ to
$N_{\lambda} \times N_{\lambda}$.

To implement this reduction, we perform a basis transformation. In practice, this can be done with a change of basis tailored for each eigenvalue:
\begin{align}
    \vecF\supbrk{\lambda}
    &=
    \sum_{k=0}^{\infty}\vecF_{k}\supbrk{\lambda}z_{1}^{k} =  M\supbrk{\lambda}\cdot\vecJ\supbrk{\lambda}\,,
    \quad
    \text{with}
    \hspace{.7em}
    M\supbrk{\lambda} =
    \NiceMatrixOptions{cell-space-limits = 3pt}
    \begin{bNiceMatrix}
        {B\supbrk{\lambda}}
        \\
        {R\supbrk{\lambda}}
        \CodeAfter
            \tikz \draw[
                line width=.4pt, gr,
            ]
                (1|-2) -- (2)
            ;
    \end{bNiceMatrix}
    \,.
    \label{eq:restriction_basis_change}
\end{align}
where $B\supbrk{\lambda}$ can be any $N_{\lambda} \times \nu$ full-rank
rectangular matrix, such that the full $M\supbrk{\lambda}$ matrix is invertible.
The new solution vector $\vecF\supbrk{\lambda}$ satisfies
\begin{align}
    \bigbrk{\vecF_{0}\supbrk{\lambda}}_{i}=0\,,\quad  i > N_{\lambda}\,.
\end{align}
This effectively restricts the remaining degrees of freedom to a $N_{\lambda}\times N_{\lambda}$ subsystem.
The differential equations in directions $i = 2, \ldots, n$ for $\vecF_{0}\supbrk{\lambda}$ will 
take the following block-triangular form:
\vspace{.5cm}
\begin{align}
    \partial_{i}\vecF_{0}\supbrk{\lambda} =
    \bigbrk{
        \partial_{i}M\supbrk{\lambda}+M\supbrk{\lambda}\cdot\Omega_{i}\supbrk{0}
    }
    \cdot \bigbrk{M\supbrk{\lambda}}^{-1}
    \cdot \vecF_{0}\supbrk{\lambda}
    =
    \hspace{2em}
    \NiceMatrixOptions{
        xdots = {horizontal-labels, line-style = <->},
        cell-space-limits = 3pt
    }
    \begin{bNiceArray}{cw{c}{1cm}c c}[margin = 2pt]
        \Block{3-3}{
            \widetilde{\Omega}\supbrk{0}_{i} 
        } & & & \Block{3-1}{\ast}
        \\
        & & &
        \\
        & & &
        \\
        \Block{1-3}{\gr{\mathbb{0}}} & & & \ast
        \CodeAfter
            \tikz \draw[
                line width=.4pt, gr,
            ]
                (1|-4) -- (5|-4)
                (1-|4) -- (5-|4)
            ;
            \tikz \draw[<->, gr, shorten <> = .5em]
                (1) ++(-.3, 0) -- ($(1|-4) + (-.3, 0)$)
                node [midway, rotate=0, anchor = east] {\scriptsize$N_{\lambda}$}
            ;
            \tikz \draw[<->, gr, shorten <> = .2em]
                (1) ++(0, .2) -- ($(1-|4) + (0, .2)$)
                node [midway, anchor = south] {\scriptsize$N_{\lambda}$}
            ;
    \end{bNiceArray}
    \cdot \vecF_{0}\supbrk{\lambda}\,,
\end{align}
where we should only solve the new $N_{\lambda} \times N_{\lambda}$ subsystem 
described by the top left $\widetilde{\Omega}\supbrk{0}_{i}$ matrix block.
Once this has been achieved, we use the inversion formula:
\begin{align}
\vecJ_{0}\supbrk{\lambda} = \bigbrk{M\supbrk{\lambda}}^{-1}\cdot\vecF_{0}\supbrk{\lambda}\, ,
\end{align}
to determine the leading order contribution to the full solution for the given 
eigenvalue $\lambda$.

This procedure is repeated for each eigenvalue $\lambda \in S$, 
thus yielding the complete set of leading order solutions $\vecJ_{0}\supbrk{\lambda}$.

\subsubsection{Higher order recursion\label{subsubsec:recursive}}
Once the leading order solutions $\vecJ_{0}\supbrk{\lambda}$ are known, 
we return to the $z_{1}$ differential equation, 
substituting the full expansion~\labelcref{eq:sol_eig}:
\begin{align}
    \partial_{1}\Bigbrk{
        \sum_{\lambda\in S}z_{1}^{\lambda}\sum_{k=0}^{\infty}\vecJ_{k}\supbrk{\lambda}z_{1}^{k}
    }
    &=
    \sum_{m=-1}^{\infty}\Omega_{1}\supbrk{m}z_{1}^{m}
    \Bigbrk{
        \sum_{\lambda\in S}z_{1}^{\lambda}
        \sum_{k=0}^{\infty}\vecJ_{k}\supbrk{\lambda}z_{1}^{k}
    }\,,
\end{align}
to construct the subleading solutions $\vecJ\supbrk{\lambda}_k$ for $k \ge 1$.
Extracting the coefficients of the $z_{1}^{\lambda+k}$ monomial we get
\begin{align}
    \label{eq:subleading}
    (k+\lambda+1)\vecJ_{k+1}\supbrk{\lambda}
    &=
    \sum_{m=-1}^{k}\Omega_{1}\supbrk{m}\cdot\vecJ_{k-m}\supbrk{\lambda}\,.
\end{align}
From eq.~\eqref{eq:subleading} we can obtain a recursive formula for the subleading 
corrections $\vecJ_{k}\supbrk{\lambda}$ that reads
\begin{align}
    \vecJ_{k+1}\supbrk{\lambda}
    &=
    \Bigbrk{
        \brk{k+\lambda+1} \, \mId_{\nu}-\Omega_{1}\supbrk{-1}
    }^{-1}\cdot\sum_{m=0}^{k}\Omega_{1}\supbrk{m}\cdot\vecJ_{k-m}\supbrk{\lambda}
    \,.
    \label{eq:recursive}
\end{align}
Therefore, starting from the solutions $\vecJ_{0}\supbrk{\lambda}$ at leading order 
in the expansion $z_{1}\rightarrow0$, we can obtain all orders in this series using a simple matrix multiplication procedure.

\subsubsection{Symmetry relations\label{subsubsec:symmetry}}

In some cases (see Section 5.4 of~\cite{Chestnov:2023kww}), we find that as $z_{1}\rightarrow0$, 
certain symmetry relations arise within our system.
For example, it may be clear that some of the elements of our solution vector $\vecJ\brk{0, z_{2}, \ldots, z_{n}}$ 
are related or equal to each other in the limit.
We can characterize these relations using the nullspace of an $r\times\nu$ matrix $R$, as in~\eqref{eq:symm}, 
where this time $r$ is the number of symmetry relations.
Thus, we may once again write a suitable basis transformation to encode this information
\begin{align}
    \vecF = M\cdot\vecJ \quad \text{with} \hspace{.7em}
    M =
    \hspace{1em}
    \NiceMatrixOptions{
        xdots = {horizontal-labels, line-style = <->},
        code-for-first-row = \scriptstyle,
        code-for-first-col = \scriptstyle,
        cell-space-limits = 1pt
    }
    \begin{bNiceMatrix}[first-row, first-col, margin = 3pt]
        & \Hdotsfor{1}[shorten = -2pt]^{\nu}
        \\
        \hspace{5pt} & {B}
        \\
        \Vdotsfor{1}[shorten-end = -1pt, shorten-start = 4pt]_{r} & {R}
        \CodeAfter
            \tikz \draw[
                line width=.4pt, gr,
            ]
                (1|-2) -- (2)
            ;
    \end{bNiceMatrix}
    \,,
\end{align}
where, as before, $B$ is an arbitrary $(\nu-r)\times\nu$ matrix is chosen in such a way as to ensure that $M$ is invertible.
The resulting vector will have zero components:
\begin{equation}
    \bigbrk{\vecF\brk{0, z_{2}, \ldots, z_{n}}}_{i} = 0\,,\quad i > \brk{\nu-r}\,.
\end{equation}
in the limit.
Furthermore, by definition, $B$ will extract the independent solutions modulo the symmetry relations, 
which will correspond to the first $\brk{\nu - r}$ entries of the $\vecF\brk{0, z_{2}, \ldots, z_{n}}$ solution vector.
This restricts to a $\brk{\nu-r} \times \brk{\nu-r}$ Pfaffian system, and we will show this in practice in \secref{subsec:PNLO}.

\subsection{A ``pen and paper'' example: rediscovering the \texorpdfstring{$_2F_1$}{dum} hypergeometric function}
In order to showcase the main features of the method presented above,
we apply it to the simple univariate case of a $_2F_1$ hypergeometric function, which satisfies
a second-order differential equation. In this case the function depends on a single kinematic variable $z$. The restriction of the differential system to a given kinematic limit does not depend on any kinematic variable and it is just a constant. Univariate cases have already appeared in the context of Feynman integrals to fix boundary conditions, see for example ~\cite{Dulat:2014mda, Mastrolia:2017pfy}.

Let us consider the Gau\ss{} hypergeometric function $y(z) \equiv {}_2 F_1\brk{a, b, c; z}$, which we take to be \emph{defined} by the integral
\begin{align}
    y\brk{z} \equiv \frac{\Gamma\brk{c}}{\Gamma\brk{b} \Gamma\brk{-b + c}}
    \int_0^1 \dd t
    \> t^{b-1} \brk{1 - t}^{- b + c - 1} \brk{1 - z t}^{-a}
    \stackrel{?}{=}
    \sum_{k = 0}^\infty
    \frac{\brk{a}_k \, \brk{b}_k}{k! \, \brk{c}_k} \, z^k
    \,,
    \label{eq:example_Gauss}
\end{align}
where $\brk{a}_k = a \, \brk{a + 1} \ldots \brk{a + k - 1}$ denotes the Pochhammer symbol.
This function solves the celebrated hypergeometric differential equation given by
\begin{align}
    z \brk{1 - z} \> \partial_z^2 y
    + \bigbrk{c - \brk{a + b + 1} z} \> \partial_z y
    - a b \> y = 0
    \,.
    \label{eq:example_hypergeometric_ode}
\end{align}
Below, we show how each step of the restriction procedure outlined in~\secref{subsec:restriction} can be applied to ``rediscover'' the $z \to 0$ series expansion in the last equality of \eqref{eq:example_Gauss}. As before, we procede in three steps:
\begin{enumerate}
    \item
        \textit{Splitting into regions}: Let us introduce a vector of unknowns $\vecJ$ to rewrite the second order differential equation~\labelcref{eq:example_hypergeometric_ode} as a Pfaffian system 
        \begin{align}
            \vecJ \equiv
            \begin{bNiceMatrix}
                y \\ \brk{1 - z} \, \partial_z y
            \end{bNiceMatrix}
            \,,
            \quad
            \partial_z \vecJ = \Omega_z \cdot \vecJ
            \,,
            \label{eq:example_Pfaffian}
        \end{align}
        where the connection matrix is
        \begin{align}
            \Omega_z =
            \begin{bNiceMatrix}[columns-width = 2em]
                \mZero & \mZero
                \\
                a b & -c
            \end{bNiceMatrix}
            \> \frac{1}{z}
            +
            \begin{bNiceMatrix}[columns-width = 2em]
                \mZero & 1
                \\
                \mZero & a + b - c
            \end{bNiceMatrix}
            \> \frac{1}{1 - z}
            \>.
        \end{align}
        The matrix contains only simple poles in $z$ variable, thanks to our choice of basis normalization in~\eqref{eq:example_Pfaffian}.
        The $z \to 0$ expansion of the connection matrix reveals that the Pfaffian system is already in normal form:
        \begin{align}
        \hspace{-0.2cm}
            \Omega_z = \sum_{m = -1}^\infty \Omega_z\supbrk{m} z^m
            =
            \begin{bNiceMatrix}[columns-width = 2em]
                \mZero & \mZero
                \\
                a b & -c
            \end{bNiceMatrix}
            \, z^{-1}
            +
            \begin{bNiceMatrix}[columns-width = 2em]
                \mZero & 1
                \\
                 \mZero & a + b - c + 1
            \end{bNiceMatrix}
            \, \brk{1 {+} z {+} z^2 {+} \ldots}
            \,.
            \label{eq:example_residue_matrix}
        \end{align}
        In particular, we see that there are two regions corresponding to the two eigenvalues of the residue matrix
        \begin{align}
            S = \mathrm{Eigenvalues}\bigbrk{\Omega_{z}\supbrk{-1}} &= \brc{0, -c}\,.
        \end{align}
        Thus, in the case of system~\labelcref{eq:example_Pfaffian}, the general form of the asymptotic expansion~\labelcref{eq:sol_eig} reduces to 
        \begin{align}
            \vecJ =
            \sum_{k = 0}^\infty z^k \> \vecJ\supbrk{0}_k
            +
            \sum_{k = 0}^\infty z^{-c + k} \> \vecJ\supbrk{-c}_k
            \,.
            \label{eq:example_expansion_full}
        \end{align}
    \item
        \textit{Leading order solution}:
        To determine the exact form of the series expansion~\labelcref{eq:example_expansion_full}, we first impose linear constraints~\eqref{eq:symm} on the leading order coefficients
        \begin{align}
            R\supbrk{0}
            \cdot
            \vecJ\supbrk{0}_0
            = 0
            \,,
            \quad
            \text{and}
            \quad
            R\supbrk{-c}
            \cdot
            \vecJ\supbrk{-c}_0
            = 0
            \,,
        \end{align}
        where the matrices $R\supbrk{0}$ and $R\supbrk{\lambda}$ are derived from independent rows of the respective \brk{shifted} residue matrices~\labelcref{eq:shifted_residue_matrix}
        \begin{align}
            R\supbrk{0} =
            \begin{bNiceMatrix}[columns-width = 1.5em]
                -a b & c
            \end{bNiceMatrix}
            \,,
            \quad
            \text{and}
            \quad
            R\supbrk{-c} =
            \begin{bNiceMatrix}[columns-width = 1.5em]
                -c & \mZero
            \end{bNiceMatrix}
            \,,
        \end{align}
        which essentially imply that the leading coefficients $\vecJ\supbrk{0}_0$ and $\vecJ\supbrk{-c}_0$ are the corresponding eigenvectors of the residue matrix~\labelcref{eq:example_residue_matrix}.
        These constraints fix the form of the leading order solutions up to overall constants to be
        \begin{align}
            \vecJ\supbrk{0}_0 \propto
            \begin{bNiceMatrix}
                1 \\ a b /c
            \end{bNiceMatrix}
            \,,
            \quad
            \text{and}
            \quad
            \vecJ\supbrk{-c}_0 \propto
            \begin{bNiceMatrix}
                \mZero \\ 1
            \end{bNiceMatrix}
            \,.
        \end{align}
        We may fix these constants from the analysis of the integrand in the two regions. In the $\lambda = 0$ region we may simply take $z \to 0$ under the integral sign:
        \begin{align}
            y|_{z \to 0}
            = \frac{\Gamma\brk{c}}{\Gamma\brk{b} \Gamma\brk{-b + c}} \int_0^1 \, \dd t \> t^{b-1} \brk{1 - t}^{c - b - 1} = 1
            \,,
        \end{align}
        and similarly $\partial_z y|_{z \to 0} = ab / c\,$. The same analysis can be carried out for the $\lambda = -c$ region upon the change of integration variable $t \mapsto t / z$, which shows that its contribution is vanishing for the integral~\eqref{eq:example_Gauss}
        leaving only the first region $\lambda = 0$ to contribute to the expansion~\labelcref{eq:example_expansion_full}:
        \begin{align}
            \vecJ =
            \begin{bNiceMatrix}
                1 \\ a b /c
            \end{bNiceMatrix}
            +
            \sum_{k = 1}^\infty z^k \> \vecJ\supbrk{0}_k
            \label{eq:example_expansion_leading}
            \,.
        \end{align}
    \item
        \textit{Higher order recursion}: With the leading expansion order coefficients nailed down, we may directly apply the general recursive formula~\labelcref{eq:recursive} 
        \begin{align}
            \vecJ_{k+1}\supbrk{0}
            &=
            \NiceMatrixOptions{cell-space-limits = 3pt}
            \begin{bNiceMatrix}[columns-width = 1.5em]
                \mZero & \frac{1}{k + 1}
                \\
                \mZero & \frac{a b}{c \brk{k + 1}} - \frac{\brk{a - c}\brk{b - c}}{
                    c \brk{c + k + 1}
                }
            \end{bNiceMatrix}
            \cdot
            \sum_{m=0}^{k} \vecJ_{k-m}\supbrk{\lambda}
            \,,
        \end{align}
        whose form is drastically simplified due to the special factorized form of the connection matrix coefficients~\labelcref{eq:example_residue_matrix}.
        A straightforward computation then reveals the first few orders in the expansion~\labelcref{eq:example_expansion_leading} 
        \begin{align}
            \vecJ =
            \begin{bNiceMatrix}
                1 \\ \frac{a b}{c}
            \end{bNiceMatrix}
            {+}
            \begin{bNiceMatrix}
                \frac{a b}{c} \\ \frac{a b \bigbrk{\brk{a + 1}\brk{b + 1} - \brk{c + 1}}}{c \brk{c + 1}}
            \end{bNiceMatrix}
            z
            {+}
            \begin{bNiceMatrix}
                \frac{a \brk{a + 1} b \brk{b + 1}}{2 c \brk{c + 1}}
                \\
                \frac{a \brk{a + 1} b \brk{b + 1} \bigbrk{\brk{a + 2}\brk{b + 2} - 2 \brk{c + 2}}}{2 c \brk{c + 1} \brk{c + 2}}
            \end{bNiceMatrix}
            z^2
            {+} \mathcal{O}\brk{z^3}
            \,,
        \end{align}
        which can be resummed in terms of Pochhammer symbols and, at the next step, back in terms of the hypergeometric function
        \begin{align*}
            \vecJ = \sum_{k = 0}^\infty
            \begin{bNiceMatrix}
                \frac{\brk{a}_{k} \, \brk{b}_{k}}{k! \, \brk{c}_{k}}
                \\
                \frac{a b}{c} \frac{\brk{a + 1}_{k} \, \brk{b + 1}_{k}}{k! \, \brk{c + 1}_{k}}
                - \frac{\brk{a}_{k} \, \brk{b}_{k}}{\brk{k - 1}! \, \brk{c}_{k}}
            \end{bNiceMatrix}
            \,
            z^k
            &=
            \begin{bNiceMatrix}
                {}_2F_1\brk{a, b, c; z}
                \\
                \frac{a b}{c} \brk{1 - z} \> {}_2F_1\brk{a + 1, b + 1, c + 1; z}
            \end{bNiceMatrix}
            \\
            &=
            \begin{bNiceMatrix}
                {}_2F_1\brk{a, b, c; z}
                \\
                \brk{1 - z} \> \partial_z \bigbrk{{}_2F_1\brk{a, b, c; z}}
            \end{bNiceMatrix}
            \,,
        \end{align*}
        which agrees with our starting point in~\cref{eq:example_Pfaffian}.
\end{enumerate}

This concludes our review of the restriction theory and its applications to the asymptotic expansion of solutions of systems of differential equations. The same three main computational steps presented in~\secref{subsec:restriction} and showcased in the example above can be applied to systems of partial differential equations, making restriction theory a valuable tool for the study of gravitational waveforms in kinematic expansions, which we turn to next.

%% file: Sections/Soft_expansion.tex
{\section{Soft expansion}\label{sec:soft_expansion}}
The soft expansion of the waveform probes the regime in which the emitted-graviton frequency
is much smaller than all other dynamical scales,
\begin{equation}
    k^\mu=\omega n^\mu \,, \qquad \omega \to 0 \, , \quad n^2=0 \, .
\end{equation}
In this limit, amplitudes obey universal factorization formulas--\emph{soft theorems}--first derived by Weinberg~\cite{Weinberg:1964ew,Weinberg:1965nx}.
These results have been revisited and extended for PM waveforms at
LO~\cite{Mougiakakos:2021ckm,Jakobsen:2021smu,Aoude:2023dui},
NLO~\cite{Georgoudis:2023eke,Bini:2024rsy},
and NNLO~\cite{Georgoudis:2025vkk},
and they provide direct control of the infrared structure of radiation, including the
gravitational memory effect.
We parametrize the kinematic invariants as:
\begin{equation}\label{eq:soft_vars}
    \begin{aligned}
        z_{1} = \hat{w}_{1}\,, \quad z_{2}=\frac{\hat{w}_{2}}{\hat{w}_{1}}\,, \quad z_{3} = \gamma\,, \quad z_{4} = \frac{i\hat{b}\cdot k}{\hat{w}_{1}}\,,
    \end{aligned}
\end{equation}
allowing us to make the change of variables from $    (\hat{w}_{1},\hat{w}_{2},\gamma,\hat{b}\cdot k)$ to $(z_{1},z_{2},z_{3},z_{4})$,
where we consider the limit:
\begin{equation}
    z_{1}\rightarrow0\, .
\end{equation} 
Under this change of variables, the Pfaffian system ~\eqref{eq:DE} is pulled 
back to the $\mathbf{z}$-variables via the chain rule, yielding
\begin{equation}
    \partial_{i}\mathbf{J} = \Omega_{i}\cdot\mathbf{J}, \quad i=1,\dots,4\,.
\end{equation}
We study the Pfaffian system in the vicinity of $z_{1}=0$,
following the restriction-theory workflow of \secref{subsec:restriction}:
bring the system to normal form by a gauge transformation, extract the residue matrix,
decompose into regions, restrict, and solve recursively.
In the soft limit the master integrals and all kinematic coefficients remain 
finite as $\varepsilon\to0$,
so it suffices to determine the solution through $\mathcal{O}(\varepsilon^{0})$.

\subsection{Splitting into regions}

We first perform a local Laurent expansion of the system of DEs near $z_{1}=0$,
\begin{equation}
    \Omega_{i}(z_{1},\mathbf{z})=\sum_{k=-2}^{\infty}\Omega_{i}^{(k)}(\mathbf{z})\,z_{1}^{k}, \qquad i=1,\dots,4\,,
\end{equation}
which exhibits second-order poles in $z_{1}$ and is therefore not in the normal form 
discussed in \secref{subsec:restriction}. 
To enforce normal form--i.e.,  to eliminate the $z_{1}^{-2}$ terms and 
leave at most simple poles in $z_{1}$--we perform a rational gauge transformation
\begin{equation}
    \label{eq:gauge_soft}
    \mathbf{J} = G\cdot\mathbf{J}', \quad G = \begin{bNiceMatrix}
        \mzero&\mzero&\mzero&\mzero&-\frac{2(1+z_{2}^{2}-2z_{2}z_{3})}{z_{4}}&\mzero\\
        \mzero&\mzero&1&\mzero&\mzero&\mzero\\
        \mzero&\mzero&\mzero&\mzero&\mzero&1\\
        \mzero&1&\mzero&\mzero&\mzero&\mzero\\
        \frac{1}{z_{1}(1+2\varepsilon)}&\mzero&\mzero&\frac{1}{z_{1}^{2}}&\frac{1}{z_{1}}&\mzero\\
        \frac{1}{z_{1}}&\mzero&\mzero&\mzero&\frac{1}{z_{1}}&\mzero
    \end{bNiceMatrix}\,.
\end{equation}
With this gauge choice the system is in normal form (\eqref{eq:normal_singular} and \eqref{eq:normal_rest}), with residue matrix
\begin{equation}
    \Omega_{1}^{(-1)} = \begin{bNiceMatrix}
        \mzero&\frac{z_{4}}{2X}&\frac{2\varepsilon z_{2}z_{4}(z_{3}-z_{2})}{X}&\mzero&-\frac{\varepsilon z_{4}}{X}\\
        \mzero&\mzero&\mzero&\mzero&\mzero&\mzero\\
        \mzero&\mzero&\mzero&\mzero&\mzero&\mzero\\
        \mzero&\mzero&\mzero&-2\varepsilon&\mzero&\mzero\\
        \mzero&\mzero&-\frac{z_{4}}{2X}&\mzero&-2\varepsilon&\mzero\\
        \mzero&1&\mzero&\mzero&\mzero&-2\varepsilon
    \end{bNiceMatrix} \, ,
\end{equation}
where we have the common denominator:
\begin{equation}
    X=1+z_{2}^{2}-2z_{2}z_{3}\,.
\end{equation}
This matrix has Jordan form
\begin{equation}
    \mathrm{JordanDecomposition}\bigbrk{\Omega_{1}^{(-1)}} = \mathrm{diag}\brk{0,0,0,-2\varepsilon,-2\varepsilon,-2\varepsilon}\,.
\end{equation}
This reveals that there are two distinct eigenvalues (each with multiplicity three):
\begin{equation}
    S=\{0^{\times 3},\,(-2\varepsilon)^{\times 3}\}\,,
\end{equation}
and hence the local Frobenius solution splits into the corresponding sectors:
\begin{equation}
    \mathbf{J}' = \mathbf{J}^{(0)}+z_{1}^{-2\varepsilon}\mathbf{J}^{(-2\varepsilon)}, \quad \mathbf{J}^{(\lambda)} = \sum_{k=0}^{\infty}\mathbf{J}_{k}^{(\lambda)}z_{1}^{k}\,.
\end{equation}
Accordingly, each eigenvalue sector is three-dimensional. 
In each sector the Pfaffian system can be restricted to an independent 
$3\times3$ subsystem by an appropriate rational rotation $M^{(\lambda)}$, 
whose upper block $B^{(\lambda)}$ spans the null space of the relevant residue 
($\Omega_{1}^{(-1)}$ for $\lambda=0$ and $\Omega_{1}^{(-1)}+2\varepsilon\,\mathbb{1}_{6}$ for $\lambda=-2\varepsilon$).

Physically, the $\lambda=0$ sector is the hard region, 
where the exchanged graviton is hard relative to the emitted soft graviton, 
$\abs{q}\gg\abs{k}$. The $\lambda=-2\varepsilon$ sector is the soft region, 
where the exchanged graviton is comparable in energy to the emission, 
$\abs{q}\sim\abs{k}$. We analyze these two regions in turn.

\subsection{Leading order solution}
\paragraph{Hard region \texorpdfstring{$\lambda=0$}{dum}.}
In the hard sector we restrict the Pfaffian system to a $3\times3$ 
subsystem by the rational rotation
$\mathbf{F}^{(0)} = M^{(0)}\cdot \mathbf{J}^{(0)}$, where $M^{(0)}$ reads as:
\begin{equation}
    M\supbrk{0} =
    {
        \NiceMatrixOptions{cell-space-limits = 3pt}
        \begin{bNiceMatrix}
            {B\supbrk{0}}
            \\
            {R\supbrk{0}}
            \CodeAfter
                \tikz \draw[
                    line width=.4pt, gr,
                ]
                    (1|-2) -- (2)
                ;
        \end{bNiceMatrix}
    }
    =
    \begin{bNiceMatrix}
        \mzero & 2\varepsilon&\mzero&\mzero&\mzero&1 
        \\
        \mzero & \mzero&-\frac{4\varepsilon X}{z_{4}}&\mzero&1&\mzero
        \\
        1&\mzero&\mzero&\mzero&\mzero&\mzero
        \\
        \mzero&1&\mzero&\mzero&\mzero&-2\varepsilon
        \\
        \mzero&\mzero&1&\mzero&\frac{4\varepsilon X}{z_{4}}&\mzero
        \\
        \mzero&\mzero&\mzero&1&\mzero&\mzero
        \CodeAfter
            \tikz \draw[
                line width=.4pt, gr,
            ]
                (1|-4) -- (7|-4)
            ;
    \end{bNiceMatrix}
    \,.
\end{equation}
The upper block $B^{(0)}$ spans the null space of $\Omega_{1}^{(-1)}$, 
while the lower block $R^{(0)}$ completes $M^{(0)}$ with linearly independent 
rows of $\Omega_{1}^{(-1)}$. 
In the limit $z_{1}\to 0$ one has $(\mathbf{F}_{0}^{(0)})_{4,5,6}=0$, 
so the dynamics reduces to a closed $3\times3$ subsystem for the first 
three components of $\mathbf{F}_{0}^{(0)}$:
\begin{equation}
    \partial_{i}\mathbf{F}_{0}^{(0)}
    =
    \begin{bNiceArray}{cw{c}{1cm}cc}[margin]
        \Block{3-3}{
            \Omega_{i}^{(3\times3)}
        } & & & \Block{3-1}{{\ast}}
        \\
        & & &
        \\
        & & &
        \\
        \Block{1-3}{\gr{\mathbb{0}}} & & & {\ast}
        \CodeAfter
            \tikz \draw[
                line width=.4pt, gr,
            ]
                (1|-4) -- (5|-4)
                (1-|4) -- (5-|4)
            ;
    \end{bNiceArray}
    \cdot \mathbf{F}_{0}^{(0)}
    \,.
\end{equation}
where
\begin{align}
        \Omega_{2}^{(3\times3)} &= \begin{bNiceMatrix}
        \mzero&\mzero&\mzero\\
        \mzero&\frac{2(z_{3}-z_{2})(z_{4}^{2}-16\varepsilon^{2}X^{2})}{X(z_{4}^{2}+16\varepsilon^{2}X^{2})}&\mzero\\
        \frac{(1+2\varepsilon)(z_{3}-z_{2})z_{4}}{2(1+4\varepsilon^{2})X(X-z_{4}^{2})}&\frac{2\varepsilon(z_{3}-z_{2})z_{4}^{2}}{X(z_{4}^{2}+16\varepsilon^{2}X^{2})}&\frac{(z_{3}-z_{2})(X-2(1+\varepsilon)z_{4}^{2})}{X(X-z_{4}^{2})}
    \end{bNiceMatrix}\, , \notag\\
    \Omega_{3}^{(3\times3)} &= \begin{bNiceMatrix}
        \mzero&\mzero&\mzero\\
        \mzero&\frac{2z_{2}(z_{4}^{2}-16\varepsilon^{2}X^{2})}{X(z_{4}^{2}+16\varepsilon^{2}X^{2})}&\mzero\\
        \frac{(1+2\varepsilon)z_{2}z_{4}}{2(1+4\varepsilon^{2})X(X-z_{4}^{2})}&\frac{2\varepsilon z_{2}z_{4}^{2}}{X(z_{4}^{2}+16\varepsilon^{2}X^{2})}&\frac{z_{2}(X-2(1+\varepsilon)z_{4}^{2})}{X(X-z_{4}^{2})}
    \end{bNiceMatrix}\, , \\
    \Omega_{4}^{(3\times3)} &= \begin{bNiceMatrix}
        \mzero&\mzero&\mzero\\
        \mzero&\frac{z_{4}^{2}-16\varepsilon^{2}X^{2}}{z_{4}(z_{4}^{2}+16\varepsilon^{2}X^{2})}&\mzero\\
        \frac{1+2\varepsilon}{2(1+4\varepsilon^{2})(X-z_{4}^{2})}&\frac{2\varepsilon z_{4}}{z_{4}^{2}+16\varepsilon^{2}X^{2}}&-\frac{(1+2\varepsilon)z_{4}}{X-z_{4}^{2}}
    \end{bNiceMatrix}\, . \notag
\end{align}
Restricting to the upper-left $3\times3$ block, we find
the system to admit an exact solution in $\varepsilon$.
The boundary vector at the reference point $\mathbf{z}^{*}=(0,1,\tfrac{3}{2},0)$, 
derived in \appref{app:boundarySoft}, is:
\begin{equation}
    \mathbf{F}_{0}^{(0)}\eval_{\mathbf{z}=\mathbf{z}^{*}} = C_{\varepsilon}^{\text{hard}}M^{(0)}\cdot G^{-1}\cdot\left[\begin{array}{c}
        1\\
        2\varepsilon\\
        1\\
        2\varepsilon\\
        \frac{1}{4(1+\varepsilon)}\\
        \frac{1}{2}
    \end{array}\right]\eval_{\mathbf{z}=\mathbf{z}^{*}} = C_{\varepsilon}^{\text{hard}}\left[\begin{array}{c}
        1\\
        \gr{0}\\
        \gr{0}\\
        \noalign{\color{gray!60}\hrule height 0.5pt}
        \gr{0}\\
        \gr{0}\\
        \gr{0}
    \end{array}\right]+\mathcal{O}(\varepsilon)\,,
\end{equation}
where $C_{\varepsilon}^{\text{hard}} = \frac{i(-1)^{\varepsilon}\Gamma(-\varepsilon)}{4\pi^{1-\varepsilon}}$. 
Since the hard-region boundary constant $C_{\varepsilon}^{\text{hard}}$ 
carries a simple pole at $\varepsilon=0$, we retain the DE solution through 
$\mathcal{O}(\varepsilon)$ to extract the finite piece. 
The resulting expression is given by:
\begin{equation}
    \mathbf{F}_{0}^{(0)} = C_{\varepsilon}^{\text{hard}}\left[\begin{array}{c}
        1\\
        -\frac{z_{4}}{2X}\\
        \frac{z_{4}}{2X}\\
        \noalign{\color{gray!60}\hrule height 0.5pt}
        \gr{0}\\
        \gr{0}\\
        \gr{0}
    \end{array}\right]+\mathcal{O}(\varepsilon), \implies \mathbf{J}_{0}^{(0)} = \left[\begin{array}{c}
         \frac{z_{4}}{2X}\\
         2\varepsilon\\
         2\varepsilon\\
         \gr{0}\\
         -\frac{z_{4}}{2X}\\
          1
    \end{array}\right]\,.
\end{equation}
With the leading solution in the $\lambda=0$ sector fixed at $z_{1}\to 0$, 
higher orders in the $z_{1}$ expansion are generated recursively by:
\begin{equation}
    \mathbf{J}_{k+1}^{(0)} = ((k+1)\mathbb{1}_{6}-\Omega_{1}^{(-1)})^{-1}\cdot\sum_{m=0}^{k}\Omega_{1}^{(m)}\cdot\mathbf{J}_{k-m}^{(0)}\,.
\end{equation}
\paragraph{Soft region \texorpdfstring{$\lambda=-2\varepsilon$}{dum}.}

Analogously to the hard sector, we restrict the Pfaffian system in the 
soft sector to a $3\times3$ subsystem by the rational rotation 
$\mathbf{F}^{(-2\varepsilon)} = M^{(-2\varepsilon)}\cdot \mathbf{J}^{(-2\varepsilon)}$, 
where $M^{(-2\varepsilon)}$ reads:
\begin{equation}
    M^{(-2\varepsilon)} = \left[\begin{array}{c}
        B^{(-2\varepsilon)} \\
        \noalign{\color{gray!60}\hrule height 0.5pt}
        R^{(-2\varepsilon)}
    \end{array}\right] = \left[\begin{array}{cccccc}
        1&\mzero&\mzero&\mzero&\mzero&\frac{2X}{z_{4}}
        \\
        \mzero&\mzero&\mzero&1&\mzero&2z_{2}(z_{3}-z_{2})
        \\
        \mzero&\mzero&\mzero&\mzero&1&\mzero
        \\
        \noalign{\color{gray!60}\hrule height 0.5pt}
        1&\mzero&\mzero&\frac{z_{2}z_{4}(z_{3}-z_{2})}{X}&\mzero&-\frac{z_{4}}{2X}
        \\
        \mzero&1&\mzero&\mzero&\mzero&\mzero
        \\
        \mzero&\mzero&1&\mzero&\mzero&\mzero 
    \end{array}\right]\,.
\end{equation}
The upper block $B^{(-2\varepsilon)}$ spans the null space of the shifted residue
$(-2\varepsilon\,\mathbb{1}_{6}-\Omega_{1}^{(-1)})$, while the lower block
$R^{(-2\varepsilon)}$ completes $M^{(-2\varepsilon)}$ with linearly independent rows of $(-2\varepsilon\,\mathbb{1}_{6}-\Omega_{1}^{(-1)})$.
As in the hard sector, in the limit $z_{1}\to 0$ one has
$(\mathbf{F}_{0}^{(-2\varepsilon)})_{4,5,6}=0$, so the dynamics closes on the first three
components and reduces to a closed $3\times3$ subsystem for
$\mathbf{F}_{0}^{(-2\varepsilon)}$:
\begin{equation}
    \partial_{i}\mathbf{F}_{0}^{(-2\varepsilon)} = \left[
    \begin{array}{ccc!{\color{gray!60}\vrule}c} 
    &&& \\
    &\Omega^{(3\times3)}_{i}&&\ast\\
    &&&\\
    \noalign{\color{gray!60}\hrule height 0.5pt} 
        &\gr{\mathbb{0}}&&\ast
    \end{array}
    \right]\cdot\mathbf{F}_{0}^{(-2\varepsilon)}\,.
\end{equation}
The soft boundary condition  computed at the same boundary point 
$\mathbf{z}^{*}$ (see \appref{app:boundarySoft}), reads as:
\begin{equation}
\mathbf{F}_{0}^{(-2\varepsilon)}\eval_{\mathbf{z}=\mathbf{z}^{*}} = C_{\varepsilon}^{\text{soft}}M^{(-2\varepsilon)}\cdot G^{-1}\cdot
\left[
\begin{array}{cccccc}
    1 & \gr{0} & 1 & \gr{0} &
    -\dfrac{4\varepsilon \log\varphi}{\sqrt{5}\,z_{1}^{2}} &
    \gr{0}
\end{array}
\right]^{\!\top}\,,
\end{equation}
where $C_{\varepsilon}^{\text{soft}} = \frac{i(-1)^{\varepsilon}\Gamma(\varepsilon)}{(4\pi)^{1-\varepsilon}}$ and $\varphi = \frac{1 + \sqrt{5}}{2} = 1.618034...$ is the golden ratio.
In contrast to the hard sector, the restricted subsystem admits no
closed-form solution in $\varepsilon$.
We therefore work in an $\varepsilon$-expansion and 
retain terms up to order $\mathcal{O}(\varepsilon)$, which is sufficient to capture the simple pole carried
by the boundary constant $C_{\varepsilon}^{\text{soft}}$.
To organize the series we first bring the system to an 
$\varepsilon$-factorized form via a rotation matrix:
$\mathbf{F}_{0}^{(-2\varepsilon)}=\mathcal{P}\cdot\mathbf{f}$.
The rotation (period) matrix $\mathcal{P}$ is fixed by the
$\varepsilon=0$ solution of the subsystem, and reads as:
\begin{equation}
    \mathcal{P} = \begin{bNiceMatrix}
         \frac{z_4 (z_2-z_3)}{3 \kappa X} & -\frac{X}{z_4}-\frac{z_4}{4
   X} & \mzero \\
 \frac{1}{3 z_2 \kappa} & z_{2}(z_{2}-z_{3}) & \mzero \\
 \mzero & \mzero & \frac{z_4}{X} \\
    \end{bNiceMatrix}\,,
\end{equation}
where $\kappa=\sqrt{z_{3}^{2}-1}$. The rotated system is in dlog form:
\begin{equation}
    \begin{aligned}
    \dd\mathbf{f} &= \varepsilon\dd\Omega_{\mathbf{f}}\cdot\mathbf{f}\,,\\
    \Omega_{\mathbf{f}} &= \left[
    \begin{array}{ccc!{\color{gray!60}\vrule}c} 
    \log\left(\frac{X}{z_{2}^{2}\kappa^{2}}\right)&-\frac{3}{4}\log\left(\frac{(z_{2}-z_{3}-\kappa)(-1+2z_{3}(z_{3}-\kappa))}{z_{2}-z_{3}+\kappa}\right)&3\log\left(\frac{z_{2}-z_{3}-\kappa}{z_{2}-z_{3}+\kappa}\right)& \\
    \mzero&\mzero&\mzero&\ast\\
    \mzero&\mzero&-2\log(z_{2})&\\
    \noalign{\color{gray!60}\hrule height 0.5pt}
        &\gr{\mathbb{0}} &&\ast
    \end{array}
    \right]\,.
    \end{aligned}
\end{equation}
The solution up to $\mathcal{O}(\varepsilon)$ is thus given by
\begin{equation}
    \mathbf{f} = (\mathbb{1}_{6}+\varepsilon\Omega_{\mathbf{f}})\cdot\mathbf{f}_{0}\,,
\end{equation}
where $\mathbf{f}_{0}$ is fixed from the boundary conditions above. 
Putting all of this together we find that the leading term in the soft region is given by:
\begin{equation}
    \mathbf{J}_{0}^{(-2\varepsilon)} = C_{\varepsilon}^{\text{soft}}\left[\begin{array}{c}
         \frac{z_{4}^{2}+4X^{2}}{2X z_{4}}+\frac{\varepsilon(z_{2}-z_{3})z_{4}}{2X \kappa}\log(-1+2z_{3}(z_{3}-\kappa))\\
         2z_{2}(z_{3}-z_{2})+\frac{\varepsilon}{2z_{2}\kappa}\log(-1+2z_{3}(z_{3}-\kappa))\\
         -\frac{z_{4}}{2X}+\frac{\varepsilon z_{4}}{X}\log(z_{2})\\
         \gr{0}\\
         \gr{0}\\
         \gr{0}
    \end{array}\right]+\mathcal{O}(\varepsilon^{2})\,.
\end{equation}
Higher orders in the $z_{1}$ expansion are generated recursively by:
\begin{equation}
    \mathbf{J}_{k+1}^{(-2\varepsilon)} = ((k+1-2\varepsilon)\mathbb{1}_{6}-\Omega_{1}^{(-1)})^{-1}\cdot\sum_{m=0}^{k}\Omega_{1}^{(m)}\cdot\mathbf{J}_{k-m}^{(-2\varepsilon)}\,.
\end{equation}

\subsection{Tree-level waveform in the soft expansion}

The full solution to the system of differential equations is then given by combining the two sectors:
\begin{equation}
    \mathbf{J} = G\cdot(\mathbf{J}^{(0)}+z_{1}^{-2\varepsilon}\mathbf{J}^{(-2\varepsilon)})\,.
\end{equation}
where $G$ is the gauge transformation defined in ~\eqref{eq:gauge_soft}.
In the final answer the pole in $\varepsilon$ cancel and we get a finite result.

The tree-level waveform in the soft expansion is then given by
combining the solution for the master integrals with the
IBP reduction matrix expanded in the soft limit $\mathcal{C}$ as:
\begin{equation}
    \cW_{h,\text{soft}}^{(0)}= \mathbf{T}\cdot \mathcal{C}\cdot\mathbf{J}\,.
\end{equation}
\paragraph{Checks with literature.}
We expand the IBP reduction matrix $\mathcal{C}$ in the soft limit 
and combine it with the sector solutions, 
reconstructing the tree-level waveform through 
$\mathcal{O}(\varepsilon^{0},\omega^{2})$. 
This is sufficient to extract the non-analytic soft behavior:
\begin{equation}
    \cW_{h,\text{soft}}^{(0)}= \kappa^3 m_1 m_2\left(c_1 \frac{1}{\omega}+ c_2 \log (\omega)+c_3 \omega \log(\omega)\right)\, , 
\end{equation}
where:
\begin{align}
            c_1
        &  
        =\frac{i \left(2 \gamma^2-1\right)(\varepsilon_k\cdot u_2 w_1-\varepsilon_k\cdot u_1 w_2)}{16 \pi  w_1^2 w_2^2 \sqrt{\gamma^2-1}}  (-2
   \varepsilon_k\cdot b w_1 w_2+\varepsilon_k\cdot u_1 k\cdot b w_2+\varepsilon_k\cdot u_2
   k\cdot b w_1)\,, \notag\\
   c_2 
   & 
   =
   \frac{ \gamma \left(2 \gamma^2-3\right) (\varepsilon_k\cdot u_2 w_1-\varepsilon_k\cdot u_1 w_2)^2}{16 \pi
    w_1 w_2 \left(\gamma^2-1\right)^{3/2}}\,,
    \\
    c_3 
    & 
    = 
    \frac{-i \gamma \left(2 \gamma^2{-}3\right) (\varepsilon_k{\cdot} u_2 w_1{-}\varepsilon_k{\cdot} u_1 w_2)}{16 \pi  w_1 w_2
   \left(\gamma^2-1\right)^{3/2}}
   [b_2{\cdot}k (\varepsilon_k{\cdot}u_2 w_1{-}\varepsilon_k{\cdot} u_1 w_2)
   {+}w_2(\varepsilon_k{\cdot}b w_1 {-}\varepsilon_k{\cdot} u_1 k{\cdot}b)]\,. \notag
\end{align}
This result is in perfect agreement with previous results at 
leading order~\cite{Mougiakakos:2021ckm,Jakobsen:2021smu,Aoude:2023dui,Georgoudis:2023eke,Bini:2024rsy}.

%% file: Sections/PN_expansion.tex
{\section{Post-Newtonian expansion}\label{sec:PN_expansion}}
The Post-Newtonian (PN) regime corresponds to the low-velocity, weak-field classical limit.
In the center-of-mass (CM) frame we organize all quantities in a Laurent expansion
in a small, dimensionless parameter \(p_\infty\ll1\), with
\begin{equation}
  \gamma=\sqrt{1+p_\infty^{2}},\qquad
  k^\mu = p_\infty\,\omega\, n^\mu,\qquad n^2=0\, .
\end{equation}
In this frame the single-particles CM energies are given by:
\begin{equation}
    E_i = \sqrt{m_i^2 + P_{cm}^2} \, , 
\end{equation}
where the total CM energy and momentum read
\begin{equation}
   E=\sqrt{m_{1}^{2}+m_{2}^{2}+2\gamma m_{1}m_{2}} \, , 
   \quad   P_{\text{cm}} = \frac{m_1 m_2 p_{\infty}}{E}\, .
\end{equation}
where $m_{i}$ are the black hole masses. 
We fix the impact-parameter direction by a unit spacelike vector \(e^\mu\) and write
\begin{equation}
    b_1^\mu  =  \frac{E_2}{E}b^\mu \, , \quad b_2^\mu = -\frac{E_1}{E}b^\mu\,, \quad b^\mu = \sqrt{-b^2}\, e^\mu \,.
\end{equation}
In this frame, 
the scalar products between the four velocities 
and the emitted graviton momentum are given by:
\begin{equation}
    w_{1} = \frac{\omega p_{\infty}}{E}(m_{1}+m_{2}(\gamma+p_{\infty}n\cdot e)), \quad w_{2} = \frac{\omega p_{\infty}}{E}(m_{2}+m_{1}(\gamma-p_{\infty}n\cdot e))\,.
\end{equation}
Specializing to four-dimensional external kinematics, 
the four-vectors take the form:
\begin{equation}
    n = (1,\sin\theta\cos\phi,\sin\theta\sin\phi,\cos\theta), \quad b = (0,1,0,0), \quad e = (0,0,1,0)\,.
\end{equation}
where $\theta$ and $\phi$ are the spherical angles of the emitted graviton.

At this stage, it is convenient to trade the dependence on spherical coordinates,
with the complex conjugated variables\footnote{Note that our normalization convention, which includes an extra factor of $i$ for $z_3$ and $z_4$, is non-standard; typically, this factor is omitted.} 
\begin{equation}\label{eq:stereo}
    z_{3} =i\, \e^{i\phi}\tan{\theta/2}, \quad 
    z_{4} =i\, \e^{-i\phi}\tan{\theta/2}\,,
\end{equation}
mapping the two-sphere to the Riemann sphere via ``stereographic projection.'' Note that \eqref{eq:stereo} can be explicitly inverted as:
\begin{equation}
    \cos\theta = \frac{1+z_{3} z_{4}}{1-z_{3}z_{4}}, \quad \cos\phi = \frac{z_{3}+z_{4}}{2\sqrt{z_{3}z_{4}}}\,.
\end{equation}
We also introduce the mass ratio $\nu=\frac{m_{1}}{m_{2}}$ and define:
\begin{equation}
    z_{1}=p_{\infty}\, , \quad z_{2}=\omega \, . 
\end{equation}
Next, we change variables from $(\hat{w}_{1},\hat{w}_{2},\gamma,\hat{b}\cdot k)$ to $(z_{1},z_{2},z_{3},z_{4})$.
We are interested in the limit:
\begin{equation}
    z_1 \to 0 \, .
\end{equation}
After that, the four-velocities and helicity polarizations are explicitly given by
\begin{equation}
    u_{1} = \frac{1}{E}(\gamma+\nu,0,p_{\infty},0), \quad 
    u_{2} = \frac{1}{E}(1+\gamma\nu,0,-p_{\infty}\nu,0), \quad 
    \varepsilon^{\pm} = \frac{1}{\sqrt{2}}\left(\partial_{\theta}n\pm \frac{i}{\sin\theta}\partial_{\phi}n\right)\,,
\end{equation}
allowing all tensor structures to be written explicitly in terms of 
$\mathbf{z}=(z_1,\dots,z_4)$.
With these variables the tree-level waveform takes the compact form
\begin{equation}
    \cW_{h}^{(0)}(\omega,\hat{\mathbf{n}}) = \mathbf{C}\cdot\mathbf{J}, \quad \mathbf{C} = \mathbf{T}\cdot\mathcal{C}\,,
\end{equation}
and the basis $\mathbf{J}$ obeys first-order differential equations:
\begin{equation}
    \partial_{i}\mathbf{J} = \Omega_{i}\cdot\mathbf{J}, \quad i=1,\dots,4\,.
\end{equation}
Since no $\varepsilon$-poles arise, we set \(\varepsilon=0\) henceforth.
\subsection{Splitting into regions}
Analogous to the soft expansion, 
we begin by analyzing the residue matrix. 
After applying the change of variables described above,
the $z_{1}$-component of the Pfaffian system takes the form
\begin{equation}
    \Omega_{1} = \sum_{k=-1}^{\infty}\Omega_{1}^{(k)}z_{1}^{k}\,.
\end{equation}
The residue matrix $\Omega_{1}^{(-1)}$ reads:
\begin{equation*}
       \begingroup
       \begin{bNiceMatrix}
        \mzero&\mzero&\mzero&\mzero&\mzero&\mzero\\
        \mzero&\mzero&\mzero&\mzero&\mzero&\mzero\\
        \mzero&\mzero&\mzero&\mzero&\mzero&\mzero\\
        \mzero&\mzero&\mzero&\mzero&\mzero&\mzero\\
        \mzero&\frac{1}{2(1+\nu)z_{2}^{2}}&\mzero&\frac{\nu}{2(1+\nu)z_{2}^{2}}&-1&\mzero\\
        \frac{1+z_{3}^{2}(z_{4}^{2}-\nu)+2(\nu-1)z_{3}z_{4}-\nu z_{4}^{2}}{2(1+\nu)(z_{3}^{2}-1)(z_{4}^{2}-1)}&\frac{(\nu-1)(z_{3}^{2}-z_{4}^{2})}{2(1+\nu)z_{2}(z_{3}^{2}-1)(z_{4}^{2}-1)}&\frac{(\nu z_{4}^{2}-1)z_{3}^{2}+\nu-2(\nu-1)z_{3}z_{4}-z_{4}^{2}}{2(1+\nu)(z_{3}^{2}-1)(z_{4}^{2}-1)}&\frac{(z_{4}^{2}-z_{3}^{2})(\nu-1)}{2(1+\nu)z_{2}(z_{3}^{2}-1)(z_{4}^{2}-1)}&0&-1
    \end{bNiceMatrix}
    \endgroup
\end{equation*}
and has Jordan form:
\begin{equation}
    \mathrm{JordanDecomposition}\bigbrk{\Omega_{1}^{(-1)}} = \mathrm{diag}\brk{-1,-1,0,0,0,0}\,.
\end{equation}
At first sight the two distinct eigenvalues of the $\Omega_{1}^{(-1)}$ matrix,
$-1$ and $0$, suggest two regions. 
However, because they differ only by integers, 
the system is resonant and these blocks do not correspond 
to independent asymptotic regions. A single analytic sector suffices: 
we remove the apparent $-1$ exponents by the gauge transformation
\begin{equation}
\mathbf{J}=G\cdot \mathbf{J}', \qquad 
G=\text{diag}(1,1,1,1,z_{1}^{-1},z_{1}^{-1})\,,
\end{equation}
which shifts the two $-1$ eigenvalues to $0$ and yields a 
Pfaffian system regular in $z_{1}$
\begin{equation}
    \Omega_{i} = \sum_{k=0}^{\infty}\Omega_{i}^{(k)}z_{1}^{k}\,, \quad \mathbf{J}' = \sum_{k=0}^{\infty}\mathbf{J}_{k}z_{1}^{k} \, ,
\end{equation}
which is finite as $z_{1}\rightarrow0$. 
After the gauge transformation we have $\Omega_{1}^{(-1)}=0$, 
so all eigenvalues are zero and there is a single analytic sector: 
each component admits a regular Taylor expansion in $z_{1}$ with no 
constraint of the form $\Omega_{1}^{(-1)}\mathbf{J}_{0}=0$. 
To reduce the independent boundary data we instead exploit 
symmetry relations, as described in \secref{subsubsec:symmetry}. 
In the limit $z_{1}\to0$ ($p_{\infty}\to0$) we have:
\begin{equation}
\hat{\omega}_{1}=\hat{\omega}_{2} \, , \quad \gamma=1 \, , \quad \hat{b}\cdot k=0\, ,
\end{equation}
leading to:
\begin{equation}
    J_{1}\eval_{z_{1}=0}=J_{3}\eval_{z_{1}=0}, \quad J_{2}\eval_{z_{1}=0}=J_{4}\eval_{z_{1}=0}\,.
\end{equation}
These relations can be captured by the equations:
\begin{equation}
    R\cdot\mathbf{J}_{0} = 0, \quad R = \left[\begin{array}{cccccc}
        1&\mzero&-1&\mzero&\mzero&\mzero\\
        \mzero&1&\mzero&-1&\mzero&\mzero
    \end{array}\right]\,,
\end{equation}
We now rotate to a basis where these two linear relations are imposed explicitly.
In this basis the corresponding difference components are isolated and, 
by construction, vanish in the $z_{1}\to0$ limit:
\begin{equation}
    M = \left[\frac{B}{R}\right] = \left[\begin{array}{cccccc}
        1&\mzero&\mzero&\mzero&\mzero&\mzero\\
        \mzero&1&\mzero&\mzero&\mzero&\mzero\\
        \mzero&\mzero&\mzero&\mzero&1&\mzero\\
        \mzero&\mzero&\mzero&\mzero&\mzero&1\\
        1&\mzero&-1&\mzero&\mzero&\mzero\\
        \mzero&1&\mzero&-1&\mzero&\mzero
    \end{array}\right]\,,
\end{equation}
The master integrals in this basis become:
\begin{equation}
    \mathbf{F}
    = M \cdot \mathbf{J}'
    =
    \begin{bNiceMatrix}
        J_{1} & \hspace{2pt} &
        J_{2} & \hspace{2pt} &
        z_{1}J_{5} & \hspace{2pt} &
        z_{1}J_{6} & \hspace{2pt} &
        J_{1}-J_{3} & \hspace{2pt} &
        J_{2}-J_{4}
    \end{bNiceMatrix}^{\!\top}
    \implies
    \mathbf{F}_{0}
    = \left[\begin{array}{cccccc}
        * & * & * & * & \gr{0} & \gr{0}
    \end{array}\right]^{\!\top}.
\end{equation}
and we can restrict our Pfaffian system to a 
$4\times4$ one as $z_{1}\rightarrow0$.
\subsection{Leading order solution\label{subsec:PNLO}}
In the $z_{1}\to0$ limit the restricted Pfaffian system becomes:
\begin{equation}
    \partial_{i}\mathbf{F}_{0} = \left[
    \begin{array}{ccc!{\color{gray!60}\vrule}c} 
    &&& \\
    &\Omega^{(4\times4)}_{i}&&\ast\\
    &&&\\
    \noalign{\color{gray!60}\hrule height 0.5pt}
        &\gr{\mathbb{0}}&&\ast
    \end{array}
    \right]\cdot\mathbf{F}_{0}\,.
\end{equation}
The $4\times 4$ sub-system reads explicitly as:
\begin{align}
            \Omega_2^{(4\times4)} & = \begin{bNiceMatrix}
            \mzero & \frac{1}{z_2} & \mzero & \mzero \\
 z_2 & \mzero & \mzero & \mzero \\
 \mzero & \mzero & -\frac{2}{z_2} & \frac{1}{z_2} \\
 \mzero & \mzero & \frac{z_2 \left(z_3^2+1\right)
   \left(z_4^2+1\right)}{\left(z_3^2-1\right) \left(z_4^2-1\right)} &
   -\frac{1}{z_2}+\frac{2}{z_3^2-1}+\frac{1}{1-z_4}+\frac{1}{z_4+1} \\
        \end{bNiceMatrix} \, , \notag \\
        \Omega_3^{(4\times4)} & = \begin{bNiceMatrix}
             \mzero & \mzero & \mzero & \mzero \\
 \mzero & \mzero & \mzero & \mzero \\
 \mzero & \mzero & -\frac{2 z_2 z_3 \left(z_4^2+1\right)}{\left(z_3^2-1\right)
   \left(z_3^2 z_4^2-1\right)} & -\frac{2 z_3
   \left(z_4^2-1\right)}{\left(z_3^2-1\right) \left(z_3^2 z_4^2-1\right)} \\
 \mzero & \mzero & -\frac{2 z_2 z_3 \left(z_4^2+1\right) \left((z_2+1)
   z_3^2+z_2-1\right)}{\left(z_3^2-1\right)^2 \left(z_3^2
   z_4^2-1\right)} & -\frac{2 z_3 \left(z_4^2-1\right) \left((z_2+1)
   z_3^2+z_2-1\right)}{\left(z_3^2-1\right)^2 \left(z_3^2
   z_4^2-1\right)} \\
        \end{bNiceMatrix} \, , 
        \\
        \Omega_4^{(4\times4)} & = \begin{bNiceMatrix}
         \mzero & \mzero & \mzero & \mzero \\
 \mzero & \mzero & \mzero & \mzero \\
 \mzero & \mzero & \frac{2 z_2 \left(z_3^2+1\right) z_4}{\left(z_4^2-1\right)
   \left(z_3^2 z_4^2-1\right)} & -\frac{2 \left(z_3^2-1\right)
   z_4}{\left(z_4^2-1\right) \left(z_3^2 z_4^2-1\right)} \\
 \mzero & \mzero & -\frac{2 z_2 \left(z_3^2+1\right) z_4 \left((z_2-1)
   z_4^2+z_2+1\right)}{\left(z_4^2-1\right)^2 \left(z_3^2
   z_4^2-1\right)} & \frac{2 \left(z_3^2-1\right) z_4 \left((z_2-1)
   z_4^2+z_2+1\right)}{\left(z_4^2-1\right)^2 \left(z_3^2
   z_4^2-1\right)} \, , \notag
        \end{bNiceMatrix}
\end{align}
and admits a closed‑form solution in generic dimensions $\D$.
Specializing to $\D=4$ and imposing the boundary data of \appref{app:boundarySoft},
the solution is:
\begin{equation}
    \mathbf{F}_{0} =
    \begin{bNiceMatrix}
        -\frac{1}{2\pi}K_{0}(z_{2}) & \hspace{2pt} &
        \frac{z_{2}}{2\pi}K_{1}(z_{2}) & \hspace{2pt} &
        \gr{0} & \hspace{0pt} &
        \gr{0} & \hspace{0pt} &
        \gr{0} & \hspace{0pt} &
        \gr{0} & \hspace{0pt}
    \end{bNiceMatrix}
    ^{\!\top}\,,
\end{equation}
where $K_{n}(z)$ is the modified Bessel function of the second kind. 
The leading solution involves only the two 
modified Bessel functions $K_{0}(z_{2})$ and $K_{1}(z_{2})$, 
and the $z_{1}$-Taylor recursion relation has purely 
rational (in $z_{i}$) coefficients.
This implies that no additional transcendental functions can appear. 
We thus factor out this minimal transcendental basis and keep 
all remaining structure in a rational coefficient matrix.
\begin{equation}
    \mathbf{F}_{0} = F_{0}\cdot\mathbf{B}, \quad \mathbf{B} = \frac{1}{2\pi}\left[\begin{array}{c}
    K_{0}(z_{2})\\
    K_{1}(z_{2})
    \end{array}\right], \quad 
     F_{0} = \left[\begin{array}{cccccc}
        -1 & \mzero & \mzero & \mzero & \mzero & \mzero \\
        \mzero & z_{2} & \mzero & \mzero & \mzero & \mzero
    \end{array}\right]^{\top}
    \,.
\end{equation}
\subsection{Tree-level waveform in the PN expansion}
The waveform in the PN limit is given by:
\begin{equation}
    \cW^{(0)}_{h} = \mathbf{C}\cdot\mathbf{J} = z_{1}^{-2}\mathbf{C}'\cdot G\cdot M^{-1}\cdot F\cdot\mathbf{B} = z_{1}^{-2}\mathbf{W}^{(0)}\cdot\mathbf{B}\,,
\end{equation}
where we factor out (i) the Bessel basis from the form factors and 
(ii) the explicit second-order pole carried by the IBP coefficient vector. Since we are interested in computing the expansion of $\mathbf{W}^{(0)}$ to a very high order, we utilize the framework of \textit{companion tensor algebra} \cite{Brunello:2024tqf}, together with \texttt{FiniteFlow}~\cite{Peraro:2016wsq,Peraro:2019svx}, as detailed in~\appref{app:method}. This allows us to obtain all PN coefficients of $\mathcal{W}_{h}^{(0)}$ to arbitrary order by iterating a rational linear recursion with fixed transcendental basis $\mathbf{B}$. We derived the PN expansion of the tree-level waveform to order $p_{\infty}^{30}$, and it is provided in the ancillary file \texttt{PN\_expansion\_30.tar.gz} in the repository~\cite{repo}. For completeness, we are also attaching the expansion up to $p_{\infty}^{10}$, provided in the ancillary file \texttt{PN\_expansion\_10.m}.
\paragraph{Checks with literature.}
The PN expansion of the tree-level waveform has been computed explicitly via
Multipolar-Post-Minkowskian methods~\cite{Blanchet:1985sp,Blanchet:1989ki,%
Blanchet:2013haa}, and have been recently rederived using scattering-amplitude methods~\cite{Mougiakakos:2021ckm,Jakobsen:2021smu,Georgoudis:2023eke,Bini:2024rsy}.
We checked explicitly that our PN expansion of the tree-level waveform
agrees with the results of~\cite{Bini:2024rsy} up to the order they have been computed,
which is $p_{\infty}^{4}$.
\subsection{Radiative multipoles from PN waveforms}
The PN-expanded waveform can be decomposed into spin-weighted 
spherical harmonics:
\begin{equation}
    \cW^{(0)}_{h}(\omega,\hat{\mathbf{n}}) = \sum_{\ell\ge2}\sum_{|m|\le\ell}\cW^{(0)}_{\ell m}(\omega)\,_{-2}Y_{\ell m}(\hat{\mathbf{n}})\,,
\end{equation}
where mode coefficients are extracted by orthonormality,
\begin{equation}\label{eq:ortho}
  \cW^{(0)}_{\ell m}(\omega)
  = \int \d\Omega\,\cW^{(0)}_{h}(\omega,\hat{\mathbf{n}})\,
    {}_{-2}\!Y_{\ell m}^{*}(\hat{\mathbf{n}})\,.
\end{equation}
The multipole modes can be expressed in terms of the radiative multipoles 
$U_{\ell m}$ and $V_{\ell m}$ as~\cite{Blanchet:2013haa}:
\begin{equation}
    \cW^{(0)}_{\ell m}(\omega) 
= \eta^{\,\ell-2}\,U_{\ell m}(\omega) + \eta^{\,\ell-1}\,V_{\ell m}(\omega)\,,
\qquad \eta \equiv \frac{1}{c}=1\,.
\end{equation}
Using the PN-expanded waveform, we extract \(U_{\ell m}\) and \(V_{\ell m}\) to high PN order,
finding perfect agreement with \cite{Bini:2024rsy} up to the order currently available,
\(p_{\infty}^{4}\).
An proof-of-concept code to extract $(\ell,m)$ multipoles from the PN expanded waveform is provided in the ancillary file \texttt{PN\_multipoles.m} in the repository~\cite{repo}.
We explain next how we efficiently evaluate the integrals in \eqref{eq:ortho} using Stokes' theorem and Hermite reduction.

\paragraph{Efficient integration over the celestial sphere.}
The angular integration over the celestial sphere
\begin{equation}
  \int \d\Omega\, f(\theta,\phi)
  = \int_{0}^{2\pi}\dd\phi
    \int_{-1}^{1}\dd(\cos\theta)\, f(\theta,\phi)\,,
\end{equation}
can be streamlined by introducing complex coordinates from \eqref{eq:stereo}:
$z = \e^{i\phi}\tan\!\frac{\theta}{2}$ and
$\bar{z} = \e^{-i\phi}\tan\!\frac{\theta}{2}$,
which map the two-sphere to the Riemann sphere via stereographic projection.
The Jacobian of this change of variables is
\begin{equation}
  \d\Omega = \frac{2i\,\dd z\,\dd \bar{z}}{(1+z \bar{z})^{2}}\,,
\end{equation}
so that
\begin{equation}
  \int \d\Omega\, f(\theta,\phi)
  = \int \dd z\,\dd \bar{z}\,
    \frac{2i}{(1+z \bar{z})^{2}}\, f(z,\bar{z})\,.
\end{equation}

When $f(z,\bar{z})$ is a rational function, the integral can be evaluated using a complex analogue of Stokes’ theorem~\cite{Mastrolia:2009dr}, which effectively reduces the surface integral on the sphere to a sum of line integrals along contours in the complex $z$ plane. The idea is then to first integrate with respect to $z$, treating $\bar{z}$ as a constant parameter, and to keep only the rational part $F_{\text{rat}}(z,\bar{z})$ of the primitive function; this is the essence of the Hermite reduction algorithm \cite{hermite1872integration,bronstein1997symbolic}. For practical computations like ours, this reduction is efficiently implemented in \textsc{Mathematica}~\cite{WolframDemoHermite}, where the algorithm returns the rational part of the primitive automatically. The remaining one-dimensional contour integral over $z_4$ is then evaluated by summing the residues at the poles of the reduced integrand. 

We schematically summarize the procedure as follows
\begin{equation}
  \begin{aligned}
    \int \d\Omega\, f(z,\bar{z})
      &= \int \dd z\,\dd \bar{z}\,
          \frac{2i}{(1+z\bar{z})^{2}}\, f(z,\bar{z}) \\
      &= \int \dd \bar{z}\, F_{\text{rat}}(z,\bar{z}) \\
      &= 2\pi i \sum_{p}
         \Res_{\bar{z} = p,\, z = p^{*}}
         F_{\text{rat}}(z,\bar{z})\,.
  \end{aligned}
\end{equation}

From a geometric viewpoint, this procedure identifies the poles of $f(z,\bar{z})$ as punctures on the Riemann sphere, and the integral reduces to the sum of their local contributions. The Hermite reduction step corresponds to projecting the integrand onto its cohomologically non-trivial part, much like the Griffiths--Dwork pole-reduction methods used in multi-dimensional residue theory~\cite{griffiths1,dwork1,Bostan_2013}. It would be interesting and useful to implement this multi-variate generalization in the future.

%% file: Sections/Canonical_form.tex
{\section{On the \texorpdfstring{$\varepsilon$}--factorized differential equation}\label{sec:canonical}}

In the previous sections, we presented a method to obtain the tree-level waveform 
in impact-parameter space through series expansions of intermediate Fourier integrals 
belonging to the family defined in~\eqref{eq:integral_family}. 
Although this approach is remarkably effective in the physically relevant 
kinematic limits considered earlier, it is also of broader interest to seek a 
representation valid for arbitrary configurations.

As in the case of Feynman integrals~\cite{Henn:2013pwa}, 
obtaining an $\varepsilon$-factorized differential equation satisfied by the 
family of Fourier integrals in~\eqref{eq:integral_family} is one way to 
construct such solutions to all orders in the dimensional-regularisation 
parameter $\varepsilon$. In this form, the analytic structure and underlying 
function space of the integral become manifest, and the solution reduces to 
iterated integrations over the functions appearing in the differential equation.

A main challenge, therefore, lies in identifying the transformation that 
brings a given system into this $\varepsilon$-form. Over the past few years, 
this strategy has been instrumental in the analytic evaluation of increasingly 
complicated Feynman integrals, from polylogarithmic multi-scale 
processes~\cite{Mistlberger:2018etf,Chicherin:2020oor,Agarwal:2021vdh,Herrmann:2021tct,Bargiela:2021wuy,Bargiela:2022lxz,Abreu:2024fei,Abreu:2024yit,Henn:2025xrc} 
to genuinely elliptic and even Calabi--Yau geometries~\cite{Adams:2018yfj,Bogner:2019lfa,Bern:2021dqo,Muller:2022gec,Pogel:2022vat,Giroux:2022wav,Gorges:2023zgv,Giroux:2024yxu,Duhr:2024bzt,Frellesvig:2024rea,e-collaboration:2025frv,Duhr:2025lbz,Becchetti:2025qlu,Chaubey:2025adn,Duhr:2025kkq,Pogel:2025bca,Coro:2025vgn,Duhr:2025xyy}. 
These developments have shown that finding such a transformation is a crucial step 
toward the systematic computation of Feynman integrals and, ultimately, 
of the associated scattering amplitudes.

In this section, we provide the necessary background and tools to 
derive an $\varepsilon$-factorized differential equation for the 
class of integrals in~\eqref{eq:integral_family} contributing to the waveform. 
We will find that the associated function space is governed by Bessel functions, 
introduced through the transformation that brings the system into $\varepsilon$-form 
via intermediate period matrices of the \textit{maximal cut} integrands paired with 
homology cycles chosen following the theory of rapid-decay homology. 

\subsection{Generalities: \texorpdfstring{$\varepsilon$}{dum}-form and rapid-decay homology}
As we just advertised, the aim in this section is to find a transformation matrix 
$\mathcal{U}$ such that a given starting basis:
\begin{equation}
    \mathbf{J} = \mathcal{U}\cdot\mathbf{J}'\,,
\end{equation}
satisfies an $\varepsilon$-factorized differential equation:
\begin{equation}
    \partial_{a}\mathbf{J}' = \varepsilon\,\Omega_{a}'\cdot\mathbf{J}'\,,
\end{equation}
where:
\begin{equation}\label{eq:DEtransform}
    \Omega_{a}' = \mathcal{U}^{-1}\cdot(\Omega_{a}\cdot\mathcal{U}-\partial_{a}\mathcal{U})\,.
\end{equation}
Assuming the choice of initial basis in \eqref{eq:startingBasis}, whose differential equation depends linearly on $\varepsilon$ (that is, $\Omega_a=\Omega_{a,0}+\varepsilon\,\Omega_{a,1}$ for all $a$), we explain how the \emph{period matrix} $\mathcal{P}$ from the independent solutions of the maximal-cut integrals in $\D=4$ can be used to build $\mathcal{U}$ above.

Independent solutions can be expressed as \emph{twisted period integrals}, 
written as the intersection number~\cite{Mastrolia:2018uzb}
\begin{equation}\label{eq:pairing1}
J_{i}\equiv\bra{\varphi_i}\Gamma] \equiv \int_{\Gamma} u(\mathbf{z})\,\varphi_i(\mathbf{z})\, \qquad i\in \{1,\ldots,\nu\}\,,
\end{equation}
between the integration cycle $\Gamma$ and the algebraic, 
dimension-independent $n$-form $\varphi(\mathbf{z})$ 
in the Baikov variables $\mathbf{z} = (z_1, \ldots, z_n)$. 
All the multi-valuedness of the integrand is absorbed into the
 \emph{twist} $u(\mathbf{z})$. Note that any Feynman or Fourier integral 
 can be written in this form using the \emph{Baikov representation}~\cite{Baikov:1996iu} 
 (see also~\cite{Correia:2025yao} for a recent derivation, as well as~\cite{Frellesvig:2024ymq,Correia:2025yao} 
 for \textsc{Mathematica} implementations in the context of Feynman integrals).

An important distinction in~\eqref{eq:pairing1} between Feynman and Fourier integrals 
arises because, for the latter, $u(\mathbf{z}) \propto \e^{z_3}$ with $z_3 \equiv i q_\perp \cdot \hat{b}$. 
When $u(\mathbf{z})$ contains such exponential factors, 
the integral is generally not absolutely convergent on the real domain. 
To define the pairing~\eqref{eq:pairing1} unambiguously, 
one must complexify the integration variables and view the integral as 
living on a complex manifold $X\subset \mathbb{C}^n$ obtained after removing 
the singular hypersurfaces of the integrand. The admissible integration cycles are 
then required to extend to infinity only through directions in which the 
exponential factor decays.

In mathematics, this refinement is naturally encoded in the 
\emph{rapid-decay homology} group $H^{\mathrm{rd}}_{n}(X,u)$, 
whose elements are precisely those chains along which $u(\mathbf{z})$ 
tends to zero exponentially at infinity. 
Two such chains are considered equivalent if they can be continuously deformed 
into one another without breaking the decaying property. In this way, 
rapid-decay homology provides the correct framework to make the 
twisted intersection number well defined, ensuring both convergence 
and deformation invariance of the resulting periods. 
We refer the interested reader to~\cite{Pham1985LaDD,b2b3fecd-cab3-35d4-8f84-a0b479a710ac,Hien:2009,MatsubaraHeo2017OnTR} 
for general properties of $H^{\mathrm{rd}}_{n}$, 
and we will limit ourselves to discussing it explicitly in simple 
univariate tree-level examples later in this section. 

We anticipate this theory to play a much less trivial role 
for the impact-parameter space waveform at one loop and beyond, 
for which the maximal cut contours discussed below become 
higher-dimensional and thus less tractable intuitively.

The $\nu$ independent solutions of the maximal-cut integrals 
in $\D = 4$ can thus be decomposed into $\nu$ independent 
\emph{periods} $J_{i}|_{\D=4}^{\text{mc}} = \sum_{j=1}^\nu c_j \mathcal{P}_{ij}=[\mathcal{P}\cdot\mathbf{c}]_i$, 
where $c_j\in \mathbb{C}$ and
\begin{equation}\label{eq:periodMatrix}
\mathcal{P}_{ij} \equiv \int_{\gamma_j} u(\mathbf{z})\big|_{\D=4}\,\varphi_i^{\text{mc}}(\mathbf{z})\,,
\end{equation}
with $\{\gamma_j\}_{j=1}^\nu$ forming a basis of admissible cycles in $H^{\mathrm{rd}}_{n}(X,u)$.
Each entry of the matrix $\mathcal{P}$ is thus an intersection number between a 
chosen form and cycle, and the collection of all such pairings forms the 
period matrix itself. 

For concreteness, we record explicitly the Baikov representations for the basis elements in $\mathbf{J}$ on their respective maximal cuts. 
Thanks to~\eqref{eq:relationB1B2}, it is sufficient to do so for the basis elements $J_{i\in\{1,2,5,6\}}$
\begin{align}\label{eq:basisform}
\hspace{-0.5cm}
    J_{i \in \{1,2\}}\big|_{\D=4}^{\text{mc}}
    &= \int u(z_3)\,\varphi_{i \in \{1,2\}}^{\text{mc}} \,,\,\,
    \text{where } 
    u(z_{3}) = \frac{\e^{z_{3}}}{\sqrt{\hat{w}_{2}^{2} - z_{3}^{2}}}
    \,\, \text{and} \,\,
    \varphi^{\text{mc}}_{i \in \{1,2\}} =\{1,z_3\} \d z_{3}\,,
    \notag
    \\
    \varepsilon^{-1}J_{i \in \{5,6\}}\big|_{\D=4}^{\text{mc}}
    &= \int v(z_3)\,\varphi_{i \in \{1,2\}}^{\text{mc}} \,,
    \,\, 
   \text{where } 
    v(z_{3}) = 
    \frac{\sqrt{(\hat{b} \cdot k_\perp)^{2} + k_{\perp}^{2}}\, \e^{z_{3}}}
         {(z_{3} - z_{+})(z_{3} - z_{-})} \,,
\end{align}
with
\begin{equation}\label{eq:zPmAndLambda}
    z_{\pm} 
    = \frac{\hat{w}_{2}\big(i\hat{b} \cdot k_\perp (\hat{w}_{2} - \gamma\hat{w}_{1})
       \pm i\hat{w}_{1}\Lambda\big)}{k_{\perp}^{2}} \, \, \qquad \text{and} \, \, \qquad
    \Lambda = \sqrt{(\gamma^{2} - 1)\big((\hat{b} \cdot k_\perp)^{2} + k_{\perp}^{2}\big)}\,.
\end{equation}
Note that we have included a pole in $\varepsilon$ for $J_{5}$ and $J_{6}$ as these vanish in exactly $\D=4$.

We conclude this subsection with a few comments on the utility 
of~\eqref{eq:periodMatrix} and the role it plays in various algorithms 
aiming to reach an $\varepsilon$-form basis (see, e.g.,~\cite{Frellesvig:2021hkr,Gorges:2023zgv,Giroux:2022wav}). 

The most direct approach is to first take $\mathcal{P}$ as the period matrix itself,
so that $\mathbf{J}' = \mathcal{P}^{-1}\!\cdot\!\mathbf{J}$. 
On the $\D=4$ maximal cut, one then has 
$\mathbf{J}|_{\D=4}^{\text{mc}} = \mathcal{P}\cdot\mathbf{c}$, 
so that $\mathbf{J}'|_{\D=4}^{\text{mc}} = \mathbf{c}$ is constant. 
Hence, the differential equation on the cut reduces to 
\begin{equation}
    \partial_a\mathbf{J}'|_{\D=4-2\varepsilon}^{\text{mc}} = \varepsilon\, \Omega_{a,1}'|_{\text{mc}}\cdot \mathbf{J}'|_{\D=4-2\varepsilon}^{\text{mc}}\,.
\end{equation}
This observation is useful because the differential equation obeyed by the cut 
integrals coincides with that of the uncut ones, 
up to non-homogeneous terms arising from lower-sector integrals that have 
no support on the cut under consideration (typically, but not necessarily, the maximal cut). Consequently, the uncut integrals also satisfy the differential equation $\varepsilon\,\Omega'|_{\text{mc}}$, modulo a possible \emph{lower}-triangular $\mathcal{O}(\varepsilon^0)$ non-homogeneous contribution. These residual terms can typically be eliminated by a secondary transformation, $\mathbf{J}' = \mathcal{V}\cdot\mathbf{J}''$, where $\mathcal{V}$ is determined systematically by direct integration (see, e.g.,~\cite{Giroux:2022wav,Giroux:2024yxu}), leaving the uncut system in the desired form $\mathrm{d}\mathbf{J}'' = \varepsilon\,\Omega''\cdot\mathbf{J}''$.

The drawback of this seemingly natural choice, $\mathcal{U} = \mathcal{P}\cdot\mathcal{V}$, 
lies in the fact that the (non-vanishing) strictly \emph{upper}-triangular part of $\mathcal{P}$ 
can generate spurious poles in the differential equation. Through this mechanism, 
for instance, non-simple poles are introduced in the differential equation 
satisfied by the elliptic sunrise integral~\cite{Giroux:2022wav}.

To avoid this unnecessary complication, we therefore adopt a refined variant 
of the above procedure~\cite{Gorges:2023zgv}, in which $\mathcal{P}$ is first 
decomposed into an \emph{upper}-triangular unipotent and \emph{lower}-triangular 
semi-simple (i.e., diagonalizable) parts, $\mathcal{P} = \mathcal{P}^{(ss)}\cdot\mathcal{P}^{(u)}$, 
and the rotation is performed with the inverse of the semi-simple part, 
$\mathcal{U} = \mathcal{P}^{(ss)}\cdot\mathcal{V}$.

\subsection{Constructing the \texorpdfstring{$\varepsilon$}{dum}-form} 

Our starting point is the basis (defined in~\eqref{eq:startingBasis}) for the integral family in~\eqref{eq:integral_family}, which, 
as noted earlier, satisfies a differential equation that is linear in $\varepsilon$ and organizes itself into three \emph{independent} $2\times 2$ diagonal-blocks (see ~\eqref{eq:DEstructure}). 
These blocks involve, respectively, the integrals subfamilies:
\begin{subequations}\label{eq:Iblocks}
    \begin{align}
            I(1,0,n_{3}) &= \int_{\mathcal{M}}\hat{\rd}^{\D-2}q_{\perp}\frac{(iq_{\perp}\cdot\hat{b})^{-n_{3}}}{(q_{\perp}^{2}-\hat{w}_{2}^{2})}\e^{iq_{\perp}\cdot \hat{b}}\,,\label{eq:Iblock1}
            \\
            I(0,1,n_{3}) &= \int_{\mathcal{M}}\hat{\rd}^{\D-2}q_{\perp}\frac{(iq_{\perp}\cdot\hat{b})^{-n_{3}}}{((q_{\perp}-k_{\perp})^{2}-\hat{w}_{1}^{2})}\e^{iq_{\perp}\cdot\hat{b}}\,,\label{eq:Iblock2}
            \\
            I(1,1,n_{3}) &= \int_{\mathcal{M}}\hat{\rd}^{\D-2}q_{\perp}\frac{(iq_{\perp}\cdot\hat{b})^{-n_{3}}}{(q_{\perp}^{2}-\hat{w}_{2}^{2})((q_{\perp}-k_{\perp})^{2}-\hat{w}_{1}^{2})}\e^{iq_{\perp}\cdot\hat{b}}\,.\label{eq:Iblock3}
    \end{align}
\end{subequations}

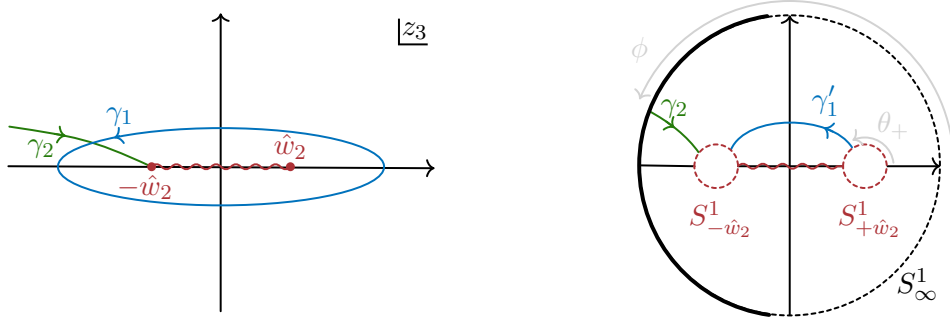
\begin{figure}
    \centering
\tikzset{every picture/.style={line width=0.75pt}} 
\begin{tikzpicture}[x=0.75pt,y=0.75pt,yscale=-1,xscale=1]
\draw [color={rgb, 255:red, 65; green, 117; blue, 5 }  ,draw opacity=1,line cap=round]   (98.5,139) .. controls (131.5,144) and (139.5,145) .. (170.25,159.25) ;
\draw[->, color={rgb, 255:red, 65; green, 117; blue, 5 },line cap=round]  (121.33,142.67) -- (127,143.67) node[above,left,yshift=-5]{$\gamma_2$};
\draw  [draw opacity=0] (420.5,130.82) .. controls (431.59,136.69) and (440.46,143.98) .. (446.28,152.19) -- (352.5,175) -- cycle ; \draw[color={rgb, 255:red, 65; green, 117; blue, 5 }]   (420.5,130.82) .. controls (431.59,136.69) and (440.46,143.98) .. (446.28,152.19) ;  
\draw[->,color={rgb, 255:red, 65; green, 117; blue, 5 }]    (429.67,136.33) -- (434,139.67) node[above]{$\gamma_2$};
\draw[<-][line cap=round]    (205.08,81.81) -- (205.08,232.19);
\draw[->][line cap=round]    (98,159) -- (313,160) ;
\draw[<-]    (492.08,82.81) -- (492.08,234.31) ;
\draw [dash pattern={on 1.5pt off 1.5pt}, line cap=round]  (416.33,158.56) .. controls (416.33,116.72) and (450.24,82.81) .. (492.08,82.81) .. controls (533.91,82.81) and (567.83,116.72) .. (567.83,158.56) .. controls (567.83,200.4) and (533.91,234.31) .. (492.08,234.31) .. controls (450.24,234.31) and (416.33,200.4) .. (416.33,158.56) node[left,xshift=117,yshift=-45]{$S^1_\infty$} -- cycle ;
\draw  [draw opacity=0][line width=1.5]  (481.67,234.03) .. controls (444.67,228.94) and (416.16,197.09) .. (416.16,158.56) .. controls (416.16,120.37) and (444.17,88.74) .. (480.7,83.24) -- (492.08,158.56) -- cycle ; \draw  [line width=1.5][line cap=round]  (481.67,234.03) .. controls (444.67,228.94) and (416.16,197.09) .. (416.16,158.56) .. controls (416.16,120.37) and (444.17,88.74) .. (480.7,83.24) ;  
\draw  [draw opacity=0][fill=Maroon  ,fill opacity=1, ] (168,159.25) .. controls (168,158.01) and (169.01,157) .. (170.25,157) .. controls (171.49,157) and (172.5,158.01) .. (172.5,159.25) .. controls (172.5,160.49) and (171.49,161.5) .. (170.25,161.5) .. controls (169.01,161.5) and (168,160.49) .. (168,159.25) -- cycle ;
\draw  [draw opacity=0][fill=Maroon  ,fill opacity=1, ] (238,159.25) .. controls (238,158.01) and (239.01,157) .. (240.25,157) .. controls (241.49,157) and (242.5,158.01) .. (242.5,159.25) .. controls (242.5,160.49) and (241.49,161.5) .. (240.25,161.5) .. controls (239.01,161.5) and (238,160.49) .. (238,159.25) -- cycle ;
\draw [color=Maroon  ,draw opacity=1 ]  (170.25,159.25)  .. controls (171.92,157.58) and (173.58,157.58) .. (175.25,159.25) .. controls (176.92,160.92) and (178.58,160.92) .. (180.25,159.25) .. controls (181.92,157.58) and (183.58,157.58) .. (185.25,159.25) .. controls (186.92,160.92) and (188.58,160.92) .. (190.25,159.25) .. controls (191.92,157.58) and (193.58,157.58) .. (195.25,159.25) .. controls (196.92,160.92) and (198.58,160.92) .. (200.25,159.25) .. controls (201.92,157.58) and (203.58,157.58) .. (205.25,159.25) .. controls (206.92,160.92) and (208.58,160.92) .. (210.25,159.25) .. controls (211.92,157.58) and (213.58,157.58) .. (215.25,159.25) .. controls (216.92,160.92) and (218.58,160.92) .. (220.25,159.25) .. controls (221.92,157.58) and (223.58,157.58) .. (225.25,159.25) .. controls (226.92,160.92) and (228.58,160.92) .. (230.25,159.25) .. controls (231.92,157.58) and (233.58,157.58) .. (235.25,159.25) .. controls (236.92,160.92) and (238.58,160.92) .. (240.25,159.25) -- (240.25,159.25) node[pos=1, above,yshift=-1]{$\hat{w}_2$} node[xshift=-55, below,yshift=1]{$-\hat{w}_2$} ;
\draw  [color= RoyalBlue ,draw opacity=1, line cap=round] (123,159.25) .. controls (123,148.48) and (159.82,139.75) .. (205.25,139.75) .. controls (250.68,139.75) and (287.5,148.48) .. (287.5,159.25) .. controls (287.5,170.02) and (250.68,178.75) .. (205.25,178.75) .. controls (159.82,178.75) and (123,170.02) .. (123,159.25) -- cycle ;
\draw[<-,color= RoyalBlue]    (151.33,144.67) -- (154.33,143.67) node[above]{$\gamma_1$};
\draw[line cap=round]    (416.33,158.56) -- (444,159) ;
\draw[line cap=round]    (466,159) -- (519,159) ;
\draw [color=Maroon  ,draw opacity=1 , line cap=round]   (466,159) .. controls (467.67,157.33) and (469.33,157.33) .. (471,159) .. controls (472.67,160.67) and (474.33,160.67) .. (476,159) .. controls (477.67,157.33) and (479.33,157.33) .. (481,159) .. controls (482.67,160.67) and (484.33,160.67) .. (486,159) .. controls (487.67,157.33) and (489.33,157.33) .. (491,159) .. controls (492.67,160.67) and (494.33,160.67) .. (496,159) .. controls (497.67,157.33) and (499.33,157.33) .. (501,159) .. controls (502.67,160.67) and (504.33,160.67) .. (506,159) .. controls (507.67,157.33) and (509.33,157.33) .. (511,159) .. controls (512.67,160.67) and (514.33,160.67) .. (516,159) -- (519,159) -- (519,159) ;
\draw[->]    (541,159) -- (568,159) ;
\draw [color=RoyalBlue  ,draw opacity=1, line cap=round]   (462.5,150.5)  .. controls (470.5,132.5) and (512.75,132.75)  .. (523.5,149.5) node[above,yshift=8,xshift=-10]{$\gamma_1'$};
\draw[<-,color=RoyalBlue]    (509.25,139.5) -- (514,141.67) ;
\draw   (309.5,97.5) -- (295,97.5) -- node[above,xshift=6,yshift=-7]{$z_3$} (295,84.5) ;
\draw [dash pattern={on 1.5pt off 1.5pt}, color=Maroon,line cap=round]   (519,159) .. controls (519,152.92) and (523.92,148) .. (530,148) .. controls (536.08,148) and (541,152.92) .. (541,159) .. controls (541,165.08) and (536.08,170) .. (530,170) .. controls (523.92,170) and (519,165.08) .. (519,159) node[yshift=-20,xshift=10]{$S^{1}_{+\hat w_{2}}$} -- cycle ;
\draw [dash pattern={on 1.5pt off 1.5pt}, color=Maroon,line cap=round]   (444,159) .. controls (444,152.92) and (448.92,148) .. (455,148) .. controls (461.08,148) and (466,152.92) .. (466,159) .. controls (466,165.08) and (461.08,170) .. (455,170) .. controls (448.92,170) and (444,165.08) .. (444,159) node[yshift=-20,xshift=10]{$S^{1}_{-\hat w_{2}}$} -- cycle ;
\draw  [draw opacity=0][line width=0.75]  (416.32,125.27) .. controls (429.1,96.02) and (458.21,75.6) .. (492.08,75.6) .. controls (537.39,75.6) and (574.19,112.16) .. (574.76,157.49)  -- (492.08,158.56) -- cycle ; \draw[color=gray!35]  [line width=0.75, line cap = round][<-]  (416.32,125.27) node[above,yshift=10]{$\phi$} .. controls (429.1,96.02) and (458.21,75.6)  .. (492.08,75.6) .. controls (537.39,75.6) and (574.19,112.16) .. (574.76,157.49) ;  
\draw  [draw opacity=0] (523.66,145.95) .. controls (525.58,145.02) and (527.73,144.5) .. (530,144.5) .. controls (537.4,144.5) and (543.5,150.04) .. (544.39,157.2) -- (530,159) -- cycle ; \draw[color=gray!35, line cap=round][<-]  (523.66,145.95) .. controls (525.58,145.02) and (527.73,144.5) .. (530,144.5) .. controls (537.4,144.5) and (543.5,150.04) .. (544.39,157.2) node[above,yshift=5]{$\theta_+$} ;  
\end{tikzpicture}
    \caption{
(Left) The twist $u(z_{3})=\e^{z_{3}}(\hat w_{2}^{2}-z_{3}^{2})^{-1/2}$ has branch points at $\pm\hat w_{2}$, connected by a red wavy branch cut.  
The exponential factor introduces an essential singularity at $z_{3}=\infty$, growing for $\Re(z_{3})\to +\infty$ and decaying for $\Re(z_{3})\to -\infty$.  
The rapid-decay homology group $H^{\mathrm{rd}}_{1}(X,u)$, with $X=\mathbb{C}\setminus\{\pm\hat w_{2}\}$, is two-dimensional and spanned by
$\gamma_{1}$ (blue, a small loop encircling the cut) and
$\gamma_{2}$ (green, a ray extending from $-\infty$ to $-\hat w_{2}$).
(Right) The corresponding basis $\{\gamma_1',\gamma_2\}$ in the relative homology group $H_{1}^{\mathrm{rel}}(\tilde X,Q)$.  
The dashed circles represent the boundary components $S_{p}^1$ at the blow-up points $p=\infty$ (black) and $p=\pm \hat{w}_2$ (red) in the oriented blow-up $\tilde X\subset \mathbb{CP}^1$. 
Their union defines the boundary set $Q$, where cycles can ends while preserving closedness under the boundary map $\partial_{\text{rel}}$.  
The decaying directions for $\gamma_{2}$ ($\tfrac{\pi}{2}<\phi<\tfrac{3\pi}{2}$) are indicated by the thick semicircle at complex infinity.
    }
    \label{fig:RapidDecayCycles1}
\end{figure}

\paragraph{The transformation for the first block.} 
The twist $u(z_3)$ in \eqref{eq:basisform} has apparent branch points at $z_{3}=\pm\hat{w}_{2}$, joined by a branch cut (see the wavy red line in \figref{fig:RapidDecayCycles1}).
Since $|\e^{z_{3}}|\!\to\!0$ for $\Re (z_{3})\!\to\!-\infty$ and diverges in the opposite direction, we restrict our attention to \emph{admissible} cycles approaching $-\infty$.
This guarantees convergence of the period integrals
\begin{equation}\label{eq:genPer}
\bra{\varphi^{\text{mc}}}\Gamma]
=\int_{\Gamma}u(z_{3})\,\varphi^{\text{mc}}(z_{3}),
\qquad 
\varphi^{\text{mc}}\in{\rm span}\{\varphi^{\text{mc}}_1,\varphi^{\text{mc}}_2\}\,.
\end{equation}
As hinted earlier, a convenient language for organizing these admissible contours 
is provided by the rapid-decay homology group $H^{\mathrm{rd}}_{1}\!\big(X=\mathbb{C}\setminus\{\pm\hat{w}_{2}\},u\big)$,
whose elements are one-chains whose ends (if any) lie at $-\infty$ or at the \emph{integrable} singularities $\pm \hat{w}_2$.
In this simple univariate case, the space is clearly two-dimensional, with spanning cycles
\begin{subequations}\label{eq:rd-basis}
    \begin{align}
         \gamma_1&:\ \text{a small positively oriented loop encircling the cut }[-\hat{w}_{2},+\hat{w}_{2}]\,, \label{eq:cycle1}
\\
 \gamma_2&:\ \text{a ray from $-\infty$ to $-\hat{w}_{2}$}\,,\label{eq:cycle2}
    \end{align}
\end{subequations}
as illustrated in \figref{fig:RapidDecayCycles1} (left).
Although $\gamma_2$ is \emph{not} closed in the usual sense (since it manifestly ``escapes'' to infinity and never returns to close upon itself) it is nonetheless closed in rapid-decay homology.  
There are two complementary ways to see this.  
First, by viewing the cut $z_3$-plane as a two-sheeted sphere, one recognizes that $\gamma_2$ effectively corresponds to ``twice'' (see \eqref{eq:twice} for an analogous statement) the contour running from $-\infty$ to $-\hat{w}_2$, crossing the branch cut, and returning to $-\infty$.  
Alternatively, $\gamma_2$ may be regarded as a \emph{relative} one-cycle, which makes its closure property manifest on a single sheet at the price of introducing boundaries.  
This follows directly from the isomorphism that can be deduced from Pham’s work~\cite{Pham1985LaDD}
\begin{equation}
H^{\mathrm{rd}}_{1}(X,u) \cong H_{1}^{\mathrm{rel}}\bigl(\tilde X,Q\bigr)\,,
\end{equation}
which identifies the rapid-decay homology group with a relative homology group with boundary $Q$ (given in~\eqref{eq:Q-explicit}) on the oriented blow-up of $\mathbb{CP}^{1}$ at the points $p\in \{\pm\hat{w}_{2},\infty\}$,
\begin{equation}
    \tilde X = \bigl\{(z_3,\theta_p)\mid z_3\in \mathbb{CP}^1,\ p\in\{\pm\hat w_2,\infty\},\ \arg(z_3-p)=\theta_p\bigr\}\,.
\end{equation}
Here, each $\theta_p$ parameterizes the circle of possible directions from 
which $p$ can be approached while keeping~\eqref{eq:genPer} finite 
(see \figref{fig:RapidDecayCycles1} (right)). Introducing the local coordinates
\begin{equation}
z_{3}-\hat w_{2}=r_{+}\e^{i\theta_{+}}\,,\qquad
z_{3}+\hat w_{2}=r_{-}\e^{i\theta_{-}}\,,\qquad
w=\tfrac{1}{z_{3}}=\rho\,\e^{i\phi}\,,
\end{equation}
these boundary circles are $S^{1}_{+\hat w_{2}}=\{r_{+}=\delta\}$, 
$S^{1}_{-\hat w_{2}}=\{r_{-}=\delta\}$, 
and $S^{1}_{\infty}=\{\rho=\delta\}$, with $\delta\to 0$, 
and the corresponding set of admissible endpoints is thus:
\begin{equation}
\hspace{-0.4cm}
Q
=
\underbracket[0.4pt]{\{(\rho=\delta,\phi)\mid \phi\in(\tfrac{\pi}{2},\tfrac{3\pi}{2})\}}_{\text{decaying arcs at }-\infty}
 \cup 
\underbracket[0.4pt]{\{(r_{+}=\delta,\theta_{+})\mid \theta_{+}\in S^{1}\}
 \cup 
\{(r_{-}=\delta,\theta_{-})\mid \theta_{-}\in S^{1}\}}_{\text{blow-ups at integrable singularities }\pm\hat w_{2}}\,.
\label{eq:Q-explicit}
\end{equation}
In this relative picture, ``closedness'' is simply the statement that 
the \emph{relative boundary} of $\gamma_2$ in $Q$ vanishes, 
namely that
\begin{equation}
\partial_{\mathrm{rel}}\gamma_2 \equiv \partial\gamma_2 \bmod Q = 0\, \iff\, 
\text{its endpoints lie in } Q\,.
\end{equation}
Hence, since the contour $\gamma_2$ is a ray extending 
from $-\infty$ (approached along a decaying direction) 
to $-\hat w_2$, its boundary in $\tilde X$ is
\begin{equation}
\partial\gamma_2 = (-\hat w_2,\theta_-) - (\infty,\phi)\,,
\end{equation}
for some $\theta_-\in S^1$ and $\phi\in(\tfrac{\pi}{2},\tfrac{3\pi}{2})$. 
Here, both endpoints are within $Q$ by construction, so
\begin{equation}
\partial_{\mathrm{rel}}\gamma_2 = 0\,,
\end{equation}
and $\gamma_2$ defines a closed class in $H^{\mathrm{rel}}_{1}(\tilde X,Q)$, 
and hence in $H^{\mathrm{rd}}_{1}(X,u)$.

To summarize what we have learned so far, 
since the spanning relative cycles must have endpoints in $Q$, 
we chose them as $\gamma_2$ (in~\eqref{eq:cycle2}) and
\begin{equation}
\gamma_1':\ \text{a ray from }[+\hat{w}_{2},-\hat{w}_{2}]\,.
\end{equation}
Both cycles (as well as many of the objects discussed above) 
are illustrated in \figref{fig:RapidDecayCycles1} (right).

Of course, there is no loss of generality in choosing $\gamma_1$ over $\gamma_1'$ and vice versa: if $u_\pm(z_3)\equiv \pm u(z_3)$ denote the values above/below the branch cut, then
\begin{equation}\label{eq:twice}
\int_{\gamma_1}u\,\varphi^{\text{mc}}_i
=\int_{-\hat{w}_2}^{\hat{w}_2}[u_+-u_-]\,\varphi^{\text{mc}}_i
=\int_{-\hat{w}_2}^{\hat{w}_2}[u_+-(-u_+)]\,\varphi^{\text{mc}}_i
=2\int_{\gamma_1'}u\,\varphi^{\text{mc}}_i\,,
\end{equation}
where, in the first equality, we drop the infinitesimal semicircles around $\pm\hat w_2$, whose contributions are bounded from above by $\sim \sqrt{\delta}$ and thus vanish as the contour `kisses' the branch cut.

Now that the homology cycles are understood, the period matrix for this $2\times2$ block is given by
\begin{equation}
    \mathcal{P}|_{\text{block 1}}= \left[\begin{array}{cc}
    \frac{1}{\pi}\int_{ \gamma_1}u(z_{3})\dd z_{3}&i\int_{ \gamma_2}u(z_{3})\dd z_{3}\\
    \frac{1}{\pi}\int_{ \gamma_1}u(z_{3})z_{3}\dd z_{3}&i\int_{ \gamma_2}u(z_{3})z_{3}\dd z_{3}
    \end{array}\right] = \left[\begin{array}{cc}
        I_{0}(\hat{w}_{2}) & K_{0}(\hat{w}_{2}) \\
        \hat{w}_{2}I_{1}(\hat{w}_{2}) & -\hat{w}_{2}K_{1}(\hat{w}_{2})
    \end{array}\right]\,,
\end{equation}
where we introduced additional \emph{constant} normalizations to make our expressions nicer 
and where $I_{n}(z)$ and $K_{n}(z)$ are modified Bessel functions of the first and second kind, respectively. For the reasons we discussed above, we next extract the semi-simple part $\mathcal{P}|_{\text{block 1}}^{(ss)}$, leaving behind a unipotent piece $\mathcal{P}|_{\text{block 1}}^{(u)}$
\begin{equation}
    \mathcal{P}|_{\text{block 1}} = \mathcal{P}|_{\text{block 1}}^{(ss)}\cdot\mathcal{P}|_{\text{block 1}}^{(u)}, \quad \mathcal{P}|_{\text{block 1}}^{(ss)} = \left[\begin{array}{cc}
        I_{0}(\hat{w}_{2}) & 0 \\
        \hat{w}_{2} I(\hat{w}_{2}) & -\frac{1}{I_{0}(\hat{w}_{2})} 
    \end{array}\right]\,.
\end{equation}
From there, we first rotate the basis with the semi-simple part transforming the differential equation matrices according to  \ref{eq:DEtransform}. The only non-zero differential equation for this $2\times2$ block is for $\hat{w}_{2}$ 
\begin{equation}
    \Omega_{\hat{w}_{2}}'|_{\text{block 1}}=\left[\begin{array}{cc}
        -\frac{2\varepsilon}{\hat{w}_{2}}&-\frac{1}{\hat{w}_{2}I_{0}(\hat{w}_{2})^{2}}\\
        -2\varepsilon I_{0}(\hat{w}_{2})I_{1}(\hat{w}_{2})&0
    \end{array}\right]\,.
\end{equation}
In order to eliminate the $\varepsilon^{0}$ term, we combine an $\varepsilon$-scaling on the second integral with an ansatz that is chosen to eliminate the resulting lower-triangular non-factorized piece. This ansatz takes the form
\begin{equation}
   \mathcal{V}|_{\text{block 1}}=\left[\begin{array}{cc}
        1&0\\
        \varepsilon G(\hat{w}_{2})&\varepsilon
    \end{array}\right]\,,
\end{equation}
and by direct integration of the vanishing condition
\begin{equation}
   ( \mathcal{V}|_{\text{block }1})^{-1}\cdot(\Omega_{\hat{\omega}_{2}}|_{\text{block }1}\cdot\mathcal{V}|_{\text{block }1}-\partial_{\hat{w}_{2}}\mathcal{V}|_{\text{block }1})|_{\varepsilon=0}=0\implies \dd G+\dd I_{0}(\hat{w}_{2})^{2}=0\,,
\end{equation}
we find, up to an irrelevant integration constant, $G(\hat{w}_{2})=-I_{0}(\hat{w}_{2})^{2}$. Combining everything together, we get
\begin{equation}\label{eq:Ublock1}
    \mathbf{J}'|_{\text{block 1}}=\Bigg(\mathcal{U}|_{\text{block 1}}\equiv \left[\begin{array}{cc}
        I_{0}(\hat{w}_{2}) & 0 \\
        \hat{w}_{2}I_{1}(\hat{w}_{2})+\varepsilon I_{0}(\hat{w}_{2}) & -\frac{\varepsilon}{I_{0}(\hat{w}_{2})}
    \end{array}\right]\Bigg)^{-1}\cdot \mathbf{J}|_{\text{block 1}}\,,
\end{equation}
such that
\begin{equation}\label{eq:deqB1}
    \d\mathbf{J}'|_{\text{block 1}}=-\varepsilon\,\d\log(\hat{w}_{2})\left[\begin{array}{cc}
        1 & \frac{1}{I_{0}(\hat{w}_{2})^{2}} \\
        I_{0}(\hat{w}_{2})^{2} &1 
    \end{array}\right]\cdot\mathbf{J}'|_{\text{block 1}}\,.
\end{equation}
\paragraph{The transformation for the second block.} 
 Next, we examine the second block corresponding to the integral subfamily in \eqref{eq:Iblock2}. Its transformation to $\varepsilon$-form can be directly derived from the first block thanks to \eqref{eq:relationB1B2}. 

With this in mind, it is clear that the transformation to $\varepsilon$-form basis takes the form
\begin{equation}\label{eq:Ublock2}
    \mathcal{U}|_{\text{block 2}}\equiv\e^{i\hat{b}\cdot k_\perp}\left[\begin{array}{cc}
        1 & 0 \\
        i\hat{b}\cdot k_\perp & 1
    \end{array}\right]\cdot \mathcal{U}|_{\text{block 1}}\eval_{\hat{w}_{2}\rightarrow\hat{w}_{1}}\,.
\end{equation}
where $\mathcal{U}|_{\text{block 1}}$ is given in \eqref{eq:Ublock1}.

The differential equation for the resulting basis, $\mathbf{J}'|_{\text{block 2}}=\mathcal{U}|_{\text{block 2}}^{-1}\cdot \mathbf{J}|_{\text{block 2}}$, thus follows from \eqref{eq:Ublock2} and \eqref{eq:deqB1}, and explicitly reads
\begin{equation}\label{eq:deqB2}
    \d\mathbf{J}'|_{\text{block 2}}=-\varepsilon\,\d\log(\hat{w}_{1})\left[\begin{array}{cc}
        1 & \frac{1}{I_{0}(\hat{w}_{1})^{2}} \\
        I_{0}(\hat{w}_{1})^{2} &1 
    \end{array}\right]\cdot\mathbf{J}'|_{\text{block 2}}\,.
\end{equation}

\paragraph{The transformation for the third block.} Although the integral~\eqref{eq:Iblock3} appears more complicated, its period matrix is in fact easier to construct.

In this case, there are no branch points in the twist $v(z_3)$ defined in~\eqref{eq:basisform}, so we may choose the independent contours to encircle each of the two poles at $z_3 = z_{\pm}$. 
The integral associated with each contour is then obtained simply by evaluating the residue at the corresponding pole. Note that, since it is equivalent to a linear combination of these contours under contour deformation, one could alternatively (but less conveniently) choose an analogue of the relative contour in~\eqref{eq:cycle2}, as $v(z_3) \propto \e^{z_3}$. 

Evaluating the residues, we obtain the transformation to $\varepsilon$-form 
\begin{equation}
\label{eq:Ublock3}
    \mathbf{J}'|_{\text{block 3}}=\Bigg(    \mathcal{U}|_{\text{block 3}} \equiv 
    \frac{1}{2i\hat{w}_{1}\hat{w}_{2}\sqrt{\gamma^{2}-1}}\left[\begin{array}{cc}
       -\e^{z_{-}}  & \e^{z_{+}}\\
       -z_{-}\e^{z_{-}}& z_{+}\e^{z_{+}}
    \end{array}\right]\Bigg)^{-1}\cdot \mathbf{J}|_{\text{block 3}}\,,
\end{equation}
where the differential equation matrices $\partial_{a}\mathbf{J}'|_{\text{block 3}}=\varepsilon\Omega'_{a}|_{\text{block 3}}\cdot\mathbf{J}'|_{\text{block 3}}$ reads as:
\begin{subequations}
    \begin{align}
      \Omega_{\hat{w}_{1}}'|_{\text{block 3}}& =\left[\begin{array}{cc}
      \frac{\hat{w}_1-\gamma  \hat{w}_2}{\lambda}-\frac{1}{\hat{w}_1} & 
      \e^{z}\left( (\hat{w}_1{-}\gamma  \hat{w}_2)\left(\frac{1}{\lambda}-\frac{2}{k_{\perp}^2}\right)+\frac{1}{\hat{w}_1}\right) \\
 \e^{-z}\left( (\hat{w}_1{-}\gamma  \hat{w}_2)\left(\frac{1}{\lambda}{-}\frac{2}{k_{\perp}^2}\right)+\frac{1}{\hat{w}_1}\right) & \frac{\hat{w}_1-\gamma 
   \hat{w}_2}{\lambda}-\frac{1}{\hat{w}_1} \\
    \end{array}\right]\,, \label{eq:firstDE}
    \\
    \Omega_{\hat{w}_{2}}'|_{\text{block 3}}& =\eqref{eq:firstDE}|_{\hat{w}_1\leftrightarrow \hat{w}_2}\,,
    \\
    \Omega_{\gamma}'|_{\text{block 3}}& =\left[\begin{array}{cc}
      \frac{\gamma }{1-\gamma ^2}-\frac{\hat{w}_1 \hat{w}_2}{\lambda} & 
      \e^{z}\left(\frac{\gamma  }{\gamma ^2-1}-\hat{w}_1 \hat{w}_2\left(\frac{1}{\lambda}-\frac{2}{k_{\perp}^2}\right)\right) \\
 \e^{-z}\left(\frac{\gamma  }{\gamma ^2-1}-\hat{w}_1 \hat{w}_2 \left(\frac{1}{\lambda}-\frac{2}{k_{\perp}^2}\right)\right) & \frac{\gamma }{1-\gamma
   ^2}-\frac{\hat{w}_1 \hat{w}_2}{\lambda} \\
    \end{array}\right]\,, \\
     \Omega_{\hat{b}\cdot k_{\perp}}'|_{\text{block 3}}& =\frac{\hat{b}\cdot k_\perp}{\lambda}\left[\begin{array}{cc}
     1 & \e^{z} \\
 \e^{-z}& 1 \\
    \end{array}\right]\,,      
    \end{align}
\end{subequations}
where we used the shorthands:
\begin{equation}\label{eq:lambdadef}
    \lambda=(\hat{b}\cdot k_\perp)^2+k_{\perp}^{2} \quad \text{and} \, \quad z=\frac{2i\hat{w}_{1}\hat{w}_{2}\Lambda}{k_{\perp}^{2}}\,,
\end{equation}
as well as $k_{\perp}^{2}=\hat{w}_{1}^{2}+\hat{w}_{2}^{2}-2\gamma\hat{w}_{1}\hat{w}_{2}$ as in~\eqref{eq:newkin}.

\paragraph{Full $\varepsilon$-transformation and anticipated practical value.} At the level of the complete initial basis in \eqref{eq:startingBasis}, the transformation  we constructed above is summarized as 
\begin{equation}
    \mathbf{J}'=\left(\mathcal{U}\equiv \left[\begin{array}{ccc}
        \mathcal{U}|_{\text{block 1}} & \cdot & \cdot \\
        \cdot & \mathcal{U}|_{\text{block 2}} & \cdot\\
        \cdot & \cdot & \mathcal{U}|_{\text{block 3}}
    \end{array}\right]\right)^{-1}\cdot \mathbf{J}\,.
\end{equation}
As desired, this system of master integrals now satisfies a $\varepsilon$-factorized block-diagonal differential equation:
\begin{equation}
    \begin{aligned}
    \dd\mathbf{J}' &= \left[\Omega_{0}'+\varepsilon\Omega_{1}'\right]\cdot\mathbf{J}'\,,
    \quad \text{where} \quad
    \Omega_{0}'  = \begin{bNiceArray}[margin = 5pt]{cccccc}
        \CodeBefore
        \Body
            \mzero & \mzero & \mzero & \mzero & \mzero & \mzero
            \\
            \mzero & \mzero & \mzero & \mzero & \mzero & \mzero
            \\
            \mzero & \mzero & \mzero & \mzero & \mzero & \mzero
            \\
            \mzero & \mzero & \mzero & \mzero & \mzero & \mzero
            \\
            \red{\bigcdot} & \mzero & \red{\bigcdot} & \mzero & \mzero & \mzero
            \\
            \red{\bigcdot} & \mzero & \red{\bigcdot} & \mzero & \mzero & \mzero
            \CodeAfter
            \tikz{
            }
    \end{bNiceArray}\,, &\quad \Omega_{1}' = \begin{bNiceArray}[margin = 5pt]{cccccc}
        \CodeBefore
            \rectanglecolor{red!20}{1-1}{2-2}
            \rectanglecolor{red!20}{3-3}{4-4}
            \rectanglecolor{red!20}{5-5}{6-6}
        \Body
            \red{\bigcdot} & \red{\bigcdot} & \mzero & \mzero & \mzero & \mzero
            \\
            \red{\bigcdot} & \red{\bigcdot} & \mzero & \mzero & \mzero & \mzero
            \\
            \mzero & \mzero & \red{\bigcdot} & \red{\bigcdot} & \mzero & \mzero
            \\
            \mzero & \mzero & \red{\bigcdot} & \red{\bigcdot} & \mzero & \mzero
            \\
            \red{\bigcdot} & \red{\bigcdot} & \red{\bigcdot} & \red{\bigcdot} & \red{\bigcdot} & \red{\bigcdot}
            \\
            \red{\bigcdot} & \red{\bigcdot} & \red{\bigcdot} & \red{\bigcdot} & \red{\bigcdot} & \red{\bigcdot}
            \CodeAfter
            \tikz{
            }
    \end{bNiceArray}\,.
    \end{aligned}
\end{equation}
The explicit form of $\Omega_0'$ and $\Omega_1'$ are given in the ancillary file \texttt{DE\_linear\_eps.m} in the repository \cite{repo}.
To remove the non-$\varepsilon$-factorized \emph{lower}-triangular contribution $\Omega'_{0}$, we further transform $\mathbf{J}'=\mathcal{V}\cdot \mathbf{J}''$, where
\begin{equation}\label{eq:Vmatrix2}
    \mathcal{V}\equiv
    \begin{bNiceMatrix}
        1 & \mzero&\mzero&\mzero&\mzero&\mzero \\
        \mzero & 1&\mzero&\mzero&\mzero&\mzero \\
        \mzero & \mzero&1&\mzero&\mzero&\mzero \\
        \mzero & \mzero&\mzero&1&\mzero&\mzero \\
        G_{1}(\hat{w}_{1},\hat{w}_{2},\gamma,\hat{b}\cdot k_\perp) & \mzero&G_{2}(\hat{w}_{1},\hat{w}_{2},\gamma,\hat{b}\cdot k_\perp)&\mzero&1&\mzero \\
        G_{3}(\hat{w}_{1},\hat{w}_{2},\gamma,\hat{b}\cdot k_\perp)& \mzero&G_{4}(\hat{w}_{1},\hat{w}_{2},\gamma,\hat{b}\cdot k_\perp)&\mzero&\mzero&1 \\
      \end{bNiceMatrix}\,,
\end{equation}
requiring
\begin{equation}\label{eq:constraint}
    \mathcal{V}^{-1}\cdot(\Omega_{0}\cdot\mathcal{V}-\dd\mathcal{V})=0 \qquad \text{or equivalently} \qquad \d \mathbf{J}''=\varepsilon [\Omega_1''\equiv  \mathcal{V}^{-1}\cdot \Omega_1' \cdot \mathcal{V}]\cdot \mathbf{J}''\,.
\end{equation}
Using the shorthand $x = \hat{b} \cdot k_\perp$, a more explicit way to write the left constraint in~\eqref{eq:constraint} is as an integral over the closed one-form
\begin{equation}\label{eq:G1specialFunction}
    G_1(p)-G_1(p_0)
    = \int_{\mathcal{C}\,:\,p_0\to p}\,[G_{\hat{w}_{1}}\,\mathrm{d}\hat{w}_{1}
    + G_{\hat{w}_{2}}\,\mathrm{d}\hat{w}_{2}
    + G_{\gamma}\,\mathrm{d}\gamma
    + G_{x}\,\mathrm{d}x]\,,
\end{equation}
where the integration contour $\mathcal{C}$ runs from the base point $p_0$ to the point $p$ without crossing any branch cut and, consequently, $G_1$ is determined only up to an additive constant $G_1(p_0)$. Explicitly, the functions in the integrand read
\begin{subequations}
\begin{align}
G_{\hat{w}_{1}} &= \frac{\e^{-z_{-}}
\left[\begin{array}{rll}&
\left(\hat{w}_{1}\hat{w}_{2}^{2}\lambda(\gamma^{2}-1)+ix(\hat{w}_{1}-\gamma\hat{w}_{2})(k_{\perp}^2-i\hat{w}_{1}\hat{w}_{2}\Lambda)\right)I_{0}(\hat{w}_{2})
\\&
+\hat{w}_{2}\lambda\left(ix(\hat{w}_{1}-\gamma\hat{w}_{2})+i\hat{w}_{2}\Lambda\right)I_{1}(\hat{w}_{1})
\end{array}\right]
}{2k_{\perp}^2\lambda^{3/2}}\,,\\
G_{\hat{w}_{2}} &= \frac{\e^{-z_{-}}
\left[\begin{array}{rll}&
    \big(\lambda(\hat{w}_{2}^{3}-3\gamma\hat{w}_{1}\hat{w}_{2}^{2}+(2+\gamma^{2})\hat{w}_{1}^{2}\hat{w}_{2}-\gamma\hat{w}_{1}^{3})\\&-ix(\gamma\hat{w}_{1}-\hat{w}_{2})(k_{\perp}^{2}-i\hat{w}_{1}\hat{w}_{2}\Lambda)\big)I_{0}(\hat{w}_{2})\\&
    -\hat{w}_{1}\lambda\left(ix(\hat{w}_{1}-\gamma\hat{w}_{2})+i\hat{w}_{2}\Lambda\right)I_{1}(\hat{w}_{1})
\end{array}\right]
}{2k_{\perp}^2\lambda^{3/2}}\,,\\
G_{\gamma} &= \frac{\e^{-z_{-}}\hat{w}_{1}\hat{w}_{2}\left[\begin{array}{rll}&(\gamma^{2}-1)\left(\hat{w}_{1}\lambda(\hat{w}_{1}-\gamma\hat{w}_{2})+ix(k_{\perp}^{2}-i\hat{w}_{1}\hat{w}_{2}\Lambda)\right)I_{0}(\hat{w}_{2})\\&
+\lambda(ix(\gamma^{2}-1)\hat{w}_{2}+i(\hat{w}_{1}-\gamma\hat{w}_{2})\Lambda)I_{1}(\hat{w}_{2})
\end{array}\right]}{2k_{\perp}^2\lambda^{3/2}(1-\gamma^{2})}\,,\\
G_{x} &= \frac{\e^{-z_{-}}\left[(k_{\perp}^2-i\hat{w}_{1}\hat{w}_{2}\Lambda)I_{0}(\hat{w}_{2})+\hat{w}_{2}\lambda I_{1}(\hat{w}_{2})\right]}{2i\lambda^{3/2}}\,,
\end{align}
\end{subequations}
with $z_\pm$ and $\Lambda$ defined in~\eqref{eq:zPmAndLambda}, and $\lambda$ in~\eqref{eq:lambdadef}, all written in terms of the dimensionless variables introduced in~\eqref{eq:dimLessVar}.
The remaining entries of~\eqref{eq:Vmatrix2} satisfy differential equations related to $G_1$ according to
\begin{equation}
    \begin{aligned}
        \partial_{a}G_{2}(\hat{w}_{1},\hat{w}_{2},\gamma,x) &= \partial_{a}G_{1}(\hat{w}_{1},\hat{w}_{2},\gamma,x)|_{x\rightarrow-x,\hat{w}_{1}\leftrightarrow\hat{w}_{2}}\,,\\
        \partial_{a}G_{3}(\hat{w}_{1},\hat{w}_{2},\gamma,x) &= \partial_{a}G_{1}(\hat{w}_{1},\hat{w}_{2},\gamma,x)|_{\Lambda\rightarrow-\Lambda}\,,\\
        \partial_{a}G_{4}(\hat{w}_{1},\hat{w}_{2},\gamma,x) &= \partial_{a}G_{1}(\hat{w}_{1},\hat{w}_{2},\gamma,x)|_{x\rightarrow-x,\hat{w}_{1}\leftrightarrow\hat{w}_{2},\Lambda\rightarrow-\Lambda}\,.\\
    \end{aligned}
\end{equation}
From the way they are derived, the functions $G_i$ are structurally similar to the so-called $G$-functions that appear in elliptic and Calabi--Yau Feynman integrals (see, for example,~\cite{Gorges:2023zgv}), which are notoriously difficult (and perhaps sometime impossible) to express in closed form. 

However, \emph{in practice}, all that is truly required to evaluate the $G_i$’s is the differential equation they obey (e.g., $\mathrm{d}G_1 = G_x\mathrm{d}x + G_\gamma\mathrm{d}\gamma + G_{\hat{w}_{1}}\mathrm{d}\hat{w}_{1} + G_{\hat{w}_{2}}\mathrm{d}\hat{w}_{2}$).
This relation determines all derivatives of $G_i$ at a given point and thus enables the systematic construction of its local power (i.e., Frobenius) series around regular or singular points through recursive relations among the expansion coefficients (see, for example,~\cite{Coro:2025vgn} for a recent application in jet production amplitudes).

The explicit form of the $\varepsilon$-factorized differential equations $\Omega_1''$ written in terms of the $G$-functions is given in the ancillary file \texttt{DE\_eps\_factorized.m} in the repository \cite{repo}.
Hence, the most direct way to render the results of this section practical for numerical evaluation of the integrals is to expand the differential equation around a regular singular point. We leave a detailed implementation of this procedure, together with the in-dept analytic characterization of the underlying function space of iterated integrals for future work.

%% file: Sections/Conclusions.tex
{\section{Conclusion and outlook}\label{sec:outlooks}}

In this paper, we present a systematic framework for computing frequency-domain gravitational waveforms in physically relevant regimes, building on the KMOC formalism for classical observables. The method focuses on the impact-parameter-space Fourier integrals that determine the waveform, whose differential equations are solved asymptotically using restriction theory in combination with companion-matrix techniques and finite-field reconstruction methods. As a proof of principle, we apply the framework to the tree-level waveform, obtaining explicit results in both the soft and PN limits. The PN expansion is validated through an analytic series computed up to $\mathcal{O}(v^{30})$; a truncation chosen for illustration rather than imposed by the method. The framework itself is general and readily extends to higher-order corrections to the waveform.

Beyond systematizing the extraction of asymptotic solutions, we also outline a strategy, inspired by~\cite{Gorges:2023zgv}, to cast the differential equations satisfied by the master Fourier integrals into an $\varepsilon$-factorized form. This is achieved by constructing an appropriate period matrix via rapid-decay homology, which organizes the integration contours in a way analogous to the $\varepsilon$-factorized-basis construction for Feynman integrals. The resulting differential system is expressed in terms of special functions involving Bessel and exponential kernels, which, to our knowledge, have not appeared previously in this context, and thus introduces a new member to the family of (generalized) hypergeometric functions, which already included polylogarithms, elliptic integrals, and (certain) Calabi--Yau periods showing up in perturbative quantum field theory.

Several extensions follow naturally. On the precision side, extending the construction to higher orders in the PM expansion is especially compelling: at next-to-leading order (NLO), the relevant families already involve two-loop Fourier integrals~\cite{Brunello:2025todo}, while at next-to-next-to-leading order (NNLO) new functions may emerge. Nevertheless, the function space underlying the PM expansion within PN is highly constrained~\cite{Bini:2023fiz,Bini:2024rsy} and, at NNLO in PM (within PN), is already known~\cite{Bini:2024ijq}, providing sharp benchmarks for our method. A natural further direction is to incorporate spin effects, which leave the integral families unchanged but significantly complicate the associated IBP coefficients.

Regarding the $\varepsilon$-factorized differential equation obtained in this paper, 
it would be valuable to establish a systematic understanding and develop
dedicated tools to handle iterated integrals with Bessel and exponential
kernels, as well as to evaluate them efficiently by analytic or numerical means. 
Building on this foundation, a natural next step is to construct and solve an $\varepsilon$-factorized form of the NLO waveform integrals~\cite{Brunello:2025todo}. This will likely require input from rapid-decay homology to identify the relevant multi-dimensional cycles and construct the associated period matrix. 

It would also be interesting to investigate how the results obtained here, in particular the $\varepsilon$-factorized differential equation, could be generalized (perhaps through analytic continuations generalizing that discussed in \cite{Kalin:2019rwq,Kalin:2019inp}) from the waveform associated with unbound scattering to that describing bound orbits. Of course, achieving this will require a deeper conceptual understanding of the analytic properties of the waveform; yet we expect that the practical tools developed in this paper provide novel starting points for investigating such continuations explicitly, should they exist. 

Our hope is that the future developments outlined above will ultimately enable the construction of high-precision, phenomenological templates for gravitational waveforms and open a new window onto the properties of distant astrophysical objects, potentially placing useful constraints on their internal structure, such as the equation of state of neutron stars.

Applying the methods developed here to other, \emph{largely unexplored} classes of observables in the context of the S-matrix theory (see, e.g.,~\cite{Caron-Huot:2023vxl}) would also be interesting.

%% file: Sections/appA_BoundarySoft.tex
\section{Boundary conditions for soft and PN expansions\label{app:boundarySoft}}

\paragraph{Soft.}
In this section, we derive the boundary conditions used in \secref{sec:soft_expansion} in both the soft and hard regions. For this, we use the results~\cite{Peskin:1995ev} for the following Euclidean signature integrals:
\begin{equation}
    \int_{\mathbb{R}^{\D}}\hat{\rd}^{\D}q\frac{1}{(q^{2}+\Lambda)^{n}} = \frac{\Gamma\left(\frac{2n-\D}{2}\right)}{(4\pi)^{\D/2}\Gamma(n)}\Lambda^{\frac{\D-2n}{2}}, \quad \int_{\mathbb{R}^{\D}}\hat{\rd}^{\D}q\frac{\e^{iq\cdot b}}{(q^{2})^{n}} = \frac{\Gamma\left(\frac{\D-2n}{2}\right)}{4^{n}\pi^{\D/2}\Gamma(n)}(b^{2})^{\frac{2n-\D}{2}}\,.
\end{equation}
For each master integral, we are interested in calculating their value at the boundary point $\mathbf{z}^{*}=(0,1,\frac{3}{2},0)$, in both the soft region with $\abs{q_{\perp}}\sim\omega$ and the hard region with $\abs{q_{\perp}}\gg\omega$. From  \eqref{eq:soft_vars}, this corresponds to $\omega=0$, $u_{2}\cdot n=u_{1}\cdot n$, $\gamma=\frac{3}{2}$ and $\hat{b}\cdot n=0$. Starting with the first master integral, we have, using \eqref{eq:BOLDQ},
\begin{align}
        J_{1} = I(1,0,0)&= \int
    \hat{\rd}^{2-2\varepsilon}q_{\perp}\frac{\e^{iq_{\perp}\cdot\hat{b}}}{q_{\perp}^{2}-\hat{w}_{2}^{2}}\,,\\
    &= -\int\hat{\rd}^{2-2\varepsilon}\mathbf{q}_{\perp}\frac{\e^{-i\mathbf{q}_{\perp}\cdot\hat{\mathbf{b}}}}{\mathbf{q}_{\perp}^{2}+\hat{w}_{2}^{2}}\notag\,,\\
    &= -\omega^{-2\varepsilon}\left(\int
    \frac{\hat{\rd}^{2-2\varepsilon}\mathbf{q}_{\perp}}{\mathbf{q}_{\perp}^{2}+\frac{z_{1}^{2}}{\omega^{2}}}+\mathcal{O}(\omega)\right)
    -\left(\int
    \hat{\rd}^{2-2\varepsilon}\mathbf{q}_{\perp}\frac{\e^{-i\mathbf{q}_{\perp}\cdot\hat{\mathbf{b}}}}{\mathbf{q}_{\perp}^{2}}+\mathcal{O}(\omega)\right)\notag\,,\\
    &= z_{1}^{-2\varepsilon}\left(C_{\varepsilon}^{\text{soft}}+\mathcal{O}(z_{1})\right)+\left(C_{\varepsilon}^{\text{hard}}(-\hat{b}^{2})^{\varepsilon}+\mathcal{O}(z_{1})\right)\,,\notag
\end{align}
where
\begin{equation}
    C_{\varepsilon}^{\text{soft}} = -\frac{\Gamma(\varepsilon)}{(4\pi)^{1-\varepsilon}} \quad \text{and} \quad C_{\varepsilon}^{\text{hard}} = -\frac{\Gamma(-\varepsilon)}{4\pi^{1-\varepsilon}}\,.
\end{equation}
Note that the opposite signs in the $\Gamma$-functions are easy to understand physically: they lead to opposite $1/\varepsilon$  residues, making the boundary value finite when the soft and hard regions are added.
We see that, as predicted in \secref{sec:soft_expansion} from the eigenvalues, the master integrals separate into a finite part as $z_{1}\rightarrow0$ and a part proportional to $z_{1}^{-2\varepsilon}$. The second master integral can be calculated from this observation and we find
\begin{equation}
    J_{2} = I(1,0,-1) = \hat{b}^{\mu}\pdv{J_{1}}{\hat{b}^{\mu}} = z_{1}^{-2\varepsilon}\left(\mathcal{O}(z_{1})\right)+\left(2\varepsilon C_{\varepsilon}^{\text{hard}}+\mathcal{O}(z_{1})\right)\,.
\end{equation}
The third and fourth master integrals can be similarly calculated from
\begin{equation}
    \begin{aligned}
    J_{3}\eval_{\mathbf{z}=\mathbf{z}^{*}} &= I(0,1,0)\eval_{\mathbf{z}=\mathbf{z}^{*}} = \e^{i\hat{b}\cdot k}J_{1}\eval_{\hat{w}_{2}\rightarrow\hat{w}_{1},\mathbf{z}=\mathbf{z}^{*}} = J_{1}\,,\\
    J_{4}\eval_{\mathbf{z}=\mathbf{z}^{*}} &= I(0,1,-1)\eval_{\mathbf{z}=\mathbf{z}^{*}} = \hat{b}^{\mu}\pdv{J_{3}}{\hat{b}^{\mu}}\eval_{\mathbf{z}=\mathbf{z}^{*}} = J_{2}\,,
    \end{aligned}
\end{equation}
where we have used the fact that at $\mathbf{z}=\mathbf{z}^{*}$ implies $\hat{w}_{1}=\hat{w}_{2}$, and $\hat{b}\cdot k=0$.
For the fifth master integral we have
\begin{align}
            J_{5}\eval_{\mathbf{z}=\mathbf{z}^{*}} = I(1,1,0)\eval_{\mathbf{z}=\mathbf{z}^{*}} &= \int
        \hat{\rd}^{2-2\varepsilon}\mathbf{q}_{\perp}\frac{\e^{-i\mathbf{q}_{\perp}\cdot\hat{b}}}{(\mathbf{q}_{\perp}^{2}+\hat{w}_{2}^{2})((\mathbf{q}_{\perp}-\mathbf{k}_{\perp})^{2}+\hat{w}_{1}^{2})}\eval_{\mathbf{z}=\mathbf{z}^{*}}\,,\notag\\
        &= \omega^{-2-2\varepsilon}\left(\int
        \hat{\rd}^{2-2\varepsilon}\mathbf{q}_{\perp}\frac{1}{\left(\mathbf{q}_{\perp}^{2}+\frac{z_{1}^{2}}{\omega^{2}}\right)\left((\mathbf{q}_{\perp}-\mathbf{n}_{\perp})^{2}+\frac{z_{1}^{2}}{\omega^{2}}\right)}+\mathcal{O}(\omega)\right)\\
        &\quad+\left(\int
        \hat{\rd}^{2-2\varepsilon}\mathbf{q}_{\perp}\frac{\e^{-i\mathbf{q}_{\perp}\cdot\hat{b}}}{(\mathbf{q}_{\perp}^{2})^{2}}+\mathcal{O}(\omega)\right)\,,\notag\\
        &= z_{1}^{-2-2\varepsilon}\left(\varepsilon C_{\varepsilon}^{\text{soft}}\int_{0}^{1}\dd x\,(x^{2}-x-1)^{-1-\varepsilon}+\mathcal{O}(z_{1})\right)\notag\\
        &\quad+\left(\frac{C_{\varepsilon}^{\text{hard}}}{4(1+\varepsilon)}(-\hat{b}^{2})^{1+\varepsilon}+\mathcal{O}(z_{1})\right)\,,\notag
\end{align}
where we have used Feynman parameters in the soft region. We can then find the solution for the soft region as a series in $\varepsilon$, and we get
\begin{equation}
    J_{5}\eval_{\mathbf{z}=\mathbf{z}^{*}} = z_{1}^{-2-2\varepsilon}\left(-\frac{4\varepsilon\tanh^{-1}\left(\frac{1}{\sqrt{5}}\right)}{\sqrt{5}} C_{\varepsilon}^{\text{soft}}+\mathcal{O}(\varepsilon,z_{1})\right)+\left(\frac{C_{\varepsilon}^{\text{hard}}}{4(1+\varepsilon)}(-\hat{b}^{2})^{1+\varepsilon}+\mathcal{O}(z_{1})\right)\,.
\end{equation}
Finally, for the sixth master integral we have
\begin{equation}
    J_{6}\eval_{\mathbf{z}=\mathbf{z}^{*}} = \hat{b}^{\mu}\pdv{J_{5}}{\hat{b}^{\mu}}\eval_{\mathbf{z}=\mathbf{z}^{*}} = z_{1}^{-2-2\varepsilon}\left(\mathcal{O}(\varepsilon,z_{1})\right)+\left(\frac{C_{\varepsilon}^{\text{hard}}}{2}(-\hat{b}^{2})^{1+\varepsilon}+\mathcal{O}(z_{1})\right)\,.
\end{equation}

\paragraph{PN.} In this section, we derive the boundary conditions used in~\secref{sec:PN_expansion}. In this case, there are no regions, so we just need to evaluate our master integrals on some submanifold of the kinematic space in $\D=4$. For this we will use Schwinger parameters
\begin{equation}
    \frac{1}{A^{n}} = \frac{1}{\Gamma(n)}\int_{0}^{\infty}\dd\alpha\,\alpha^{n-1}\e^{-\alpha A}\,,
\end{equation}
and evaluate the integrals in the PN limit, which corresponds to:
\begin{equation}
    \gamma\rightarrow1, \quad \hat{w}_{i} \approx \omega=z_{2}, \quad \hat{b}\cdot k\rightarrow0\,.
\end{equation}
For the boundary conditions, we will consider the point $\mathbf{z}=\mathbf{z}^{*}=(0,1,z_{3},z_{4})$. Starting with the first master integral, we have, using \eqref{eq:BOLDQ},
\begin{equation}
    \begin{aligned}
    J_{1}\eval_{\mathbf{z}=\mathbf{z}^{*}} = I(1,0,0)\eval_{\mathbf{z}=\mathbf{z}^{*}} &= \int
    \hat{\rd}^{2}q_{\perp}\frac{\e^{iq_{\perp}\cdot \hat{b}}}{q_{\perp}^{2}-1}\\
    & = -\int\hat{\rd}^{2}\mathbf{q}_{\perp}\frac{\e^{-i\mathbf{q}_{\perp}\cdot \mathbf{\hat{b}}}}{\mathbf{q}_{\perp}^{2}+1}
    \\
    &
    =-\int_{0}^{\infty}\dd\alpha\,\e^{-\alpha}\int\hat{\rd}^{2}\mathbf{q}_{\perp}\,\e^{-i\mathbf{q}_{\perp}\cdot \mathbf{\hat{b}}-\alpha \mathbf{q}_{\perp}^{2}}
    \\
    &
    = -\int_{0}^{\infty}\dd\alpha\,\e^{-\alpha-\frac{\mathbf{\hat{b}}^{2}}{4\alpha}}\int\hat{\rd}^{2}\mathbf{q}_{\perp}\,\e^{-\alpha \mathbf{q}_{\perp}^{2}}
    \\
    &
    =- \frac{1}{4\pi}\int_{0}^{\infty}\dd\alpha\,\alpha^{-1}\,\e^{-\alpha-\frac{\mathbf{\hat{b}}^{2}}{4\alpha}}
    \\
    &
    =- \frac{1}{2\pi}K_{0}\left(\sqrt{-\hat{b}^{2}}\right)\,,
    \end{aligned}
\end{equation}
where here $\mathbf{\hat{b}}^2= -\hat{b}^2$.

The second master integral is then given by
\begin{equation}
    J_{2}\eval_{\mathbf{z}=\mathbf{z}^{*}} = I(1,0,-1)\eval_{\mathbf{z}=\mathbf{z}^{*}} = \hat{b}^{\mu}\pdv{J_{1}}{\hat{b}^{\mu}}|_{\mathbf{z}=\mathbf{z}^{*}} = \frac{1}{2\pi}\sqrt{-\hat{b}^{2}}K_{1}\left(\sqrt{-\hat{b}^{2}}\right)\,.
\end{equation}
We can then obtain the third and fourth master integrals from these
\begin{equation}
    \begin{aligned}
    J_{3}\eval_{\mathbf{z}=\mathbf{z}^{*}} &= I(0,1,0)\eval_{\mathbf{z}=\mathbf{z}^{*}} = \e^{i\hat{b}\cdot k}J_{1}\eval_{\hat{w}_{2}\rightarrow\hat{w}_{1},\mathbf{z}=\mathbf{z}^{*}} = J_{1}\eval_{\mathbf{z}=\mathbf{z}^{*}}\,,\\
    J_{4}\eval_{\mathbf{z}=\mathbf{z}^{*}} &= I(0,1,-1)\eval_{\mathbf{z}=\mathbf{z}^{*}} = \hat{b}^{\mu}\pdv{J_{3}}{\hat{b}^{\mu}} = J_{2}\eval_{\mathbf{z}=\mathbf{z}^{*}}\,.
    \end{aligned}
\end{equation}
The final two master integrals are also evaluated simply as
\begin{equation}
    \begin{aligned}
        J_{5}\eval_{\mathbf{z}=\mathbf{z}^{*}} = I(1,1,0)\eval_{\mathbf{z}=\mathbf{z}^{*}} &= \int
        \hat{\rd}^{2}q_{\perp}\frac{\e^{iq_{\perp}\cdot \hat{b}}}{(q_{\perp}^{2}-1)^{2}}\\
        &= \int\hat{\rd}^{2}\mathbf{q}_{\perp}\frac{\e^{-i\mathbf{q}_{\perp}\cdot \hat{\mathbf{b}}}}{(\mathbf{q}_{\perp}^{2}+1)^{2}}\\
        &= \frac{1}{4\pi}\int\dd\alpha\,\e^{-\alpha-\frac{\mathbf{\hat{b}}^{2}}{4\alpha}}\\
        &= \frac{1}{4\pi}\sqrt{-\hat{b}^{2}}K_{1}\left(\sqrt{-\hat{b}^{2}}\right)\,,\\
        J_{6}\eval_{\mathbf{z}=\mathbf{z}^{*}} = I(1,1,-1)\eval_{\mathbf{z}=\mathbf{z}^{*}} &= \hat{b}^{\mu}\pdv{J_{5}}{\hat{b}^{\mu}}\eval_{\mathbf{z}=\mathbf{z}^{*}}\\
        &= \frac{1}{4\pi}\hat{b}^{2}K_{0}\left(\sqrt{-\hat{b}^{2}}\right)\,.
    \end{aligned}
\end{equation}

%% file: Sections/appC_method.tex
\section{Computational implementation in the PN limit \label{app:method}}
In this section, we describe the computational techniques used to find the $z_{1}$-expansion of the tree-level waveform in the post-Newtonian limit, using \textit{companion tensor algebra} \cite{Brunello:2024tqf}. In this case, we are interested in reconstructing the expansion of the $\mathbf{W}^{(0)}$ functions, which are given by
\begin{equation}
    \mathbf{W}^{(0)} = \mathbf{A}\cdot F\,,
\end{equation}
where $\mathbf{A}=\mathbf{C}'\cdot G\cdot M^{-1}$ and $F$ satisfies the differential equation
\begin{equation}
    \partial_{1}F-\Omega_{F}\cdot F = 0\,, \quad \Omega_{F} = M\cdot G^{-1}(\Omega_{1}\cdot G\cdot M^{-1}-\partial_{1}(G\cdot M^{-1}))\,,
\end{equation}
with all these matrices defined in~\secref{sec:PN_expansion}. Companion tensor algebra allows us to perform such series expansions without symbolic processing. In this framework, one views the $z_{1}$-expansion of the solution $F$ as a \brk{semi-infinite} matrix
\begin{align}
    F = \sum_{k=0}^{\infty}F_{k}z_{1}^{k}
    =
    \NiceMatrixOptions{
        code-for-first-row = \scriptstyle,
        code-for-first-col = \scriptstyle,
        cell-space-limits = 3pt,
    }
    \begin{bNiceMatrix}[margin = 3pt, first-row, first-col]
        &
        & \gr{\overset{-1}{\downarrow}}
        & \gr{\overset{0}{\downarrow}}
        & \gr{\overset{1}{\downarrow}}
        & \gr{\overset{2}{\downarrow}}
        & 
        \\
        \gr{1 \rightarrow}
        & \Cdots[color = gr]
        & \mZero
        & (\mathbf{F}_{0})_{1}
        & (\mathbf{F}_{1})_{1}
        & (\mathbf{F}_{2})_{1}
        & \Cdots[color = gr]
        \\
        \gr{2 \rightarrow}
        & \Cdots[color = gr]
        & \mZero
        & (\mathbf{F}_{0})_{2}
        & (\mathbf{F}_{1})_{2}
        & (\mathbf{F}_{2})_{2}
        & \Cdots[color = gr]
        \\
        & & \Vdots[color = gr] & & & \Vdots[color = gr]
        \\
        \gr{\nu \rightarrow}
        & \Cdots[color = gr]
        & \mZero
        & (\mathbf{F}_{0})_{\nu}
        & (\mathbf{F}_{1})_{\nu}
        & (\mathbf{F}_{2})_{\nu}
        & \Cdots[color = gr]
        \CodeAfter
            \tikz \draw[
                gr,
                line cap = round,
                line width = 1.06pt,
                dash pattern= on 0pt off .18cm
            ]
                (1|-3) ++(-.5, 0)
                -- ++($(1|-4.5) - (1|-3) + (0, .2)$)
            ;
    \end{bNiceMatrix}
    \cdot
    \begin{bNiceMatrix}[margin]
        \Vdots[color = gr] \\ z_{1}^{-1} \\ 1 \\ z_{1} \\ z_{1}^{2} \\ \Vdots[color = gr, shorten-end = 5pt]
    \end{bNiceMatrix}
    \,.
\end{align}
Here, $(\mathbf{F}_{i})_{j}$ is a $2$-vector corresponding to the $j^{\text{th}}$ row of the $i^{\text{th}}$ term in the $z_{1}$-expansion of $F$, making up the two components of the Bessel function basis. In this matrix representation, the rows correspond to the $\nu$-dimensional vector space of master integrals, while the columns correspond to the space of Laurent monomials in the expansion variable $z_{1}$
\begin{align}
    \mathcal{L} = \mathrm{Span}\bigbrk{\ldots, z_{1}^{-1}, 1, z_{1}, z_{1}^{2}, \ldots}
    \,.
\end{align}
Series expansions of the solution, therefore, naturally belong to tensor products of these two vector spaces, complemented by the $2$-dimensional Bessel function space $F\in\mathbb{K}_{\nu}\otimes\mathcal{L}\otimes\mathbb{B}_{2}$.
Within this framework, various algebraic operations are replaced with \emph{companion matrix} representations of the corresponding operators on this tensor space, essentially reducing symbolic computations to matrix multiplication.

\paragraph{Companion matrices.} Our building blocks are companion matrices for $z_{1}$ and $\partial_{1}$ in the $\mathcal{L}$ vector space
\begin{align}
    \mathcal{L}\sbrk{z_{1}} =
    \NiceMatrixOptions{code-for-first-col = \scriptstyle}
    \begin{bNiceArray}{c ccc ccc ccc}[margin, nullify-dots, first-col]
        \CodeBefore
        \Body
        & \phantom{\gr{0}} & & & & & & & &
    \\
        & \phantom{1} & & & & & & & &
    \\
        & & \Ddots[shorten-end=-1mm] & \Ddots[color = gr, draw-first] & & & & & &
    \\
        \gr{-1 \rightarrow}
        & & & 1 & \gr{0} & & & & & &
    \\
        \gr{0 \rightarrow}
        & & & & 1 & \gr{0} & & & &
    \\
        \gr{1 \rightarrow}
        & & & & & 1 & \gr{0} & & &
    \\
        \gr{2 \rightarrow}
        & & & & & & 1 & \gr{0} & &
    \\
        & & & & & & & \Ddots & \Ddots[color = gr, shorten-end=0mm] &
    \\
        & & & & & & & & \phantom{1} & \phantom{\gr{0}}
    \CodeAfter
        \tikz \draw[
            gr,
            line cap = round,
            line width = 1.06pt,
            dash pattern= on 0pt off .18cm
        ]
            (1|-1.5) ++(-.5, 0)
            -- ++($(1|-4.5) - (1|-1.5) + (0, .2)$)
            ++($(1|-8) - (1|-4.5) - (0, .2)$)
            -- ++($(1|-8.5) - (1|-7) - (0, .2)$)
        ;
    \end{bNiceArray}
    \hspace{1.5cm}
    \mathcal{L}\sbrk{\partial_{1}} =
    \begin{bNiceArray}{c ccc ccc ccc}[margin, nullify-dots, first-col]
        & \phantom{\gr{0}} & \phantom{0} & & & & & & &
    \\
        & & & & & & & & &
    \\
        & & & \Ddots[color = gr] & \Ddots & & & & &
    \\
        \gr{-1 \rightarrow}
        & & & & \gr{0} & 0 & & & &
    \\
        \gr{0 \rightarrow}
        & & & & & \gr{0} & 1 & & &
    \\
        \gr{1 \rightarrow}
        & & & & & & \gr{0} & 2 & &
    \\
        \gr{2 \rightarrow}
        & & & & & & & \gr{0} & 3 &
    \\
        & & & & & & & & \Ddots[color = gr] & \Ddots
    \\
        & & & & & & & & & \phantom{\gr{0}}
    \CodeAfter
        \tikz \draw[
            gr,
            line cap = round,
            line width = 1.06pt,
            dash pattern= on 0pt off .18cm
        ]
            (1|-1.5) ++(-.5, 0)
            -- ++($(1|-4.5) - (1|-1.5) + (0, .2)$)
            ++($(1|-8) - (1|-4.5) - (0, .2)$)
            -- ++($(1|-9.5) - (1|-8) - (0, .2)$)
        ;
    \end{bNiceArray}
    \,,
    \label{eq:cmats_weyl_infinite}
\end{align}
where in both sketches the $0$'s are along the diagonal.
In practice $\mathcal{L}$ is an infinite-dimensional vector space, however we can truncate it by defining a minimum and maximum value of $z_{1}$ that we are considering.
The minimum value is determined by the start of the series expansion we are looking for, and in this case it is just $0$.
The maximum value is determined by the order we are interested in finding our series expansion up to, we call this $\ord$.
Then we can truncate this to a $\brk{\ord + 1}$-dimensional vector space
\begin{align}
    \mathcal{L}\sbrk{z_{1}}
    = 
    \NiceMatrixOptions{code-for-first-col = \scriptstyle}
    \begin{bNiceArray}{c ccc ccc}[margin, nullify-dots, first-col]
        \CodeBefore
        \Body
        \gr{0 \rightarrow}
        & \gr{0} & & & & &
        \\
        \gr{1 \rightarrow}
        & 1 & \gr{0} & & & &
        \\
        \gr{2 \rightarrow}
        & & 1 & \gr{0} & & &
        \\
        & & & \Ddots & \Ddots[color = gr, shorten-end=0mm] & &
        \\
        & & & & & \gr{0} &
        \\
        \gr{\ord \rightarrow}
        & & & & & 1 & \gr{0}
    \CodeAfter
        \tikz \draw[
            gr,
            line cap = round,
            line width = 1.06pt,
            dash pattern= on 0pt off .18cm
        ]
            (1|-4) ++(-.5, 0)
            -- ++($(1|-6.5) - (1|-4) + (0, .2)$)
        ;
        \tikz \draw[<->, gr, shorten <> = .5em]
            (1-|7.5) ++(.5, 0)
            -- ($(7-|7.5) + (.5, 0)$)
            node [midway, rotate=0, anchor = west] {\scriptsize$\ord + 1$}
        ;
        \tikz \draw[<->, gr, shorten <> = .2em]
            (7-|1) ++(0, -.2)
            -- ($(7-|7) + (0, -.2)$)
            node [midway, anchor = north] {\scriptsize$\ord + 1$}
        ;
    \end{bNiceArray}
    \hspace{2cm}
    \mathcal{L}\sbrk{\partial_{1}}
    =
    \NiceMatrixOptions{code-for-first-col = \scriptstyle}
    \begin{bNiceArray}{c ccc ccc}[margin, nullify-dots, first-col]
        \CodeBefore
        \Body
        \gr{0 \rightarrow}
        & \gr{0} & 1 & & & &
        \\
        \gr{1 \rightarrow}
        & & \gr{0} & 2 & & &
        \\
        \gr{2 \rightarrow}
        & & & \gr{0} & 3 & &
        \\
        & & & & \Ddots[color = gr, shorten-end=1mm] & \Ddots &
        \\
        & & & & & \gr{0} & \ord - 1
        \\
        \gr{\ord \rightarrow}
        & & & & & & \gr{0}
    \CodeAfter
        \tikz \draw[
            gr,
            line cap = round,
            line width = 1.06pt,
            dash pattern= on 0pt off .18cm
        ]
            (1|-4) ++(-.5, 0)
            -- ++($(1|-6.5) - (1|-4) + (0, .2)$)
        ;
        \tikz \draw[<->, gr, shorten <> = .5em]
            (1-|7.5) ++(.5, 0)
            -- ($(7-|7.5) + (.5, 0)$)
            node [midway, rotate=0, anchor = west] {\scriptsize$\ord + 1$}
        ;
        \tikz \draw[<->, gr, shorten <> = .2em]
            (7-|1) ++(0, -.2)
            -- ($(7-|7) + (0, -.2)$)
            node [midway, anchor = north] {\scriptsize$\ord + 1$}
        ;
    \end{bNiceArray}
    \hspace{.5cm}
    \label{eq:cmats_weyl}
\end{align}
\vspace{.5cm}

From these we can write any rational function in its companion matrix form, through matrix multiplication.
The only caveat is the existence of poles, for example, consider the following rational function that contains a $p^{\text{th}}$ order pole
\begin{align}
    r\brk{z_{1}}
    =
    \frac{f\brk{z_{1}}}{g\brk{z_{1}}}
    =
    \frac{
        \sum_{i=0}^{\infty}f_{i} \, z_{1}^{i}
    }{
        z_{1}^{p}\sum_{j=0}^{\infty} g_{i} \, z_{i}
    }
    \,,
\end{align}
where $g_{0}\neq0$.
Naively finding the companion matrix representation of this function would give us
\begin{align}
    \mathcal{L}\sbrk{r\brk{z_{1}}}
    =
    \frac{
        \sum_{i=0}^{\infty} f_{i} \, \mathcal{L}\sbrk{z_{1}}^{i}
    }{
        \mathcal{L}\sbrk{z_{1}}^{p}\sum_{j=0}^{\infty} g_{j} \, \mathcal{L}\sbrk{z_{1}}^{j}
    }
    =
    \mathcal{L}\sbrk{z_{1}}^{-p}
    \cdot
    \frac{
        \sum_{i=0}^{\infty} f_{i} \, \mathcal{L}\sbrk{z_{1}}^{i}
    }{
        \sum_{j=0}^{\infty} g_{j} \, \mathcal{L}\sbrk{z_{1}}^{j}
    }
    \,.
\end{align}
This raises a problem, as $\mathcal{L}[z_{1}]$ is not invertible, so we cannot raise it to negative powers.
Therefore we must define an inverse companion matrix
\begin{align}
    \mathcal{L}\sbrk{z_{1}^{-1}}
    =
    \NiceMatrixOptions{code-for-first-col = \scriptstyle}
    \begin{bNiceArray}{c ccc ccc}[margin, nullify-dots, first-col]
        \CodeBefore
        \Body
        \gr{0 \rightarrow}
        & \gr{0} & 1 & & & &
        \\
        \gr{1 \rightarrow}
        & & \gr{0} & 1 & & &
        \\
        \gr{2 \rightarrow}
        & & & \gr{0} & 1 & &
        \\
        & & & & \Ddots[color = gr, shorten-end=1mm] & \Ddots &
        \\
        & & & & & \gr{0} & 1
        \\
        \gr{\ord \rightarrow}
        & & & & & & \gr{0}
    \CodeAfter
        \tikz \draw[
            gr,
            line cap = round,
            line width = 1.06pt,
            dash pattern= on 0pt off .18cm
        ]
            (1|-4) ++(-.5, 0)
            -- ++($(1|-6.5) - (1|-4) + (0, .2)$)
        ;
    \end{bNiceArray}\,.
\end{align}
Now we replace any poles of order $p$ by $p$ powers of this matrix, and so any rational function can be written in companion matrix representation
\begin{align}
    \mathcal{L}\sbrk{r\brk{z_{1}}}
    =
    \mathcal{L}\sbrk{z_{1}^{-1}}^{p}
    \cdot
    \Bigbrk{\sum_{j=0}^{\infty}g_{j} \, \mathcal{L}\sbrk{z_{1}}^{j}}^{-1}
    \cdot
    {\sum_{i=0}^{\infty} f_{i} \, \mathcal{L}\sbrk{z_{1}}^{i}}
    \,,
\end{align}
and as long as $g_{0}\neq0$, the denominator will be invertible.

\paragraph{Waveform from companion tensors.} We can now use this companion tensor formalism to extract the coefficients in the expansion of $\mathbf{W}^{(0)}$. This is given by the system of equations
\begin{equation}
    \begin{aligned}
        \mathbf{W}^{(0)}-\mathbf{A}\cdot F &= 0\,,\\
        (\partial_{1}\mathbb{1}_{\nu}-\Omega_{F})\cdot F &=0\,,
    \end{aligned}
\end{equation}
Similarly to what is done in the framework of intersection theory \cite{Brunello:2024tqf}, we can recast this as a system of linear equations by converting these operations on $F$ into companion matrix operators
\begin{equation}
    \begin{aligned}
    \mathbf{W}^{(0)}-\mathcal{L}\sbrk{\mathbf{A}} \cdot F &= 0
    \,,
    \\
    \left(\mathcal{L}[\partial_{1}]\mathbb{1}_{\nu}-\mathcal{L}[\Omega_{1}]\right)\cdot F &=0
    \,.
    \end{aligned}
\end{equation}
We can then reformulate these conditions as

\begin{align}
    \begin{bNiceArray}{ccw{c}{5cm}c}[margin]
        \mId & \Block{1-3}{
            -\mathcal{L}\sbrk{\mathbf{A}}
        } & &
        \\
        \mZero & \Block{3-3}{
                \mathcal{L}\sbrk{\partial_{1}} \, \mId_{\nu} - \mathcal{L}\sbrk{\Omega_{1}}
        } & &
        \\
        \Vdots[color = gr] & & &
        \\
        \mZero & & &
        \CodeAfter
            \tikz \draw[
                line width=.4pt, gr,
            ]
                (2-|1) -- (2-|5)
            ;
            \tikz \draw[
                line width=.4pt, gr,
            ]
                (1-|2) -- (5-|2)
            ;
    \end{bNiceArray}
    \cdot
    \begin{bNiceMatrix}[margin]
        \mathbf{W}\supbrk{0}
        \\
        \Block{3-1}{F}
        \\
        \\
        \\
        \CodeAfter
            \tikz \draw[
                line width=.4pt, gr,
            ]
                (2-|1) -- (2)
            ;
    \end{bNiceMatrix}
    =
    0
    \label{eq:ctensor_system}
    \,.
\end{align}

The problem is then reduced to row reduction of this matrix.
In practice we formulate this whole framework within \texttt{FiniteFlow} \cite{Peraro:2019svx}, and the new package for polynomial reduction and elimination theory over finite fields, \texttt{SPQR} \cite{Chestnov:2025todo}.